\documentclass[twoside,11pt]{article}

\usepackage{jmlr2e}

\ShortHeadings{Multi-Stage Adaptive Testing}{Wang, Gang and Sun}
\firstpageno{1}

\usepackage{caption}
\usepackage{amssymb}
\usepackage{amsbsy}
\usepackage{amsthm}
\usepackage{bm}
\usepackage{bbm}
\usepackage{xcolor}
\usepackage{float}
\usepackage[nointegrals]{wasysym}
\usepackage{enumitem}
\usepackage{caption}
\usepackage{subcaption}
\usepackage{textcomp}
\usepackage{thmtools, thm-restate}
\usepackage{mathtools}
\usepackage{tikz}
\usepackage{algpseudocode}
\usepackage{algorithm}

\newtheorem{theorem}{Theorem}
\newtheorem{assumption}{Assumption}
\newtheorem{corollary}{Corollary}
\newtheorem{proposition}{Proposition}
\newtheorem{lemma}{Lemma}
\newtheorem{example}{Example}

\newtheorem{remark}{Remark}
\newtheorem{definition}{Definition}
\newcommand{\beq}{\begin{equation}}
	\newcommand{\eeq}{\end{equation}}
\newcommand{\beas}{\begin{eqnarray*}}
	\newcommand{\eeas}{\end{eqnarray*}}
\newcommand{\bea}{\begin{eqnarray}}
	\newcommand{\eea}{\end{eqnarray}}
\newcommand{\bei}{\begin{itemize}}
	\newcommand{\eei}{\end{itemize}}
\newcommand{\ben}{\begin{enumerate}}
	\newcommand{\een}{\end{enumerate}}
\newcommand{\bet}{\begin{theorem}}
	\newcommand{\eet}{\end{theorem}}
\newcommand{\bel}{\begin{lemma}}
	\newcommand{\eel}{\end{lemma}}
\newcommand{\bep}{\begin{proposition}}
	\newcommand{\eep}{\end{proposition}}
\newcommand{\bed}{\begin{definition}}
	\newcommand{\eed}{\end{definition}}
\newcommand{\bec}{\begin{corollary}}
	\newcommand{\eec}{\end{corollary}}
\newcommand{\bex}{\begin{example}}
	\newcommand{\eex}{\end{example}}

\newcommand{\EE}{\mathbb{E}}
\newcommand{\II}{\mathbb{I}}

\begin{document}

	\title{Sparse Recovery With Multiple Data Streams: An Adaptive Sequential Testing Approach}
	
	\author{\name Weinan Wang \email wwang@snap.com\\
		\addr Snap Inc.\\
		2850 Ocean Park Blvd.\\
		Santa Monica, CA, 90405, USA\\
		\AND
		\name Bowen Gang \email bgang@fudan.edu.cn\\
		\addr Department of Statistics and Data Science\\
		Fudan University\\
		Shanghai, 200433, China\\
		\AND
		\name Wenguang Sun \email wgsun@zju.edu.cn\\
		\addr Center for Data Science\\
		Zhejiang University\\
		Hangzhou, 310013, China}
	
	\editor{}
	\maketitle

	\begin{abstract}
		Multistage design has been used in a wide range of scientific fields. By allocating sensing resources adaptively, one can effectively eliminate null locations and localize signals with a smaller study budget. We formulate a decision-theoretic framework for simultaneous multi-stage adaptive testing and study how to minimize the total number of measurements while meeting pre-specified constraints on both the false positive rate (FPR) and missed discovery rate (MDR). The new procedure, which effectively pools information across individual tests using a simultaneous multistage adaptive ranking and thresholding (SMART) approach, controls the error rates and leads to great savings in total study costs. Numerical studies confirm the effectiveness of SMART. The SMART procedure is illustrated through the analysis of large-scale A/B tests, high-throughput screening and image analysis.
	\end{abstract}

	\medskip
	
	
	\section{Introduction}
	
Suppose we are interested in recovering the support of a high-dimensional sparse vector $\pmb\mu=(\mu_1, \cdots, \mu_p)\in \mathbb R^p$ based on measurements from variables $X_1$, $\cdots$, $X_p$. Let $\mathcal S=\{i: \mu_i\neq 0\}$ denote the support of $\pmb\mu$. We focus on a setup where measurements on variables are performed sequentially. 

\subsection{A multistage mixture model}\label{sec:1.1}

We assume that the data obey a multistage random mixture model, which can be described using a hierarchical approach. Denote $\mathbb I(\cdot)$ an indicator function and let $\theta_i=\mathbb{I}\{i\in \mathcal{S}\}$. Then $\theta_i$ is a Bernoulli variable obeying  
\begin{equation}\label{mix1}
			\theta_i\sim \mbox{Bernoulli}(\pi), \quad i=1,\cdots, p,
\end{equation}
where $\pi$ is the expected fraction of non-zero coordinates. The unobserved effect sizes $\mu_i$ are generated conditional on $\theta_i$: 
\begin{equation}\label{mix2}
  \mbox{$\mu_i=0$ if $\theta_i=0$\; and \; $\mu_i\sim G(\cdot)$ if $\theta_i=1$, }
\end{equation}
where $G(\cdot)$ is the non-null distribution of $\mu_i$.  We do not impose any parametric assumptions on $g$; Section \ref{estimation.subsec} discusses a nonparametric approach to estimating $g(\cdot)$, the density of $G(\cdot)$. 
The observation model is given by
		\begin{equation}\label{mix3}
			X_{ij}=\mu_i+\epsilon_{ij}, \quad \epsilon_{ij}\sim \mathcal N(0, \sigma^2), \quad i=1,\cdots, p; \; j=1, 2, 3,\cdots.
		\end{equation}
The measurements are taken in stages $j=1, 2, \cdots$, where $X_{ij}$ follows the null distribution $F_0=\mathcal N(0, \sigma^2)$ if $i\notin \mathcal S$ and the non-null distribution $F_{1i}=\mathcal N(\mu_i, \sigma^2)$ if $i\in \mathcal S$. Denote $f_0$ and $f_{1i}$ the corresponding densities. The model assumes that $F_0$ is identical for all $i\notin \mathcal S$, whereas $F_{1i}$ can vary across $i\in \mathcal S$. Allowing for heterogeneous $F_{1i}$ is desirable in applications where non-zero coordinates have different effect sizes. 
The fraction $\pi$ is usually small. 
This model has been widely used in large-scale inference problems (e.g. \citealp{efron2016deconv}) by providing a flexible framework for a range of estimation and testing problems. In this article, we study the \emph{sparse recovery} problem, which aims to recover the subset $\mathcal S\subset\{1, \cdots, p\}$ with negligible decision errors. 

	\subsection{Non-adaptive vs. adaptive designs} 
	
	Consider a setting where each measurement is associated with a fixed cost; hence collecting many repeated measurements on \emph{all} variables are prohibitively expensive in large-scale studies where $p$ is in the order of thousands or even millions. 
	
	In applications with a sequential design, it may not be necessary to measure all features at all stages. Let $\mathcal A_j \subseteq \mathcal A_0\coloneqq \{1, \cdots, p\}$, $j=1, 2, \cdots$, be a monotone sequence of nested sets, with $\mathcal A_j\subseteq A_{j-1}$ and $\mathcal A_j$ denoting  the set of coordinates on which measurements are performed at stage $j$. Denote $X_{\mathcal A_j}=(X_{ij}: i\in \mathcal A_j)$ the corresponding collection of observations. In a non-adaptive setting, $X_{ij}$ are sampled at every stage following a pre-fixed policy and each coordinate $i$ is expected to receive the same amount of measurement budget. In an \emph{adaptive sampling design} \citep{MalNow14a, MalNow14b, li2015identifying, jain2004adaptive}, the sampling scheme varies across different coordinates, with the flexibility of adjusting $\mathcal A_j$  in response to the information collected at previous stages.
	
	The adaptive sampling and inference framework provides a powerful approach to sparse  estimation and testing problems. Intuitively, the sensing resources in later stages can be allocated in a more cost-effective way to reflect our updated contextual knowledge during the course of the study; hence greater precision in inference can be achieved with the same study budgets or computational costs. A plethora of powerful multistage testing and estimation procedures have been developed under this flexible framework; some recent developments include the hierarchical testing procedures for pattern recognition \citep{BlaGem05, Meietal09, SunWei15}, distilled sensing and sequential thresholding methods for sparse detection \citep{Hauetal11, MalNow11, MalNow14a}, multi-scale search and open-loop feedback control algorithms for adaptive estimation \citep{Basetal11, WeiHer13, WeiHer15} and sequentially designed compressed sensing \citep{Hauetal12, MalNow14b}. These works demonstrate that methodologies adopting adaptive designs can substantially outperform those developed under non-adaptive settings. 
	
	Our goal is to develop a cost-effective multistage sampling and inference procedure to narrow down the focus in a sequential manner to identify the vector support reliably.  
	The proposed strategy consists of (a) a stopping rule for selecting the active coordinates, and (b) a testing rule for deciding whether a coordinate contains a signal.

	\subsection{Applications and statistical challenges}
	
	Multistage experiments have been widely used in many scientific fields including large-scale A/B testing \citep{Joh15}, environmental sciences \citep{Cor88, ThoSeb96}, microarray, RNA-seq, and protein array experiments \citep{Muletal04, RosMul13}, geostatistical analysis \citep{Roe93, Bloetal02}, genome-wide association studies \citep{Satetal04, Rotetal10} and astronomical surveys \citep{Meietal09}. 
	We first describe some applications and then discuss related statistical issues.
	
	\begin{itemize}
		
		\item \emph{High-throughput screening} [HTS, \cite{Zhaetal99, Bletal03}]. HTS is a large-scale hierarchical process that has played a crucial role in fast-advancing fields such as stem cell biology and drug discovery. In drug discovery, HTS involves testing a large number of chemical compounds in multiple stages including target identification, assay development, primary screening, confirmatory screening, and follow-up of hits. The accurate selection of useful compounds is an important issue at each aforementioned stage of HTS. For example, at the primary screening stage, a library of compounds is tested to generate an initial list of active compounds (hits). The goal of this stage is to reduce the size of the library significantly with negligible false negative rate. In the confirmatory screening stage, the hits are further investigated to generate a list of confirmed hits (leads), which will be used to develop drug candidates. As the lab costs for leads generation are very high, an important task at the confirmatory stage is to construct a subset with negligible false positive rate while keeping as many useful compounds as possible. 
		
		\item \emph{Large-scale A/B testing.} A/B testing has begun to be widely deployed by various high-tech firms for identifying features/designs that work the best among a large number of potential candidates. It provides a powerful tool for testing new ideas for a wide range of real-world decision-making scenarios \citep{kohavi2017online, xu2015infrastructure, fabijan2018online}. For example, A/B testing can offer key insights on (a) how to improve the users' experiences, (b) what new features should be incorporated into the product to boost the profits, and (c) which design is the most effective for getting more people click the sign-up button. A significant challenge in large-scale A/B testing is the multiplicity issue, i.e. we need to conduct hundreds of experiments simultaneously for a long period, and each experiment may involve thousands of metrics to be evaluated throughout the entire duration. An effective multistage design can substantially save study costs, minimize risk exposures to customers and enable faster decision making. 
		
		\item \emph{Ultra-high dimensional testing in astronomical surveys}. The fast and accurate localization of sparse signals poses a significant challenge in Astrophysics \citep{MeiRic06}. When a large number of images are taken with high frequencies, the computational cost of an exhaustive search through every single image and pixel can be prohibitively expensive as the search often involves testing billions of hypotheses. A multistage decision process can lead to great savings in total sensing  efforts by quickly narrowing down the focus to a much smaller subset of most promising spots in the images \citep{Djo12}.
		
	\end{itemize}
	
	These three applications will be discussed in detail in Section \ref{Apply.sec}. 
	
	
	\subsection{Problem formulation}
	
In the design and analysis of large-scale multistage studies, the inflation of decision errors and soaring study costs are among the top concerns. First, to identify useful signals effectively, we need to control the \emph{missed discovery rate} (MDR) to be small at all stages since missed signals will not be revisited in subsequent stages. Second, to reduce the {study costs} or increase the \emph{sampling efficiency}, it is desirable to eliminate as many null locations as possible at each stage. Finally, to avoid misleading scientific conclusions, the final stage of our analysis calls for a strict control of the \emph{false positive rate} (FPR). Next, we formulate a decision-theoretic framework that addresses the three issues (MDR control, FPR control and sampling efficiency) integrally.
	
When many coordinate-wise sequential decisions are made simultaneously, the control of inflated decision errors becomes a critical issue. Let $\pmb\delta=(\delta_1, \cdots, \delta_p)$, where $\delta_i=0/1$ indicates that $i$ is classified as a null/non-null case. 
Define the false positive rate (FPR) and missed discovery rate (MDR) as 
	\beq\label{FPR-MDR}
	\mbox{FPR}({\pmb \delta})=\frac{\mathbb{E}\left\{\sum_{i=1}^p (1-\theta_i) \delta_{i}\right\}}{\mathbb{E}(\sum_{i=1}^p \delta_{i})}\; \mbox{ and } \;  \mbox{MDR}({\pmb \delta}) =\frac{\mathbb{E}\left\{\sum_{i=1}^p \theta_i(1-\delta_{i})\right\}}{\mathbb{E}(\sum_{i=1}^p \theta_i)}.
	\eeq
	
	\begin{remark}\rm{
			The FPR is also referred to as the marginal false discovery rate (mFDR; \citealp{GenWas02, SunCai07}). In the non-sequential setting, the FPR and the widely used false discovery rate (FDR; \citealp{BenHoc95}) are shown to be asymptotically equivalent for the Benjamini-Hochberg procedure \citep{GenWas02} and a general class of FDR procedures (Proposition 7 in Appendix A.2 in \cite{Caietal19}). Such asymptotic equivalence requires further research under the sequential setting. The main consideration of using FPR (as opposed to FDR) is to develop optimality results under a decision-theoretic framework \citep{Ber85}. Theorem \ref{thm:VALID} shows that our proposed method controls both the FPR and FDR at user-specified levels.  An alternative measure to the MDR is the false negative rate (FNR; \citealp{GenWas02, Sar04}). The concepts of FNR and MDR are different, with the MDR being a more appropriate error measure in the sparse setting; see \cite{MeiRic06, Hauetal11, CaiSun17} for related discussions. }
	\end{remark}
	
We define the expected sample size (ESS) per unit as a criterion for comparing the efficiency of different multistage testing methods:
	\beq\label{ESS}
	\mbox{ESS}(\pmb N)=\mathbb{E}\left(p^{-1}\textstyle\sum_{i=1}^p N_i \right),
	\eeq
	where $\pmb N=(N_1, \cdots, N_p)$, with $N_i$ being the sample size [also referred to as the stopping time \citep{Ber85, Sie85}, defined formally in Section \ref{dt-form.subsec}]  at coordinate $i$. Let $\alpha$ and $\gamma$ be the user-specified FPR and MDR levels. We study the following constrained optimization problem:
	\begin{equation}\label{optimal-seq-proc.def}
		\mbox{ minimize } \mbox{ESS}(\pmb N) \text{ subject to } \mbox{FPR}(\pmb \delta) \leq \alpha \; \mbox{and} \; \mbox{MDR}(\pmb \delta) \leq\gamma.
	\end{equation}
The formulation \eqref{optimal-seq-proc.def} reflects the key role of adaptive sampling, which effectively reduces the ESS by allowing one to stop sampling early at certain coordinates. 

The formulation \eqref{optimal-seq-proc.def} naturally extends the classical sequential testing problem, considered in \cite{Wal45} and \cite{Sie85}, to the compound decision setting \citep{Rob51}. Specifically, the sequential sparse recovery problem, which involves adaptive sampling at many coordinates simultaneously, can be viewed as a \emph{compound decision problem}. Each component problem corresponds to the sequential testing of a single hypothesis $H_{0}: \mu=0$ vs. $H_{1}: \mu\neq 0$ based on a data stream. In the seminal work of \cite{Wal45}, the following constrained optimization problem is studied:
	\beq\label{formulation-single}
	\mbox{ minimize } \mathbb E(N) \text{ subject to } \mathbb P_{H_0}(\mbox{Reject}) \leq \alpha^\prime \; \mbox{and} \; \mathbb P_{H_1}(\mbox{Accept}) \leq\gamma^\prime.
	\eeq
Here $N$ is the stopping time, and $\alpha^\prime$ and $\gamma^\prime$ are pre-specified  Type I and Type II error rates. $\mathbb E(N)$, which represents the average sampling costs, characterizes the efficiency of a sequential procedure. 
The connection between the sparse recovery problem and sequential testing becomes clear by comparing \eqref{formulation-single} with \eqref{optimal-seq-proc.def}. 

The sequential probability ratio test (SPRT) is shown to be \emph{optimal} \citep{Wal45, Sie85} for the single sequential testing problem \eqref{formulation-single} in the sense that it has the smallest $\mathbb E(N)$ among all sequential procedures at level $(\alpha^\prime, \gamma^\prime)$. This theory will be extended to the multiple sequential testing setup in Section \ref{dt-form.subsec}. 
			
	\subsection{A preview of the proposed algorithm}

	We formulate the sparse recovery problem as a sequential testing problem with multiple data streams. A new adaptive testing procedure is developed under the compound decision-theoretic framework. The proposed procedure, which employs a simultaneous multistage adaptive ranking and thresholding (SMART) approach, not only uses all information collected through multiple stages but also exploits the compound structure of the decision problem to pool information across different coordinates. We show that SMART controls the FPR and MDR at user-specified levels and achieves the information-theoretic lower bounds. SMART is simple, fast, and capable of dealing with millions of tests. Numerical studies confirm the effectiveness of SMART for error rates control and demonstrate that it leads to substantial savings in study costs.

\begin{figure}[ht]

		\includegraphics[width=1\textwidth]{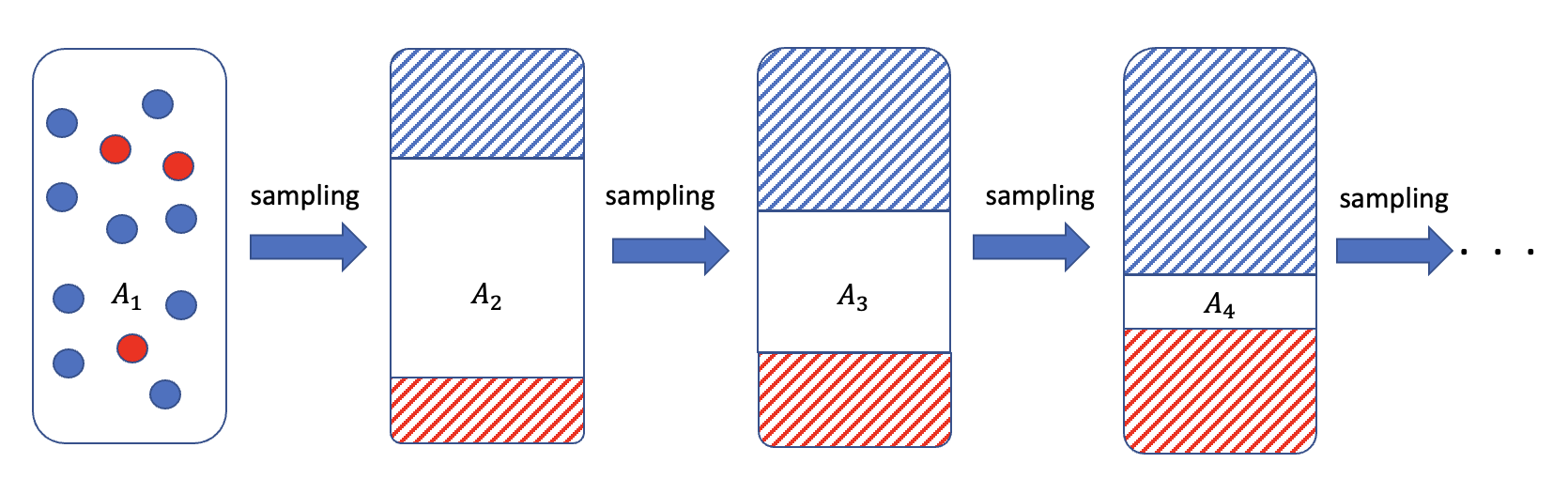} 
	
	\caption{A schematic illustration of the SMART procedure: the blue, red and white areas respectively represent the subset of acceptances, rejections and indecisions. The active sampling set $\mathcal A_j$ shrinks over time.}\label{sm.illus}
\end{figure}

A schematic illustration of the SMART algorithm is provided in Figure \ref{sm.illus}. At each stage, SMART determines two data-driven thresholds, which divide the ordered test statistics into three categories and thus lead to three types of decisions. The first type identifies non-null cases, which form the subset of rejections (the red area at the bottom); the second type eliminates null cases, which form the  subset of acceptances (the blue area on the top); and the third type selects coordinates for further measurements, which form $\mathcal A_j$, the subset of indecisions at stage $j$ (the white area in the middle). We only take new observations in the active set, which shrinks over the sampling process until it becomes empty. The shaded areas (blue and red) correspond to the savings in study costs. 
	
\subsection{Some theoretical insights}
	
We first review the limits for fixed and adaptive designs in the literature and then compare them with our new limits derived based on adaptive sampling and SMART. Consider a two-point normal mixture model under a single-stage design
	\beq\label{nmix}
	X_i \stackrel{i.i.d.}{\sim} (1-\pi_p)N(0, 1)+\pi_p N(\mu_p, \sigma^2),
	\eeq
	where $\pi_p=p^{-\beta}$, $\mu_p=\sqrt{2r\log p}$ and $0<\beta<1, 0<r<1$, $\sigma^2$ is unknown. The fundamental limits for a range of global and simultaneous inference problems have been derived under this setup \citep{DonJin04, Caietal07, CaiSun17}. 
	Of particular interest is the \emph{classification boundary} \citep{MeiRic06, Hauetal11, CaiSun17}, which demarcates the possibility of constructing a subset with both the FPR and MDR tending to zero. Under model \eqref{nmix}, the classification boundary is a straight line $r=\beta$ in the $\beta$-$r$ plane for both the homoscedastic case $\sigma=1$ \citep{MeiRic06, Hauetal11} and heteroscedastic case $\sigma\neq 1$ \citep{CaiSun17}. Hence the goal of $R^*\equiv\mbox{FPR}+\mbox{MDR}\rightarrow 0$ requires that the signal magnitude $\mu_p$ must be at least in the order of $\sqrt{\log p}$. This gives the fundamental limit of sparse recovery for fixed designs. 
	
	
	The rate $\sqrt {\log p}$ can be substantially improved in the adaptive setting. For example, \cite{Hauetal11} proposed the \emph{distilled sensing} (DS) method, which is capable of achieving the classification boundary with much weaker signals. DS is a multistage testing procedure with a total measurement budget of $2p$. It assumes that observations follow a mixture model with noise distributed as standard normal. At each stage, DS keeps locations with positive observations and obtain new observations for these locations in the next stage. It was shown in \cite{Hauetal11} that after $k=\max\{\lceil \log_2\log p\rceil, 0\}+2$ steps, the DS algorithm successfully constructs a subset with both FPR and MDR tending to zero provided that $\mu_p$ \emph{diverges at an arbitrary rate} in the dimension $p$.
	
	A more general result on the limit of sparse recovery was obtained in \cite{MalNow14a}, where the KL divergence and the average measurements per dimension $\tau$ are used in place of the growing signal amplitude $\mu_p$ to characterize the difficulty of the problem. 
	Let $s$ denote the cardinality of the support. It was shown in \cite{MalNow14a} that under fixed designs, the reliable recovery requires that $\tau$ must be at least in the order of $\log p$, whereas under adaptive designs, the required $\tau$ is in the order of $\log s$, where $s=|\mathcal S|$ is the cardinality of the support of $\mathcal \mu$. This reveals the advantage of adaptive designs, noting that $\log s=(1-\beta)\log p$ under the calibration $\pi_p=p^{-\beta}$. The single sequential thresholding method (SST)  proposed in \cite{MalNow14a} is optimal in the sense that it achieves the rate of $\log s$ asymptotically \citep{MalNow14a}. It is important to note that the theory in \cite{MalNow14a} is developed under the more stringent family-wise error rate (FWER) paradigm. However, in large-scale testing problems, it is desirable to use a more practical and powerful error rate such as the FDR \citep{BenHoc95} or its asymptotically equivalent version FPR \eqref{FPR-MDR}. 
	
We investigate the upper and lower bounds under the the FPR-MDR paradigm and show that this  paradigm requires fewer samples to guarantee that $R^*\rightarrow 0$. The result strengthens the rate obtained by  \cite{Hauetal11} under model \eqref{nmix}. SMART only requires a $\tau$ of the order $\log {\xi}(p)$, where ${\xi}(p)$ is a function that diverges to $\infty$ at any arbitrarily slow rate. This slightly improves the rate of $\log{s}$ achieved by SST. The improvement is expected as SST is developed to control the more stringent FWER. Hence our theory is revealing and novel in charactering the merits of adaptive designs under a more relevant error rate notion in large-scale inference problems.

	\subsection{Main contributions}
	
	The error control in multi-stage testing has been studied extensively in \cite{BenHel07, VicHom07, Dmietal07, GoeMan08, Yek08,  LiaNet10, BenBog14}. However, existing methods are usually designed for different types of applications and cannot be easily tailored for the problem under consideration. The biggest limitation is that existing works only focus on the control of false positive errors, whereas the sparse recovery problem also calls for the control of missed discoveries. Moreover, the ranking and stopping rules proposed in previous studies \citep{Zehetal08, Posetal09, Saretal13} are considered suboptimal. This is because the data combination methods that rely on $p$-values do not fully capitalize on the information provided by different stages, and they fail to effectively exploit the overall structure of the compound decision problem. Furthermore, recent research on sequential testing using SPRT rules \citep{BarSon13, Bar14, BarSon16} involves significant computational complexity, rendering it impractical for large-scale studies, such as those encountered in HTS and Astrophysics, where millions of tests are conducted simultaneously. For a more comprehensive discussion, please refer to Section \ref{comparisons.subsec}. The issue of optimality, in particular, has not been addressed in the aforementioned studies.
	
	SMART has several advantages over existing methods for sparse recovery such as DS \citep{Hauetal11}) and simple sequential thresholding (SST, \citealp{MalNow14a}). First, DS and SST use fixed thresholding rules that do not offer accurate error rate control. Second, although DS and SST employ a multistage design, they are inefficient because the stopping and testing rules at stage $k$ only depend on the observations at the current stage, and the observations from previous stages are abandoned. Finally, as opposed to DS and SST that ignore the compound decision structure, SMART employs the ranking and compound thresholding idea in the multiple testing literature \citep{SunCai07} to pool information across individual tests. Consequently, SMART controls the error rates more precisely with substantial efficiency gain. 	
	
This article presents several technical/theoretical contributions. Firstly, the application of compound sequential decision theory, which has received limited attention in the multi-armed bandit literature, effectively resolves the issue of overshooting and significantly improves the efficiency of sequential methods. Secondly, proving the optimality results is a challenging task. One key technique employed is the argument based on swapping, complemented by the use of proof by contradiction. Lastly, in Theorem 3, our proof leverages the relationship between the MDR and the False Negative Rate (FNR). The concept of FNR is particularly well-suited for multi-stage analysis. The relationship not only simplifies our current analysis but also serves as a valuable tool for addressing related problems.

	\subsection{Organization of the paper}
	In Section \ref{procs.sec} we first formulate a compound decision-theoretic framework for sequential sparse recovery problems, and then develop oracle rules for FPR and MDR control. Section \ref{SMART.sec} proposes the SMART procedure and discusses its implementation. Section \ref{Limits.sec} derives the fundamental limits for sparse inference and establishes the asymptotic optimality of SMART. In Section \ref{Simu.sec}, we conduct numerical experiments to compare SMART with competitive methods. Section \ref{Apply.sec} applies SMART for analyzing data from large-scale A/B testing, high-throughput screening and astronomical surveys. Section \ref{discuss:sec} concludes the article with a discussion of related works and future directions. 
	The technical derivations and proofs are given in Section \ref{derivation-thresholds.subsec} and Section \ref{Proof.sec}, respectively. 
	Additional numerical results are provided in the Supplementary Material.
	
	\setcounter{equation}{0}

	\section{Oracle Rules for Sparse Recovery}\label{procs.sec}
	
	The sparse recovery problem in the adaptive setting involves the simultaneous testing of $p$ hypotheses: 
	\beq\label{multi-testing}
	H_{0, i}: \mu_i=0  \mbox{ vs. } H_{1, i}: \mu_i\neq 0, \quad 1\leq i\leq p,
	\eeq
	based on $p$ streams of observations $\{(X_{i1}, X_{i2}, \cdots): 1\leq i\leq p\}$. 
	In this section, we first derive an oracle procedure for FPR and MDR control (Section \ref{dt-form.subsec}), then we propose practical strategies to implement the oracle procedure (Sections \ref{thresholds.sec}-\ref{approx.subsec}).

	\subsection{Oracle procedure}\label{dt-form.subsec}
	
	Let $X_{ij}$ be the measurement of variable $X_i$ at stage $j$ and $\pmb X_i^j=(X_{i1}, \cdots, X_{ij})$  the collection of measurements on $X_i$ up to stage $j$. A {multistage decision procedure} involves choosing a \emph{stopping rule} and a \emph{testing rule} for each location. At location $i$, the stopping rule  $\pmb\tau_i$ 
	consists of a series of functions $\tau_{i1}(\pmb X_i^1)$, $\tau_{i2}(\pmb X_i^2)$, $\cdots$,
	where $\tau_{ij}$ takes values in $\{0, 1\}$, with 0 and 1 standing for ``taking another
	observation'' and ``stopping sampling and making a decision'', respectively. We consider a class of multistage designs where the focus is sequentially narrowed down, i.e. the active sets satisfy $\mathcal A_1=\{1, \cdots, p\}$, and  $\mathcal A_j  \subseteq \mathcal A_{j-1}$ for $j=2, 3, \cdots$. At every coordinate $i\in\mathcal A_j$, there are three possible actions: (i) stop sampling and claim $H_{0, i}$ is true; (ii) stop sampling and claim $H_{0, i}$ is false; and (iii) do not make a decision and take another observation. We first focus on the case where one sample is collected per stage; the case with multiple samples is discussed in Section \ref{estimation.subsec}.

	The monotonicity condition on $(A_j)$ is a critical requirement in multi-stage analysis, such as distilled sensing, and plays a crucial role in resource-sensitive scenarios like drug discovery and resume screening. In drug discovery, screening steps aim to shortlist a few candidates out of thousands or millions, making the monotonicity condition essential. In HTS, promising compounds are selected for costly further testing to confirm their effectiveness. The resource-sensitive nature of HTS necessitates the avoidance of revisiting unpromising compounds to save study costs. Consequently, a monotone structure naturally emerges within the active sets $(A_j)$. 
	
	We follow the standard notations under the decision-theoretic framework detailed in Chp. 7 of \cite{Ber85}. The stopping rule $\pmb\tau_i$ can be equivalently described by the {stopping time} $N_i=\min\{n\geq 1: \tau_{in}(\pmb X_i^n)=1 \}$, the final stage at which we stop sampling and make a decision. If we collect one sample per stage, we use the terms \emph{sample size} and \emph{stopping time} interchangeably. The testing rule $\delta_i \in \{0, 1\}$ is carried out at the terminal sampling stage $N_i$, where $\delta_i=0$ (1) indicates that coordinate $i$ is classified as a null (non-null) case. A multistage decision procedure is therefore denoted $\pmb d=(\pmb N, \pmb\delta)$, where $\pmb N=(N_1, \cdots, N_p)$ and $\pmb \delta=(\delta_1, \cdots, \delta_p)$ are sample sizes and terminal decisions, respectively. Consider the following test statistic
	\beq\label{oracle-stat}
	T_{OR}^{i, j}=\mathbb{P}(\theta_i=0|\pmb X_i^j).
	\eeq
We view $T_{OR}^{i,j}$ as a significance index reflecting our confidence on claiming that $H_{0, i}$ is true based on all information available at stage $j$. 
	Let $t_l$ and $t_u$ be constants satisfying $0< t_l<t_u< 1$. Consider a class of sequential testing procedures $\pmb d^{\pi}(t_l, t_u)$ of the  form: 
	\begin{eqnarray}\label{d-proc}
		\mbox{ stop sampling for unit $i$ at }
		N_i=\min\{j\geq 1: T_{OR}^{i, j}\leq t_l \;\mbox{or} \; T_{OR}^{i,j} \geq t_u\},\nonumber\\
		\mbox{ and decide $\delta_{i}=1$ if $T_{OR}^{i, j}\leq t_l$ and $\delta_{i}=0$ if $T_{OR}^{i, j}\geq t_u$,}
	\end{eqnarray} 
where the superscript $\pi$ means we are considering the idealized case where $T_{OR}^{i, j}$ are known.	Denote by $\mathcal D_{\alpha, \gamma}$ the collection of all
	sequential decision procedures which simultaneously satisfy
	$
	(i)\; \mbox{FPR}({\pmb d})\leq\alpha \; \mbox{and}\; (ii)\; \mbox{MDR}({\pmb d})\leq \gamma.
	$
	Let $f(\pmb{X}^1_i|\theta_i)$ be the conditional density function evaluated at $\pmb{X}^1_i$ given $\theta_i$. 
	The following assumption, which is a standard condition in the sequential analysis literature (cf. \citealp{Ber85, Sie85}), requires that $f(\pmb X_i^1|\theta_i=0)$ and $f(\pmb X_i^1|\theta_i=1)$ differ with some positive probability. This ensures that the sequential testing procedure will stop within a finite number of stages almost surely.
	
	\begin{assumption}\label{stoppingtime}
		$\mathbb{P}_{\theta_i}\left(\log\left\{\frac{f\left(\pmb X_i^1|\theta_i=1\right)}{f\left(\pmb X_i^1|\theta_i=0\right)}\right\}=0\right)<1$ for $i=1, \cdots, p$.
	\end{assumption}

	The next theorem derives an oracle sequential testing procedure that provides the optimal solution to the constrained optimization problem \eqref{optimal-seq-proc.def}.
	\begin{restatable}{theorem}{theoopt}
		\label{thm:theoopt}	
		Consider the multistage model \eqref{mix1}--\eqref{mix3}, and a class of sequential testing rules $\pmb d^{\pi}(t_l, t_u)$ taking the form of \eqref{d-proc}. Denote $Q_{OR}(t_l, t_u)$ and $\tilde Q_{OR}(t_l, t_u)$ the FPR and MDR levels of $\pmb d^{\pi}(t_l, t_u)$, respectively. Then under Assumption \ref{stoppingtime}, we have
		\begin{description}
			\item (a). $Q_{OR}(t_l, t_u)$ is non-decreasing in $t_l$ for a fixed $t_u$, and $\tilde Q_{OR}(t_l, t_u)$ is non-increasing in $t_u$ for a fixed $t_l$.
			
			\item (b). Let $0<\alpha,\gamma<1$. 
			Then there exists a pair of oracle thresholds $(t_{OR}^l, t_{OR}^u)$, based on which we can define the oracle procedure 
			\beq\label{oracle-proc}
			\pmb d_{OR}\equiv\pmb d^{\pi}(t_{OR}^l, t_{OR}^u)
			\eeq 
			such that $\pmb d_{OR}$ is optimal in the sense that (i) $\mbox{FPR}({\pmb d_{OR}})\leq\alpha$; (ii) $\mbox{MDR}({\pmb d_{OR}})\leq\gamma$; and (iii) $\mbox{ESS}({\pmb d_{OR}})\leq \mbox{ESS}({\pmb d_*})$ for all $\pmb d_*\in\mathcal D_{\alpha, \gamma}$. 
		\end{description} 
	\end{restatable}
	
	\begin{remark}\rm{
			The development of a multiple testing procedure consists of two steps: deriving the optimal ranking statistic and finding a threshold along the ranking. Under the FPR criterion, the optimal ranking statistic is $T_{OR}$. The recent work by \cite{HelRos19} indicates that $T_{OR}$ is also optimal under the FDR criterion. The FPR criterion leads to fixed thresholds $t_{OR}^l$ and $t_{OR}^u$. By contrast, the theory in \cite{HelRos19} indicates that the thresholds must be functions of data when the FDR criterion is used. It is important to note that the oracle rule is only used to motivate our methodology in later sections.  Algorithm 1 in Section \ref{SMART.subsec} controls both the FDR and FPR, and employs data-driven thresholds that vary across data sets. More comparisons and discussions of the FDR and FPR criteria are provided in \cite{HelRos19} and Appendix. } 
	\end{remark}

The main goal of Theorem \ref{thm:theoopt} (as well as some later theorems) is to motivate our methodology and provide insight on the potential merits of the framework. Similar to the theories in \cite{Johetal22}, we have considered the ideal case with known $T^{i,j}_{OR}$, and the validity of the theorems does not depend on specific family of distributions. We develop data-driven procedures in Section \ref{estimation.subsec} for practical implementations where unknown quantities must be estimated.

	The oracle procedure $\pmb d_{OR}$ is a thresholding rule based on the oracle statistic $T_{OR}$ and oracle thresholds $t_{OR}^l$ and $t_{OR}^u$. 
	However, Theorem \ref{thm:theoopt} only shows the existence of $t_{OR}^l$ and $t_{OR}^u$, which are unknown in practice. In Section \ref{thresholds.sec}, we extend the classical ideas in \cite{Wal45} and \cite{Sie85} to derive approximations of $t_{OR}^l$ and $t_{OR}^u$.  
	
	\subsection{Approximation of oracle thresholds and its properties}\label{thresholds.sec}
	
	If we focus on a single location $i$, then the sequential probability ratio test (SPRT) \citep{Wal45} is a thresholding rule based on $
	\mathcal{L}_{i, j}= \frac{f(\pmb X_i^j|\theta_i=1)}{f(\pmb X_i^j|\theta_i=0)}
	$
	and of the form 
	$$
	\begin{array}{lll}
		\mbox{if $\log\mathcal{L}_{i, j} \leq a$}, & \mbox{stop sampling and decide $H_{0, i}$ is true;} \\ 
		\mbox{if $\log\mathcal{L}_{i, j} \geq b$,} &  \mbox{stop sampling and decide $H_{0, i}$ is false;} \\
		\mbox{if $a < \log\mathcal{L}_{i, j} < b$,} & \mbox{take another observation.}
	\end{array} 
	$$
	Let
	$\alpha^\prime=\mathbb P_{H_{0, i}}(\mbox{Reject $H_{0, i}$})$
	and 
	$\gamma^\prime=\mathbb P_{H_{1, i}}(\mbox{Accept $H_{0, i}$})$ be prespecified Type I and Type II error rates.
	Then by applying Wald's identity, the thresholds $a$ and $b$ can be approximated as: 
	\beq
	\tilde a = \log \frac{\gamma^\prime}{1-\alpha^\prime}\; \mbox{ and } \; \tilde b = \log \frac{1-\gamma^\prime}{\alpha^\prime}.
	\eeq
The oracle statistic 
	\begin{equation}\label{eq:torform}
	T_{OR}^{i, j}:=\mathbb{P}(\theta_i=0|\pmb{X}^j_i)=\dfrac{(1-\pi)f(\pmb X_i^j|\theta_i=0)}{(1-\pi)f(\pmb X_i^j|\theta_i=0)+\pi f(\pmb X_i^j|\theta_i=1)}=
	(1-\pi)/\{(1-\pi)+\pi \mathcal{L}_{i, j}\}
	\end{equation}
is monotone in $\mathcal{L}_{i, j}$. Therefore in the multiple hypothesis setting, we can view the oracle rule $\pmb d_{OR}=\pmb{d}^{\pi}(t_{OR}^l, t_{OR}^u)$ as $m$ parallel SPRTs. Then the problem boils down to how to obtain approximate formulas of the oracle thresholds $t_{OR}^l$ and $t_{OR}^u$ for a given pair of pre-specified FPR and MDR levels $(\alpha, \gamma)$. 
	In our derivation, the classical techniques (e.g. Section 7.5.2 in \cite{Ber85}) are used, and the relationships between the FPR and MDR levels $(\alpha, \gamma)$ and the Type I and Type II error rates $(\alpha^\prime, \gamma^\prime)$ are exploited. The derivation of approximation formulas with the FPR and MDR constraints is complicated; we provide the key steps in Section \ref{derivation-thresholds.subsec} of the Supplementary Material. The approximated thresholds are given by
	\begin{equation}\label{approx.formula}
		\tilde t_{OR}^l = \alpha \; \mbox{ and } \; \tilde t_{OR}^u =\frac{1-\pi}{\pi\gamma+1-\pi}.
	\end{equation}

	\begin{remark}
The thresholds in Equation \eqref{approx.formula} are chosen under the implicit assumption that $\alpha \leq (1-\pi)/(\pi\gamma+1-\pi)$. This assumption is relatively weak and holds true when $\pi<1/2$ and $\alpha<1/2$.
	\end{remark}

	The next theorem shows that the pair \eqref{approx.formula} is valid for FPR and MDR control.
	
	\begin{theorem}\label{threshold.thm}
Consider the multistage model \eqref{mix1}--\eqref{mix3}. 
		Denote $\tilde{\pmb d}_{OR}=\pmb{d}^{\pi}(\tilde t_{OR}^l, \tilde t_{OR}^u)$ the thresholding procedure that operates according to \eqref{d-proc} with upper and lower thresholds given by \eqref{approx.formula}. { Then we have}
		$$
		\mbox{$\mbox{FPR}(\tilde{\pmb d}_{OR})\leq\alpha$, $
			\mbox{FDR}(\tilde{\pmb d}_{OR})\leq\alpha$ and  $\mbox{MDR}(\tilde{\pmb d}_{OR})\leq\gamma.$ }
		$$
	\end{theorem}

	
\subsection{The overshooting problem}\label{approx.subsec}
	
	

	The approximations used in Equation \eqref{approx.formula} results in very conservative FPR and MDR levels. This is because of a phenomenon called ``overshooting,'' which we will explain. The technical material below only provides motivations for our SMART algorithm that leverages the compound structure of the sequential testing problem, and may be skipped during an initial read.

\begin{figure}
		\includegraphics[scale=0.5]{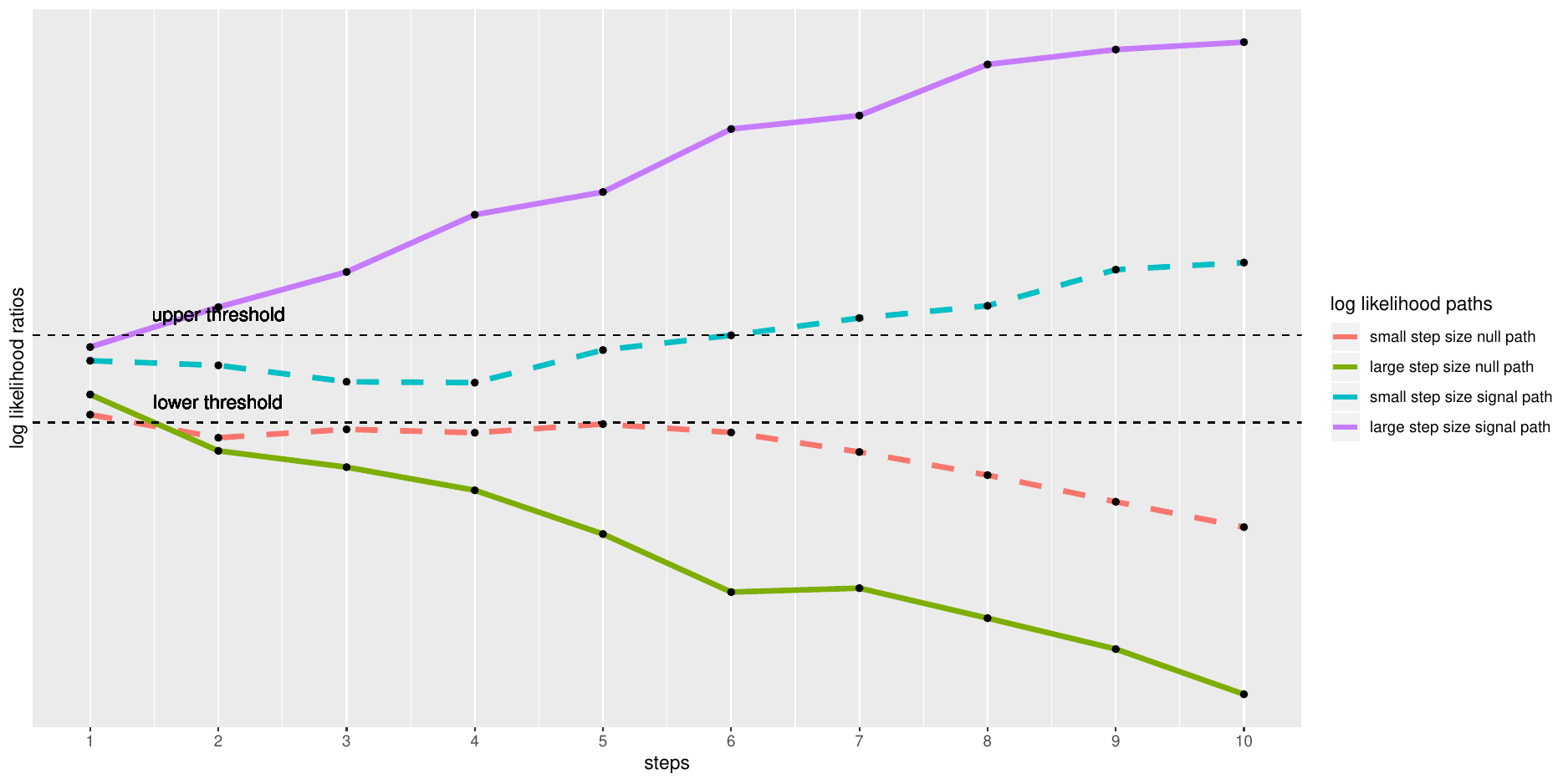}
		\caption{\small The sample paths for two SPRTs, with the null being  $N(0,1)$, and alternative being  $N(1,1)$ (dashed lines, smalle steps) and $N(2, 1)$ (solid lines, large steps), respectively. $L_{ij}$ change drastically with large steps, accentuating the overshooting issue. }\label{path.fig2}
	\end{figure}

	Concretely, the sequential probability ratio test (SPRT) process can be thought of as a Brownian motion with an average ``step size'' determined by $E_{\theta_i}[\log\{f_1(X_{ik})/f_0(X_{ik})\}]$. Figure \ref{path.fig2} shows sample paths for units with large and small step sizes, which illustrate how the SPRT statistics change under the null and alternative hypotheses. The SPRT makes a decision when $\log \mathcal L_{ij}$ exits the interval $(a, b)$. To derive $\tilde t_{OR}^l$ and $\tilde t_{OR}^u$, we use Wald's approximation, assuming that the two boundaries are exactly hit by $\log \mathcal L_{ij}$. However, this assumption is highly idealized, as shown by the sample paths in Figure \ref{path.fig2}, which exhibit ``overshooting'' (Section 7.5.3, \cite{Ber85}). This phenomenon tends to overestimate the true error probabilities and lead to conservative error control, as indicated by Theorem 12 in Section 7.5.3 of \cite{Ber85}.

Our simulations show that the FPR of SPRTs based on Wald's approximations can be as low as 0.01 when the nominal FPR level is 0.05. Such conservativeness in error rate control can result in a loss of efficiency, particularly in high-dimensional settings where most locations are only tested once or twice (i.e., the SPRTs have ``large step sizes''). These concerns motivated us to develop new methodologies in Section \ref{SMART.sec}. 
	
	\setcounter{equation}{0}

	\section{Sparse Recovery via SMART}\label{SMART.sec}
	
	In the sparse recovery problem \eqref{optimal-seq-proc.def}, the $p$ sequential decisions are combined and evaluated as a whole; this is called a \emph{compound decision problem} \citep{Rob51}. Let $\pmb d=(\pmb d_1, \cdots, \pmb d_p)$ denote a decision procedure. Then $\pmb d$ is \emph{simple} if $\pmb d_i$ depends on data from location $i$ alone, and \emph{compound} if $\pmb d_i$ also depends on data from other locations $j\neq i$. A fundamental result in \cite{Rob51} is that compound rules are in general superior to simple rules, even when the observations from different units are independent. In the context of multiple testing, Sun and Cai (2007) \cite{SunCai07} showed that more powerful FDR procedures can be constructed by pooling information from independent tests. This section develops a compound thresholding rule to show that the accuracy of single SPRTs can be greatly improved. 

This section begins by introducing a prototype algorithm under the oracle setup, which assumes the knowledge of both $\pi$ and $g$. Subsequently, we carefully develop a data-driven SMART procedure that can be implemented in real-world scenarios where $\pi$ and $g$ are unknown.
	

	\subsection{A prototype algorithm under the oracle setup}\label{SMART.subsec}
	
	The operation of SMART can be described as follows. At stage $j$, we first calculate the oracle statistics in the active set $\mathcal A_j$ and sort them from smallest to largest. Then we carry out two thresholding procedures along the ranking: part (a) chooses a lower cutoff  $\hat t_{OR}^{l, j}$ with selected locations claimed as signals; part (b) chooses an upper cutoff $\hat t_{OR}^{u, j}$ with selected locations claimed as nulls. We stop sampling on locations where definitive decisions are made, and take new observations on remaining locations for further investigation. The above steps will be iterated until convergence. The detailed operation is described in Algorithm \ref{alg:prototype_algorithm}. 
	

\begin{algorithm}
	\caption{A Prototype Algorithm}\label{alg:prototype_algorithm}
	\begin{algorithmic}[]
		\State\textbf{Input} Target FPR level $\alpha$, target MDR level $\gamma$, $\pmb{X}_i^j$ for $i=1,\ldots,p$, $j=1,\ldots$, non null proportion $\pi$, conditional densities
		$f(\pmb{X}^j_i|\theta_i)$ for $\theta_i=0/1$, $j=1,\ldots$
		\State\textbf{Preliminary Step}  Define the lower and upper thresholds $\tilde {t}_{OR}^l=\alpha$ and $ \tilde {t}_{OR}^u=\frac{1-\pi}{\pi\gamma+1-\pi}$.
		
		\State Let $\mathcal A_j$ be the active set at stage $j$,  $j=1, 2, \cdots$.
		\State Iterate Steps 1 to 3 until $\mathcal{A}_{j}=\emptyset$.
		\State \textbf{Step 1} (Ranking). For all $i\in \mathcal{A}_j$, compute $T_{OR}^{i,j}$ using \eqref{eq:torform} and sort them in ascending order
		\State \quad $T_{OR}^{(1),j} \leq T_{OR}^{(2),j} \leq \cdots \leq T_{OR}^{(k_j),j},$ where $k_j=\text{Card}(\mathcal{A}_j)$.
	\State \textbf{Step 2} (Thresholding).
		\State \quad (a) (Signal discovery). Let $k_j^s = \max\{r: r^{-1}\sum_{i=1}^r T_{OR}^{(i),j}\leq \tilde {t}_{OR}^l\}$ and $\hat t_{OR}^{l, j}=T_{OR}^{(k_j^s),j}$.
		\State \quad \quad Define $\mathcal S_j^s = \{i \in \mathcal{A}_j: T_{OR}^{i,j} \leq \hat t_{OR}^{l, j}\}$. For all  $i \in \mathcal S_j^s$, stop sampling and let $\delta_i=1$.
		\State \quad (b) (Noise elimination). Let $k_j^e=\max\{r: r^{-1}\sum_{i=0}^{r-1}(T_{OR}^{(k_j-i),j})\geq \tilde {t}_{OR}^u\}$ and $\hat t_{OR}^{u, j}=T_{OR}^{(k_j-k_j^e+1), j}$.
		\State \quad \quad Define $\mathcal S_j^e=\{i\in \mathcal{A}_j: T_{OR}^{i,j}\geq \hat t_{OR}^{u, j}\}$. For all  $i \in \mathcal S_j^e$, stop sampling and let $\delta_i=0$.
		\State \textbf{Step 3} (Updating). Let $\mathcal{A}_{j+1} = \mathcal{A}_j \setminus(\mathcal{S}_j^s\cup \mathcal{S}_j^e)$. Take new observations on $\mathcal{A}_{j+1}$.
		\State \textbf{Output} Estimated non-null positions $\{i:\delta_i=1\}$.
	\end{algorithmic}
\end{algorithm}

	The SMART algorithm uses stage-wise thresholds  $\hat t_{OR}^{l, j}$ and  $\hat t_{OR}^{u, j}$. The operation of the algorithm implies that we always have $\hat t_{OR}^{l, j}\geq \tilde t_{OR}^l$ and  $\hat t_{OR}^{u, j}\leq \tilde t_{OR}^u$ for all $j$. Hence SMART always uses fewer samples than $\tilde{\pmb d}_{OR}$. The next theorem shows that SMART is valid for FPR and MDR control.
	\begin{restatable}{theorem}{valid}
		\label{thm:VALID}
		Denote $\pmb d_{SM}$ the SMART procedure described in Algorithm 1 with pre-specified FPR and MDR levels $(\alpha, \gamma)$. { Then}
		$${\rm FPR}({\pmb d_{SM}})\leq\alpha,\; \rm FDR({\pmb d_{SM}})\leq\alpha\; \mbox{ and } \; {\rm MDR}(\pmb d_{SM})\leq\gamma.$$
	\end{restatable}

	\begin{remark} \rm{
			The FPR and MDR control would not suffer from selection effects. First, the operation of SMART implies that only one terminal decision is made for each coordinate (with no further inference); the action of ``collecting one more observation'' is an \emph{indecision} that does not play a role in the FPR or MDR analyses. Second, the theory on FPR and MDR control can be proven under the framework of multiple testing with groups \citep{Efr08a, CaiSun09} by viewing stages as groups [cf. Section \ref{proof-thm3.subsec} in the Supplementary Material].}
	\end{remark}

	\subsection{Overshooting, compound thresholding and knapsack problems}
	
	In the thresholding step, SMART chooses the lower and upper cutoffs based on the \emph{moving averages} of the selected oracle statistics. We call SMART a \emph{compound thresholding} procedure, for its stage-wise cutoffs  $\hat t_{OR}^{l, j}$ and  $\hat t_{OR}^{u, j}$ are jointly determined by data from multiple locations. By contrast, we call $\tilde{\pmb d}_{OR}$ a simple thresholding rule as the decision at location $i$ only depends on its own data. By adopting a compound thresholding scheme and pooling information across parallel SPRTs, SMART overcomes the overshoot problem of individual SPRTs. To illustrate the core idea, consider the following toy example.

	\begin{example}\rm{
			Let the FDR level $\alpha=0.05$. Suppose at stage $j$ the ordered $T_{OR}$ values are $\{0.01, 0.055, 0.07, 0.10, \cdots\}$. If we use the simple thresholding rule $\tilde{\pmb d}_{OR}$ with $\tilde {t}_{OR}^l=\alpha=0.05$, then we can reject one hypothesis with $T_{OR}^{(1), j}=0.01$; the gap between $T_{OR}^{(1), j}$ and $\alpha$ is 0.04. By contrast, the moving average of the top three statistics is $0.045<\alpha$; hence SMART rejects three hypotheses. The gap between the moving average and the threshold is only 0.005. Thus the boundary $\alpha$ can be approximated more precisely by the moving average. In large-scale testing problems, the gap is almost negligible, implying very accurate error rate control; this has been corroborated by our simulation studies. }
	\end{example}
	
	We provide further insights by viewing the decision process of SMART as a {0-1 knapsack problem with varying FPR capacities} \citep{gang2023structure}.  Let $k_j^*=\max\{1\leq i\leq m: T_{OR}^{(i), j}\leq \alpha\}$. Then the discovery step in Algorithm 1 can be written as
	\beq\label{az2.proc}
	\textstyle \sum_{i=k_j^{*}+1}^{k_j^{s}} \left( T_{OR}^{(i),j}-\alpha \right) \leq \sum_{i=1}^{k_j^{*}} \left( \alpha-T_{OR}^{(i),j}\right).
	\eeq
	If we view the nominal FDR level $\alpha$ as the initial capacity in a knapsack problem, then Equation \ref{az2.proc} shows that every time we reject a hypothesis with $T_{OR}^{i, j}<\alpha$, we will increase the capacity of the knapsack, enabling one to reject hypotheses with $T_{OR}^{i, j}>\alpha$. Hence the approximation errors (overshooting) can be effectively mitigated by 
	aggregating the gaps between $T_{OR}$ and $\alpha$ to increase the capacity and make more discoveries. 
	
	\subsection{Numerical implementation of SMART}\label{estimation.subsec}
	
	In practice the oracle statistic $T_{OR}$ is unknown and needs to be estimated from data. This subsection develops an algorithm for computing $T_{OR}$ based 
	the model in introduced in Section \ref{sec:1.1}, which is summarized in Algorithm 2. Note that the model in Section \ref{sec:1.1} can be equivalently written as
		\begin{equation}\label{multi-stage-rm.model}
			\mu_{i}\sim (1-\pi)\delta_0(\cdot)+\pi g(\cdot), \quad X_{ij}\sim N(\mu_{i},\sigma^2),\quad i=1,\ldots,p,\quad j=1, 2, \ldots
		\end{equation}
In this subsection we assume $g$ is mixture of point masses
$
g(\cdot)=\sum_{s=1}^{m_1}w_s\delta_{\eta_s}(\cdot).
$
Consequently, the oracle statistics $T_{OR}^{i,j}$ can be written as
\begin{equation}\label{eq:34}
	T^{i,j}_{OR}:=\mathbb{P}(\theta_i=0|\pmb{X}^j_i)=\dfrac{(1-\pi)\prod^j_{t=1}\phi(x_{it},0,\sigma^2) }{(1-\pi)\prod^j_{t=1}\phi(x_{it},0,\sigma^2)+\pi\sum_{s=1}^{m_1} w_s\prod_{t=1}^j\phi(x_{it},\eta_s,\sigma^2) },
\end{equation}
	where $\phi(\cdot,\mu,\sigma^2)$ is the density function of $N(\mu,\sigma^2)$.

In the preliminary estimation step, we propose to use the method in \cite{JinCai07} to estimate the unknown mixing proportion $\pi$. Next, the estimation of $(1-\pi)\delta_0+\pi g(\cdot)$, referred to as the deconvolution problem, has been studied intensively in the Empirical Bayes literature (\cite{Koenker17}, \cite{efron2016deconv}, \cite{jiang2009general}). The \textsl{GLmix} function in the R package \textbf{REBayes}, which is employed in our algorithm, is shown to deliver excellent performance in practice. \textsl{GLmix} first discretizes $(1-\pi)\delta_0+\pi g(\cdot)$ as a mixture of point masses and then assigns weights to each point mass to approximate $(1-\pi)\delta_0+\pi g(\cdot)$. To get the final estimate of $g$, we first calculate the minimum detectable signal strength using the classification boundary in \cite{Cai11}, i.e.
calibrate $\hat{\beta}=-\frac{\log{\hat{\pi}}}{\log{p}}$ and set threshold $A_{\kappa}=\max\{\sqrt{2\hat\beta\log{p}}, \kappa\}$,
where $\kappa$ is a pre-specified constant (empirically, the choice of $\kappa=1.5$ works well in practice 
and has been used across all simulations). Then we  filter out the locations with absolute signal amplitudes smaller than $A_\kappa$.  Finally, we update $\hat{g}$ by re-weighting the remaining locations so that the sum of point mass probabilities equals 1.

In later stages $j=2, 3, \cdots$, we sequentially update $\hat{T}_{OR}^{i, j}$ using $\hat{g}$ and $\hat{\pi}$ that are estimated based on data collected at stage 1.
The asymptotic properties of $\hat{\pi}$ and $\hat{g}(\cdot)$ have been investigated in \cite{JinCai07} and \cite{SahGun20}, respectively. The numerical experiments in \cite{JinCai07} and \cite{koenker2014convex} suggest these estimators perform quite well in practice. Together with the next proposition, the data-driven SMART provides a reasonable and sensible algorithm for the sparse recovery problem. 

	\begin{proposition} \label{prop1}
		Suppose $g(\cdot)=\sum_{s=1}^{m_1}w_s\delta_{\eta_s}(\cdot)$ is a finite mixture of point masses, with $supp(g)\cap [-a,a]=\emptyset$ for some $a>0$ and $w_s>\epsilon$ for all $s$ and some fixed $\epsilon$. 	Further assume that $\hat{g}(\cdot)=\sum_{s=1}^{m_2}\hat{w}_s\delta_{\hat{\eta}_s}(\cdot)$ with $\hat{w}_s>\epsilon$ for all $s$, and for every $\eta_s$ there exist $\hat{\eta}_k$ such that $|\eta_s-\hat{\eta}_k|<a/2-\epsilon$. Let $T_{OR}^{i,j}$ be as defined in \eqref{eq:34} and
		 $\hat{T}_{OR}^{i, j}$ as defined in Algorithm \ref{alg.dd2}, 
	then for any $\alpha$, there exist $K_\alpha$ such that 
	$$
	\mathbb{P}\left\{\forall j>K_\alpha, |T_{OR}^{i, j}-\hat{T}_{OR}^{i, j}|<e^{-c j}\right\}>1-\alpha,
	$$
	where $c>0$ is a constant depends on $\alpha$, $\pi$, $\hat{\pi}$, $\sigma$, $\hat{\sigma}$, $a$.
	\end{proposition}
	
The proof of Proposition 1, provided in Section \ref{Proof.sec} uses Theorem 1 from \cite{howard2021time}, which establishes a time uniform confidence sequence for the sample mean of a 1-subgaussian distribution. Our work builds upon this result, leveraging (a) the one-to-one correspondence between the oracle statistic $T^{i,j}_{OR}$ and the likelihood ratio $\mathcal{L}_{i,j}$, and (b) the fact that, given $\hat{g}$ and $\hat{\sigma}$, the likelihood ratio is a function of the sample mean.

	\begin{remark}
	We have assumed that $g(\cdot)$ is a mixture of point masses, which is not a restrictive assumption. The mixture of point masses is sufficiently complex to represent the underlying process that generates our observed data. \cite{kiefer1956consistency} proved that the nonparametric maximum likelihood estimator (NPMLE) is a consistent estimator of $G$. Furthermore, the classical Caratheodory theorem establishes that the optimal $\hat G$ must be atomic, which is a discrete distribution with fewer than $n$ atoms.  The assumption that $supp(g)\cap [-a,a]=\emptyset$ for some $a>0$ is reasonable, since the difference between the means of signal and noise must exceed a certain threshold for the problem to be meaningful. Additionally, we assume that $\hat{g}$ is a finite mixture of point masses, which is a standard assumption in the deconvolution literature \citep{efron2016deconv,koenker2014convex}.
The support of $\hat{g}$, which is represented by $\{\hat{\eta}_1,\ldots, \hat{\eta}_{m_2}\}$, is typically a pre-specified grid on the real line. If a lower bound of $a$ is known, a sufficiently dense grid can be chosen in $[\min X_{i1}, \max X_{i1}]\backslash (-\tilde{a},\tilde{a}) $ such that the condition that ``for every $\eta_s$ there exist $\hat{\eta}_k$ such that $|\eta_s-\hat{\eta}_k|<a/2-\epsilon$'' is satisfied with high probability.
	\end{remark}

The proof of Proposition \ref{prop1} establishes, with high probability, that $\hat{T}^{i,j}_{OR}$ converges to 0 or 1 at a rate of $O(e^{-cj})$. As a result, the number of iterations is at most $\log p$. Therefore, the computational complexity for Algorithm \ref{alg.dd2} does not exceed $O(p(\log p)^2)$.

\begin{algorithm}
	\caption{The data-driven SMART algorithm.}
	\label{alg.dd2}
	\begin{algorithmic}
			\State\textbf{Input} Target FPR level $\alpha$, target MDR level $\gamma$, $\pmb{X}_i^j$ for $i=1,\ldots,p$, $j=1,\ldots$.
		\State \textbf{Preliminary estimation ($j=1$):}
		\State \textbf{Step 1}
		 Given data $X_{11},\ldots, X_{p1}$, estimate $\pi$ and $\sigma^2$. Denote the estimated value as $\hat{\pi}$ and $\hat{\sigma}^2$ respectively.
		\State \textbf{Step 2} Use \textsl{GLmix} function in the R package \textbf{REBayes} to estimate $(1-\pi)\delta_0+\pi g(\cdot)$.
		Denote the estimate as $\psi(\mu)$.
		\State  \textbf{Step 3} Calibrate $\hat{\beta}=-\frac{\log{\hat{\pi}}}{\log{p}}$, then set $A_{1.5} = \max\{\sqrt{2\hat\beta\log{p}}, 1.5\}$.
		\State  \textbf{Step 4} Define $\tilde{g}(\mu)=\psi(\mu)\mathbb{I}(|\mu| \geq A_{1.5})$. Set $\hat{g}(\mu) = \dfrac{\tilde{g}(\mu)}{\int_{-\infty}^{\infty}\tilde{g}(\mu)d\mu} = \sum_{s=1}^m w_s \delta_{\eta_s}(\cdot)$.
		\State  \textbf{Step 5} Define the lower and upper thresholds $\tilde{t}^l_{OR}=\alpha$ and $\tilde{t}^u_{OR}=\dfrac{1-\hat{\pi}}{\hat{\pi}\gamma+1-\hat{\pi}}$.
		\State \textbf{SMART procedure ($j>1$):}
		\State Let $\mathcal{A}_j$ be the active set at stage $j$, $j=2,3,\ldots$. Iterate Step 6 to Step 9 until $\mathcal{A}_j = \emptyset$.
		\State \textbf{Step 6} (Estimation). For all $i \in \mathcal{A}_j$, estimate $T^{i,j}_{OR}$ using
		\State $\hat{T}^{i,j}_{OR} := \dfrac{(1-\hat{\pi})\prod^j_{t=1}\phi(x_{it},0,\hat{\sigma}^2) }{(1-\hat{\pi})\prod^j_{t=1}\phi(x_{it},0,\hat{\sigma}^2)+\hat{\pi}\prod_{t=1}^j\phi(x_{it},\hat{\mu}^j_{i},\hat{\sigma}^2)}$,
		\State where $\hat{\mu}^j_{i} := \sum_{s=1}^{m}\frac{ w_s\prod_{t=1}^{j}\phi(x_{it},\eta_s,\hat{\sigma}^2)}{\sum_{k=1}^{m}w_k\prod_{t=1}^{j}\phi(x_{it},\eta_k,\hat{\sigma}^2)}\eta_s$ is the posterior mean under $\theta_i=1$
		\State and $\phi(x_{it},\eta_s,\hat{\sigma}^2)= \frac{1}{\sqrt{2\pi}\hat{\sigma}}\exp\left(-\frac{(x_{it}-\eta_s)^2}{2\hat{\sigma}^2} \right)$.
		\State \textbf{Step 7} (Ranking). For all $i\in\mathcal{A}_j$, sort $\hat{T}^{i,j}_{OR}$ in ascending order.
		\State $\hat{T}^{(1),j}_{OR} \leq \hat{T}^{(2),j}_{OR} \leq \cdots \leq \hat{T}^{(k_j),j}_{OR}$.
		\State \textbf{Step 8} (Thresholding).
		\State (a) (Signal discovery) Let $k^d_j = \max\{r : r^{-1}\sum_{i=1}^r \hat{T}^{(i),j}_{OR} \leq \tilde{t}^l_{OR} \}$ and $\hat{t}^{l,j}_{OR} = \hat{T}^{(k^d_j),j}_{OR}$.
		\State Define $\mathcal{S}^d_j = \{i\in \mathcal{A}_j : \hat{T}^{i,j}_{OR} \leq \hat{T}^{(k^d_j),j}_{OR}\}$. For all $i\in \mathcal{S}^d_j$, stop sampling and let $\delta_i=1.$
		\State (b) (Noise elimination) Let $k^e_j = \max\{r : r^{-1}\sum_{i=1}^r \hat{T}^{(k_j-i),j}_{OR} \geq \tilde{t}^u_{OR} \}$ and $\hat{t}^{u,j}_{OR} = \hat{T}^{(k_j-k^e_j+1),j}_{OR}$. 
		\State Define $\mathcal{S}^e_j = \{i\in \mathcal{A}_j : \hat{T}^{i,j}_{OR} \geq \hat{T}^{(k^e_j),j}_{OR}\}$. For all $i\in \mathcal{S}^e_j$, stop sampling and let $\delta_i=0.$
		\State \textbf{Step 9} (Updating). Let $\mathcal{A}_{j+1} = \mathcal{A}_j \backslash (\mathcal{S}^d_j \bigcup \mathcal{S}^e_j)$. Take new observations on $\mathcal{A}_{j+1}$.
			\State \textbf{Output} Estimated non-null locations $\{i:\delta_i=1\}$.
	\end{algorithmic}
\end{algorithm}

It would be ideal to incorporate samples collected in later stages to estimate $\pi$ and $g$. However, that such a method could be impractical for several reasons. Firstly, estimating $g$ requires solving a complicated optimization problem that can be computationally intensive. Updating $g$ at each stage would result in an algorithm that is prohibitively slow. Secondly, existing deconvolution techniques \citep{koenker2014convex, efron2016deconv, jiang2009general} require the mean parameters $\mu_i$ to be iid from $g$. However, once a location is identified as signal or noise, the SMART algorithm no longer samples from that location. Consequently, the samples collected in subsequent stages only reflect a biased portion of $g$ (this issue is known as selection bias or data-snooping bias in the literature). Developing a consistent algorithm for estimating $g$ while fully exploiting the information from all samples is a non-trivial question of itself and requires further investigation.

	\subsection{Connections to distilled sensing and single sequential thresholding}
	
The distilled sensing (DS, \cite{Hauetal11}) and single sequential thresholding (SST, \cite{MalNow14a}) methods provide efficient adaptive sampling schemes for sparse recovery. Both works consider the following  setup:
$$
X_{ij}=\mu_i+\gamma_{ij}^{-1/2}\epsilon_{ij}, \quad i=1,\ldots,p, \quad j=1, 2, \ldots,
$$
where $\gamma_{ij}$ may be conceptualized as the ``sample size'' collected at location $i$ stage $j$. In the setting of our paper, $\gamma_{ij}=0$ or 1, where $\gamma_{ij}=0$ indicates that no observation is collected at $(i,j)$. Both DS and SST aim to recover the support under the budget constraint $\sum_{j=1}^{n}\sum_{i=1}^{p}\gamma_{ij}\leq R(p)$, where $R(p)$ is an increasing function of $p$ that represents the total measurement budget. A prototype DS algorithm operates as follows: it starts with the full set $\mathcal A_1=\{1, \cdots, p\}$, then performs $k$ stages of distillation, where $k$ is determined by a pre-specified $R(p)$. At stage $j$, $x_{ij}=\mu_i+\epsilon_{ij}$ is observed and the active set is updated as $A_{j+1}\leftarrow \{i\in\mathcal A_j: x_{ij}>0\}$. This simple operation indicates that only locations with positive observations are kept. The DS algorithm outputs $A_{k}$ as the final ``discovery set''. If the signals are sparse, then each stage eliminates about half of the null locations. The operation of SST is similar: at each step the procedure takes $R(p)/(2p)$ samples from each active location. The summary statistic at location $j$ is obtained as the sum of $R(p)/(2p)$ observations. If the sum is less than zero, then no observation will be collected for that coordinate in subsequent passes. In both DS and SST, the observations from previous stages are abandoned and the compound structure of the problem is ignored, which leads to efficiency loss. 

SMART employs an adaptive sampling strategy that is equipped with a simultaneous inference framework to control user-specified error rates. It improves DS and SST in several ways. First, as opposed to DS and SST, which use a fixed thresholding rule $\II(x_{ij}>0)$ at all units and through all stages, SMART uses data-driven thresholds that are \emph{adaptive} to both data and pre-specified FPR and MDR levels. Second, the sampling and inference rules in DS and SST only use observations at the current stage. By contrast, the building block of SMART is $T_{OR}^{i,j}$, which uses all observations collected up to the current stage. This can greatly increase the signal to noise ratio and lead to more powerful testing and stopping rules. Third, SMART uses the powerful compound thresholding idea to exploit the multiple hypothesis structure, which can greatly improve the accuracy of inference and lead to more precise error rates control. Finally, DS and SST only have a stopping rule to eliminate null locations. Intuitively, such a scheme is inefficient as one should also stop sampling at locations where the evidence against the null is extremely strong. The proposed scheme in Step 2 of SMART involves two algorithms that can simultaneously identify signals and eliminate noise. In other words, SMART stops sampling once a definitive decision ($\delta_i=0$ or $\delta_i=1$) is reached; this more flexible operation is desirable and would further save study budgets.

\subsection{Comparison with other methods}\label{comparisons.subsec}

We provide a description of the sequential procedure by \citep{BarSon13}, denoted BS, and contrast it with the SMART algorithm. As the BS procedure is a bit complicated, we only discuss high level ideas and omit some technical details below. 
		
Let $\Lambda^i_j$ denote the test statistic for coordinate $i$ at stage $j$, $1\leq i\leq p$, $j=1, 2, \cdots$. For each location $i$, it is assumed that there exist carefully chosen critical values
		\begin{equation}\label{seq}
			A^i_1\leq A^i_2\leq \ldots\leq A^i_p\leq B^i_p\leq B^i_{p-1}\leq \ldots \leq B^i_1.
		\end{equation}
The BS procedure conducts simultaneous tests at all coordinates in a sequential manner. It stops sampling until $\Lambda_j^i\notin (A_1^i, B_s^i)$ or $\Lambda_j^i\notin (A_s^i, B_1^i)$; a rejection/acceptance decision is made if some $\Lambda^i_j$ crosses a certain critical value: if $\Lambda_i^j\notin (A_1^j, B_s^j)$[or $\Lambda_i^j\notin (A_s^j, B_1^j)$], i.e. $\Lambda_i^j$ crosses $B_s^j$ [or $A_s^j$] first, then the null hypothesis is rejected [or accepted]. The sequence of critical values $A^i_1\leq A^i_2\leq \ldots\leq A^i_p\leq B^i_p\leq B^i_{p-1}\leq \ldots \leq B^i_1$, which must satisfy error bounds on both Type I and Type II error probabilities, are unknown in practice, and must be computed approximately using Monte Carlo simulation. \cite{BarSon13} employ a strategy that combines several techniques such as (a) the thresholds in the step-up procedure by Benjamini-Hochberg, (b) sequential (or group sequential) MC sampling methods of Gaussian data, and (c) the utilization of limiting distributions of the signed roots of complicated equations, for computing these critical values. It turns out that the computation of the sequence \eqref{seq} can be quite complicated and intensive. As the computational burden is in the order of $O(p^3)$, the simulation study in \cite{BarSon13} only considers $p=20$ data streams.

In contrast, SMART is built upon the compound decision theory and local false discovery rate (\citealp{efron2001micro, SunCai07}) ideas. Its operation is fundamentally different from existing SPRTs, and is particularly suitable for large-scale testing problems. The thresholds can be computed quickly using a simple step-up procedure that employs elegant ideas similar to those in \cite{SunCai07}. This step-up algorithm, which involves ranking and thresholding, has computational complexity of $O(p\log(p))$. Hence SMART is capable of handling millions of data streams. The thresholds lead to effective error rates control and substantially outperform competing methods in expected sample size (power).

	\section{Limits of Sparse Recovery}\label{Limits.sec}

	\setcounter{equation}{0}

	This section discusses the notion of fundamental limits in sparse inference for both fixed and adaptive settings. The discussion serves two purposes. First, as the limits reflect the difficulties of support recovery under different settings, they demonstrate the advantage of using adaptive designs over fixed designs. Second, the limit provides an optimality benchmark for what can be achieved asymptotically by any inference procedure in the form of a lower bound; hence we can establish the optimality of an inference procedure by proving its capability of achieving the limit.

	
	\subsection{The fundamental limits in adaptive designs} 
	
 Given the consideration of signal recovery with thousands, and even millions, of variables, it is assumed that $p\rightarrow \infty$ in the asymptotic analysis conducted in this work. Under this asymptotic framework, the FPR and MDR levels are denoted $\alpha_p$ and $\gamma_p$, both of which tend to zero as $p\rightarrow \infty$.
	In \cite{MalNow14a},  the fundamental limits for reliable recovery were established under the family wise error rate (FWER) criterion. This section extends their theory to the important FPR and MDR paradigm. 
	
	The basic setup assumes that the null and alternative distributions $F_0$ and $F_1$ are identical across all locations. This more restrictive model is only for theoretical analysis and our proposed SMART procedure works under the more general model \eqref{mix1}-\eqref{mix3}, which allows $F_{1i}$ to vary across testing units. Let $G$ and $H$ be two distributions with corresponding densities $g$ and $h$. Define $D(G\| H)=\int_{-\infty}^{\infty} g(x)\log{\frac{g(x)}{h(x)}}dx$. Then the Kullback-Leibler (KL) divergence, given by 
	$$
	D_{KL}(F_0, F_1)=\max\{D(F_0\|F_1), D(F_1\|F_0)\},
	$$
	can be used to measure the distance between $F_0$ and $F_1$. Let $\tau$ be the average number of measurements allocated to each unit. Consider a general multistage decision procedure $\pmb d$. The performance of $\pmb d$ is characterized by its total risk $R^*(\pmb d)=\text{FPR}(\pmb d)+\text{MDR}(\pmb d)$. Intuitively, $D_{KL}(F_0, F_1)$ and $\tau$ together would characterize the possibility of constructing a $\pmb d$ such that $R^*(\pmb d)\rightarrow 0$. The fundamental limit, described in the next theorem, gives the minimum condition under which it is possible to construct a   $\pmb d$ such that  $R^*(\pmb d)\rightarrow 0$.  
	
	
	\begin{restatable}{theorem}{limit.thm} {\bf Fundamental limits (lower bound)}. 
		\label{limit.thm}
		Let $\pmb d$ be a multistage adaptive testing rule. Assume that $\pi<\frac{1}{2}$. If $$\tau\equiv\mbox{ESS}(\pmb d) \leq \frac{\log(4\eta)^{-1}}{D_{KL}(F_0, F_1)},$$ then we must have $R^*(\pmb d)\geq \eta$ for all $\eta>0$. 
	\end{restatable}

In asymptotic analyses we typically take an $\eta$ that converges to 0 slowly. Theorem \ref{limit.thm} shows that any adaptive procedure with total risk tending to zero must at least have an ESS (or the average sample size per dimension) in the order of $\log \{(4\eta)^{-1}\}$. This limit can be used as a theoretical measure to assess the efficiency of a multistage procedure.

Theorem 4 is closely related to Theorem 1 in 	\cite{MalNow14a}. Let $\mathcal{S}=\{i :\theta_i= 1\}$ and $\hat{\mathcal{S}}=\{i:\delta=1\}$. Denote $m$ the average number of samples per index and $\mathbb{P}_e=\mathbb{P}(\mathcal{S}\neq \hat{\mathcal{S}})$. With these definitions in place, we present below a reformulation of the theory in 	\cite{MalNow14a} using our notation system.
	
	\medskip
	\noindent	\textbf{Theorem 1 in 	\cite{MalNow14a}}. Any uniform coordinate-wise sequential thresholding procedure with 
	$
	m\leq \dfrac{\log s+ \log (\frac{1}{4\delta})}{D_{KL}(F_0, F_1)}
	$
	must have $\mathbb{P}_e\geq 1-e^{-\delta}\approx \delta$. 
  \medskip
  	
This fundamental limit for signal recovery was established under the (FWER) criterion. However, the FWER criterion can be too conservative for large-scale testing problems. Our Theorem 4 expands on the prior theory that relies on the FWER criterion and introduces a more appropriate framework for modern data applications. Specifically, our extension is based on a new statistic and a new criterion, resulting in a novel bound that accurately reflects the potential of reliable signal recovery at a reduced sampling cost, under the less conservative FPR and MDR paradigm.  
	
	 Denote $\pmb d_{SM}=(\pmb N_{SM}, \pmb \delta_{SM})$ the SMART procedure described in Algorithm 1. 
	As we proceed, we need the following assumption that is essentially equivalent to Condition (8) in \cite{MalNow14a}: 
	\begin{eqnarray}
		\mathbb{E}(T_{OR}^{i,N_i}|T_{OR}^{i,N_i}>t_u) & \leq & \frac{t_u}{(1-t_u)e^{-C_1}+t_u},\label{eqn:C1} \\
		 \mathbb{E}(T_{OR}^{i,N_i}|T_{OR}^{i,N_i}<t_l)& \geq & \frac{t_l}{(1-t_l)e^{C_2}+t_l}
		\label{eqn:C2}
	\end{eqnarray} 
for all possible thresholds $t_l<t_u$. The condition is satisfied when $\log\mathcal{L}_{i,1}$ follows a bounded distribution such as Gaussian and exponential distributions. A more detailed discussion on this issue can be found in \cite{Gho60}. The following theorem derives the upper bound and establishes the asymptotic optimality of SMART.
	
	\begin{restatable}{theorem}{optimality}
	\label{optimal.thm} {\bf Asymptotic optimality (Upper bound).}
	Consider the SMART procedure described in Algorithm 1 with lower and upper thresholds 
	$$
	\tilde{t}^l_{OR}=\frac{1}{{\xi}(p)^{1+\epsilon}} \quad \mbox{and} \quad {\tilde{t}^u_{OR}}=\frac{(1-\pi){\xi}(p)^{1+\epsilon}}{\pi+(1-\pi){\xi}(p)^{1+\epsilon}},
	$$ 
	where $ \epsilon >0$ is {an arbitrarily small} constant and ${\xi}(p)$ is a function of $p$ that grows to infinity at an arbitrarily slow rate. Then under \eqref{eqn:C1}-\eqref{eqn:C2}, the SMART procedure satisfies $\lim_{p\rightarrow \infty}R^*(\pmb \delta_{SM})=0$
	and 
	$$
	\lim_{p\rightarrow \infty}\left(\mbox{ESS}(\pmb N_{SM})- \frac {(1+\epsilon)\log{{\xi}(p)}}{\min\{D({F}_0\|{F}_1),D({F}_1\|{F}_0)\}}\right)\leq 0.  
	$$ 
\end{restatable}

Consider the asymptotic regime presented in Model \eqref{nmix}, where $F_0=N(0,1)$, $F_1=N(\mu_p,\sigma^2)$, $\pi_p=p^{-\beta}$ and $\mu_p=\sqrt{2r\log p}$. Some calculations reveal that
  $$
  D(F_0\|F_1)=\log\sigma+\dfrac{1+\mu_p^2}{2\sigma^2}-\frac{1}{2}\quad \text{and}\quad  D(F_1\|F_0)=-\log\sigma+\dfrac{\sigma^2+\mu_p^2}{2}-\frac{1}{2},  
  $$
  both of which are of order $\mu_p^2$.
Theorem \ref{optimal.thm} shows that SMART is capable of achieving the limit given in Theorem \ref{limit.thm} up to a constant multiple. If we take $\eta=1/{\xi}(p)$ with ${\xi}(p)\rightarrow\infty$ slowly, then the rates in Theorems \ref{limit.thm} and \ref{optimal.thm} would match. Theorems \ref{limit.thm} and \ref{optimal.thm} together show that SMART provides a good solution to Problem \eqref{optimal-seq-proc.def} under the asymptotic setup.

	\cite{MalNow14a} showed that when DS or SST methods are considered, the minimum condition under which $R^*(\pmb{d})\rightarrow 0$ requires the expected sample size (ESS) per unit to be in the order of $\log s$, where $s$ is the number of signals. In the asymptotic setup where $s\rightarrow \infty$ as $p\rightarrow \infty$, the result from \cite{MalNow14a} has effectively imposed a lower bound on the rate of required sample size (ESS) in order for $R^*(\pmb{d})\rightarrow 0$. Our theory shows that it is possible to achieve $R^*(\pmb{d})\rightarrow 0$ with ESS being in the order of $\xi(p)$, which is allowed to converge to infinity at an arbitrarily slow rate. This rate can be, of course, slower than $\log s$. This weaker minimum condition demonstrates the theoretical advantage of SMART. Our theory is corroborated by the numerical results, which demonstrate the substantial efficiency gain of SMART over DS and SST, characterized by the expected sample size (ESS) per unit, in many settings.

	\section{Simulation}\label{Simu.sec}
	
	This section is organized as follows. Section \ref{simu1.sec} compares SMART with simultaneous SPRTs (with single thresholding) to illustrate the advantage of compound thresholding. Section \ref{simu2.sec} and \ref{simu-hunt.sec} compare the ranking efficiency of SMART vs. existing methods for sparse recovery and error rate control. We show that SMART achieves the same FPR and MDR levels with a much smaller study cost. Section \ref{simu3.sec} investigates the robustness of SMART for error rate control under a range of dependence structures. Section \ref{simu4.sec} illustrates the use of SMART as a  global testing procedure.
	
	\setcounter{equation}{0}

	\subsection{Simple thresholding vs. compound thresholding} \label{simu1.sec}
	This simulation study compares the following methods: 
	\begin{itemize}
    \item [(a)] the simple thresholding procedure $\tilde{\pmb d}_{OR}$ presented before Theorem \ref{threshold.thm}, which assumes an oracle knows the true parameters (OR.ST); 
    \item [(b)] the SMART procedure with known parameters (OR.SM); 
    \item [(c)] the simple thresholding procedure with estimated parameters (DD.ST), where DD refers to ``data-driven''; 
    \item [(d)] the SMART procedure with estimated parameters (DD.SM).
	\end{itemize}
	
We generate data according to the following mixture model 
	$$
	X_{ij} \sim (1-\pi)N(0,1)+\pi N(\mu_i,1),
	$$
for $i=1,\ldots, p$ and $j=1, 2, \ldots$, where $\pi$ and $\mu_i$ will be specified shortly. The number of locations is $p=10^5$. Let $(\alpha, \gamma)=(0.05, 0.05)$ be pre-specified FPR and MDR levels. The following settings are considered:
\begin{itemize}
		\item
		Setting 1: Set $\pi=0.01$ and  $\mu_i=\mu$ for all $i$. Vary $\mu$ from 2 to 4 with step size 0.2.
		\item
		Setting 2: Set $\pi=0.05$ and  $\mu_i=\mu$ for all $i$. Vary $\mu$ from 2 to 4 with step size 0.2.
		\item
		Setting 3: Draw $\mu_i$ randomly from a uniform distribution $\mathcal{U}(2,4)$. Vary $\pi$ from $0.05$ to $0.2$ with step size $0.01$. 
	\end{itemize}

\begin{figure}[ht]
		\begin{tabular}{c}
			\includegraphics[width=0.94\textwidth]{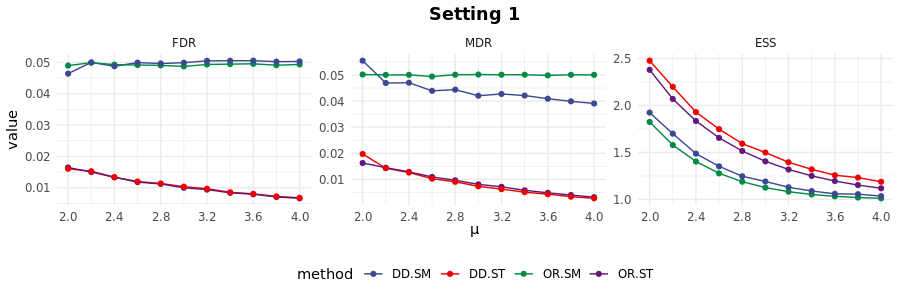} \\
			\includegraphics[width=0.94\textwidth]{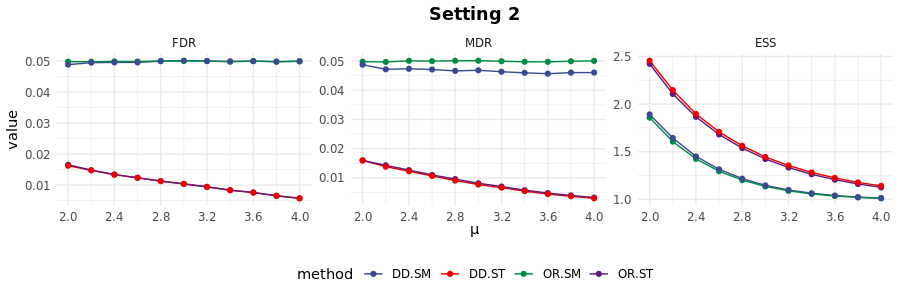} \\
			\includegraphics[width=0.94\textwidth]{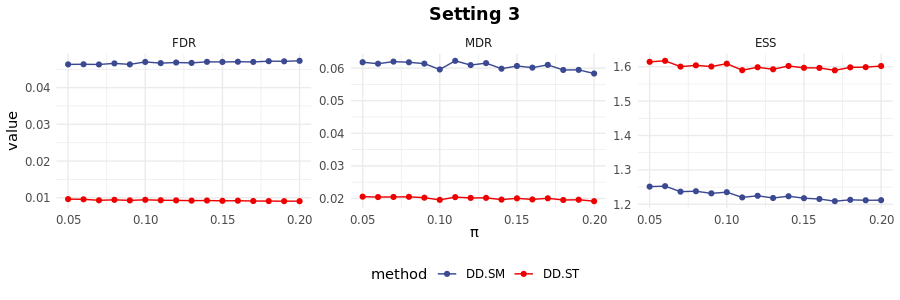}
		\end{tabular}
		
		\vspace{-0.5cm} 
		\caption{SMART vs. single thresholding: the displayed procedures are DD.SM, OR.SM, DD.ST, OR.ST.}\label{SM-ST.fig}
	\end{figure}

	We apply the four methods to the simulated data. The FPR, MDR and ESS (expected sample size) are computed based on the average of $100$ replications, and are plotted as functions of varied parameter values. The results are summarized in Figure  \ref{SM-ST.fig}. In Setting 3, the two oracle methods (OR.ST and OR.SM) are not implemented as recursive formulae for $T_{OR}^{i,j}$ are unavailable.
	
	We can see that all four methods control the FPR at the nominal level. However, the two single thresholding methods (OR.ST and DD.ST) are very conservative (the actual FPR is only about half of the nominal level). Similarly, both OR.ST and DD.ST are very conservative for MDR control. Specifically, 
	OR.ST and DD.ST operate as $p$ parallel SPRTs and suffer from the overshoot problem. The approximation errors of SPRTs can be greatly reduced by SMART that employs the compound thresholding strategy. We can see that the SMART procedures (OR.SM and DD.SM) control the error rates more accurately and require smaller sample sizes. When signals are sparse and weak (top middle panel), the MDR level of DD.SM is slightly higher than the nominal level. This is due to the estimation errors occurred at stage 1. It is of interest to develop more accurate estimation procedures in such settings. The FDR and FPR levels are similar for OR.SM and DD.SM. See Figure 2 in the supplement for details.  
	
	\subsection{SMART vs. Distilled Sensing}\label{simu2.sec}
	
	This simulation study compares the efficiency performance of SMART vs DS \citep{Hauetal11}. As DS does not provide precise error rates control, the simulation is designed in the following way to make the comparison on an equal footing: we first run  DS with up to 10 stages and record its FDR and MDR levels. Then we apply SMART at the corresponding FDR and MDR levels so that the two methods have roughly equal error rates. The ESS is used to compare the efficiency. 
	
	The data are generated from the same model as that in Section \ref{simu1.sec}. The number of locations is $p=10^5$. The following two settings are considered:
	\begin{itemize}
		\item
		Setting 1: Let $\pi=0.05$ and $\mu_i=\mu$ for all $i$. Vary $\mu$ from 2 to 4 with step size 0.2.
		\item
		Setting 2: Draw $\mu_i$ randomly from  $\mathcal{U}(2,4)$. Vary $\pi$ from $0.05$ to $0.2$. 	\end{itemize}

\begin{figure}[ht]
		\begin{tabular}{c}
			\includegraphics[width=0.95\textwidth]{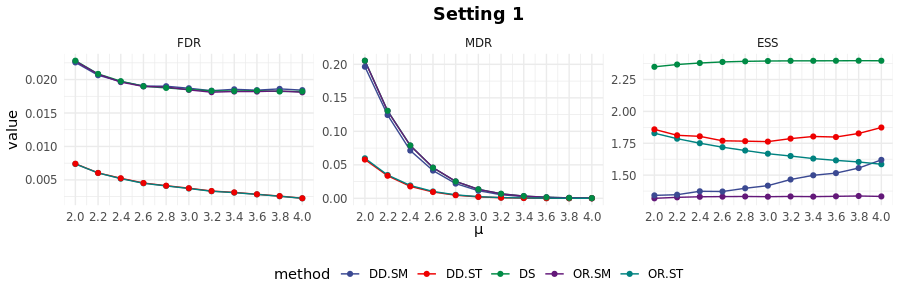} \\
			\includegraphics[width=0.95\textwidth]{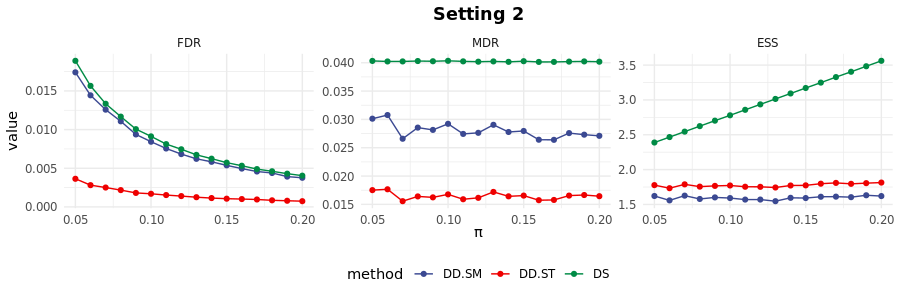}
		\end{tabular}
		\vspace{-0.5cm} \caption{Comparison with DS. SMART outperforms DS in the sense that it achieves the same FDR and MDR levels with much smaller sample sizes.}\label{SM-DS.fig}
	\end{figure}

	In both settings, the FDR, MDR and ESS are computed by averaging over $100$ replications, and are plotted as functions of varied parameter values. The results are summarized in Figure  \ref{SM-DS.fig}.
	
	We can see that the error rates of both OR.SM and DD.SM match well with those of DS, but they require fewer samples. OR.ST and DD.ST also outperform DS, achieving lower error rates with fewer samples. DD.SM controls the error rates more accurately compared to DD.ST, and requires fewer samples. 
	
	\subsection{SMART vs. the Hunt procedure}\label{simu-hunt.sec}
	
	This section conducts simulation studies to compare SMART with the multi-stage testing procedure proposed in \cite{Posetal09}, which is denoted ``Hunt'' subsequently. We set the nominal FDR and MDR levels to be 0.05 and generate $m=10^5$ observations from the model in Section \ref{simu1.sec}
	We consider two settings: 
	\begin{itemize}
		\item Setting 1: Vary $\pi$ from 0.05 to 0.2 and generate $\mu_i\sim\mathcal{U}(2,4)$.
		\item Setting 2: Fix $\pi=0.05$. Let $\mu_i=\mu$, vary $\mu$ from 2 to 4.
	\end{itemize}
	
\begin{figure}[ht]
		\begin{minipage}{\textwidth}
			\centering
			\includegraphics[scale=0.6]{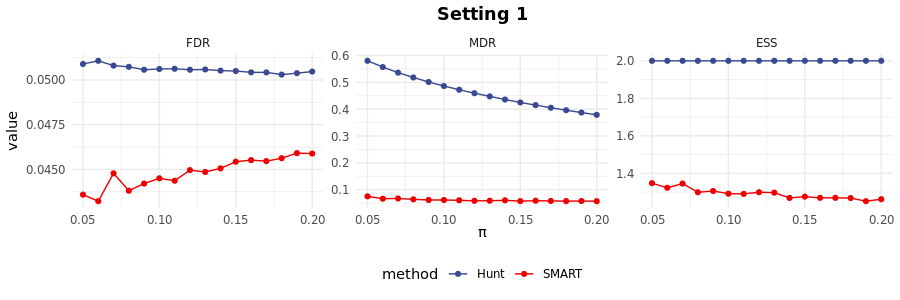} \\
			\includegraphics[scale=0.6]{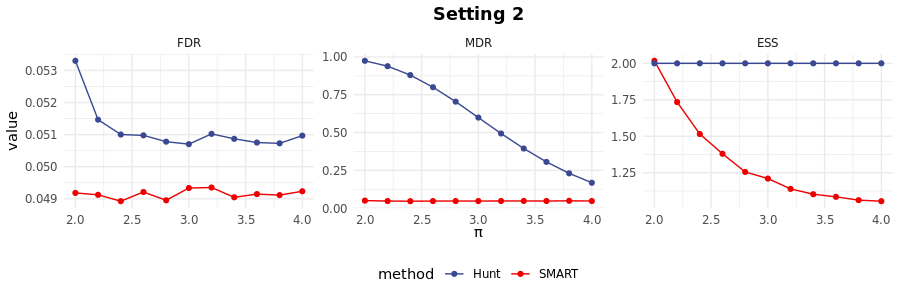}
			\vspace{-0.5cm}
			\captionof{figure}{SMART vs Hunt. Top and bottom rows correspond to Settings 1 \& 2. }
			\label{Hunt.fig}
		\end{minipage}
	\end{figure}

	We apply Hunt and data-driven SMART and compute the FDR, MDR and the total sample size by averaging results in 100 replications. The Hunt method operates under a fixed stage regime; we fix the number of stages to be 2 for illustration. The sample size required for the Hunt procedure is always $2m=2\times 10^5$. The simulation results are summarized in Figure \ref{Hunt.fig}. We can see that SMART effectively controls both the FDR and MDR. By contrast, Hunt controls the FDR but does not control the MDR. This is because Hunt is not designed to control the missed discovery rates. Moreover, SMART requires a smaller sample size to achieve effective error rate control. The intuition is that the group sequential design adopted by Hunt does not eliminate the null coordinates in the first stage; this leads to loss in efficiency and higher study costs. 
	
	\subsection{Robustness of SMART under dependence}\label{simu3.sec}
	
	Our method and theory assume independence of observations across the coordinates. However, the independence assumption may be violated in practice. This section carries out additional numerical studies with correlated error terms across locations to investigate the robustness of the proposed SMART algorithm.
	
	For the sequential testing setup, we first generate non-null effects from $\mathcal{U}(2,4)$ at $p=10^4$ locations. Then we apply the single thresholding procedure DD.ST and SMART procedure DD.SM to simulated data with correlation structures described in Settings 1-3 below. We use the computational algorithm described in Section \ref{estimation.subsec}. 
	The FPR, MDR and ESS (expected sample sizes) are then calculated based on averaging results from $200$ data sets, and finally plotted as functions of the non-null proportion $\pi$. The simulation results are summarized in Fig. \ref{Dependent.fig}.
	
	\begin{itemize}
		
		\item Setting 1 (Gaussian process): the error terms are generated from a zero-mean Gaussian process with the covariance functions  
		$$
		C(r)=e^{-r^2}, \; \mbox{with } r\geq 0\text{ being the distance between two locations.}
		$$
		
		\item Setting 2 (Tridiagonal): the error terms are generated from a multivariate normal distribution with the following covariance matrix $\Sigma$:
		\begin{eqnarray*}
			\Sigma_{i,i}&=&1,\; i=1,\cdots,p,\\
			\Sigma_{i,i+1}&=&0.5,\;i=1,\cdots,p-1,\\
			\Sigma_{i-1,i}&=& 0.5,\;i=2,\cdots,p,\\
			\Sigma_{i,i+2}&=&0.4,\;i=1,\cdots,p-2,\\
			\Sigma_{i-2,i}&= &0.4,\;i=3,\cdots,p.
		\end{eqnarray*}
		All other entries equal to zero. 
		
		\item  Setting 3 (Autoregressive model): the error terms are generated from a multivariate normal distribution with the following covariance matrix $\Sigma$:
		\begin{eqnarray*}
			\Sigma_{i,j}&=& 0.5^{|i-j|},\; i, j=1, \cdots,p.
		\end{eqnarray*}
		
	\end{itemize}

	\begin{figure}
		\begin{tabular}{c}
			\includegraphics[width=0.97\textwidth]{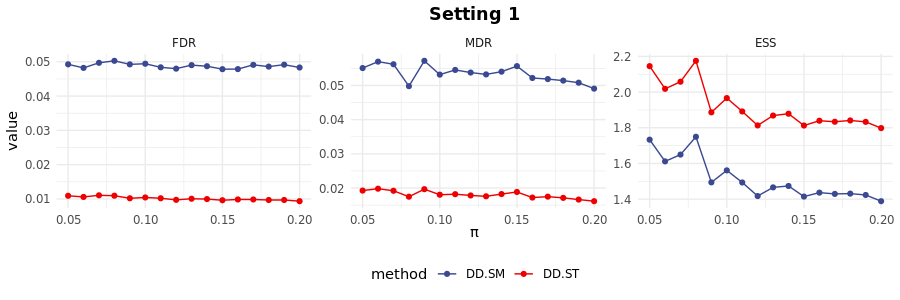} \\
			\includegraphics[width=0.97\textwidth]{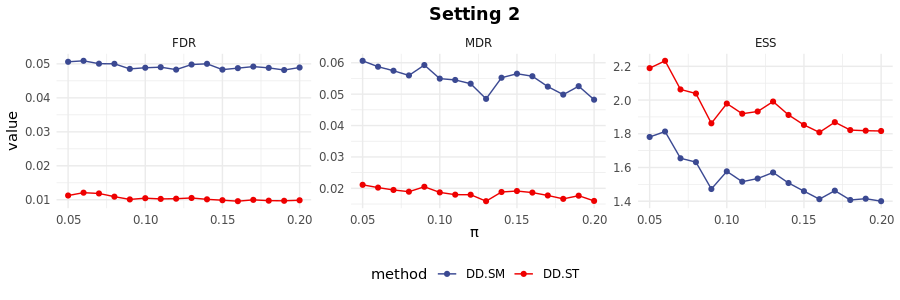} \\
			\includegraphics[width=0.95\textwidth]{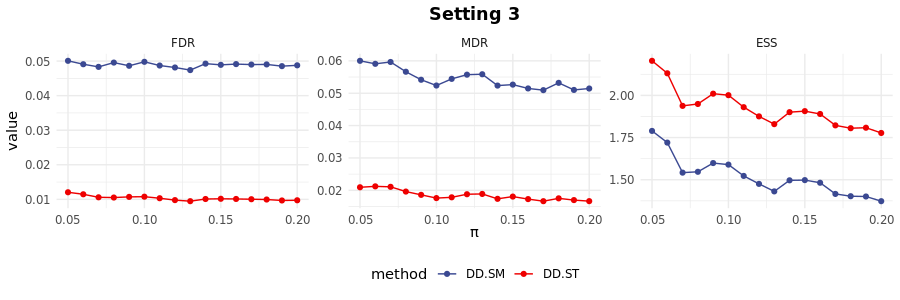} 
		\end{tabular}
		\vspace{-0.5cm} \caption{Dependent tests: Settings 1-3 are described in the main text. SMART controls the error rates at the nominal levels in most settings.}\label{Dependent.fig}
	\end{figure}

	We observe some similar patterns as before. SMART (DD.SM) roughly controls the FPR at the nominal level when moderate correlations are present between locations. The single thresholding procedures with parallel SPRTs (DD.ST) are too conservative for error rates control in all settings. The resulting ESS of DD.ST is significantly larger than that is needed by DD.SM. In Setting 2 the FPR of SMART can be higher than the nominal level 0.05 but seems to be acceptable. SMART controls the MDR under the nominal level in all settings. Our simulation studies suggest that SMART has robust performance under a range of dependence structures. However, the scope of our empirical studies is limited and more numerical and theoretical support are needed for making definitive conclusions. Finally, we stress that it is still an open issue to develop multiple testing methods that incorporates informative correlation structures to improve efficiency \cite{SunCai09, Sunetal15}. The modeling of spatial dependence in the sequential setting is complicated because we have a lot of missing data in later stages. 
	
	\subsection{SMART for signal detection}\label{simu4.sec}
	
	SMART can be employed as a method for signal detection \cite{DonJin04}. The goal is to test the global null hypothesis 
	\begin{equation}\label{global-null}
		H_0^{p}: \{X_1, \cdots, X_p\} \;{\sim}\; F_0 \mbox{ vs. } H_1^{p}: \{X_1, \cdots, X_p\} \;{\sim}\; (1-\pi)F_0+\pi F_{1i}
	\end{equation} 
	for some $\pi>0$. This is a \emph{global inference} problem that is very different from the \emph{simultaneous inference} problem we have considered in previous sections. Comparing the detection boundaries derived in \cite{DonJin04} with the upper limit in Theorem 5, we can see that SMART requires fewer samples than non-adaptive testing schemes for separating $H_0^p$ and $H_1^p$ with negligible testing errors. Similar findings have been reported by \cite{Hauetal11}. 
	
	This section carries out simulation studies to investigate the performance of SMART as a signal detection procedure. Specifically, SMART (Algorithm 1) eliminates noise locations stage by stage. If all coordinates are eliminated as noise locations eventually, then we do not reject the global null. If any coordinate is selected as signal, then we reject the global null. Under the global null $\pi=0$, the false discovery proportion ($\mbox{FDP}({\pmb \delta})=\frac{\sum_{i=1}^p (1-\theta_i) \delta_{i}}{\max\{(\sum_{i=1}^p \delta_{i}),1\}}$) only takes two possible values: 0 (if no coordinate is selected as signal) and 1 (if at least one coordinate is selected as signal). It was argued by \cite{BenHoc95} that under the global null, the FPR cannot be controlled, whereas it is always possible to control the FDR. In our simulation, the FDR is calculated as the average of the FDPs over 200 replications; the average may be conceptualized as an estimate of the Type I error rate for testing the global null, i.e. the relative frequency for incorrectly rejecting the global null when $\pi=0$ among all repeated experiments.

	\begin{figure}[h]
		\includegraphics[width=0.95\textwidth]{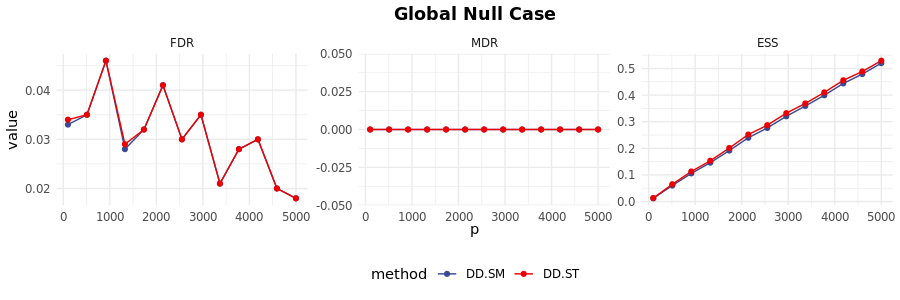} \caption{\small SMART for signal detection: the FDR is used as the Type I error rate.}\label{global-null.fig}
	\end{figure}
	
	We apply SMART by setting $\alpha=\gamma=0.05$. The data are generated using the same models as those in previous sections except that we now let $
	\pi=0$. We vary the dimension $p$ from 100 to 5,000 and plot the corresponding FDRs. It can be seen that SMART controls the FDR at the nominal level 0.05 in all settings. The MDR and ESS are also reported. (If no true signals are discovered, then the MDR will be set as 0.) We conclude that SMART controls the Type I error rate for testing the global null.

	\section{Applications}\label{Apply.sec}
	
	In this section we apply the SMART procedure to A/B testing (Section \ref{apply-ab.sec}), high-throughput screening (HTS, Section \ref{apply-hts.sec}) and satellite image analysis (Section \ref{apply-image.sec}).
	
	\setcounter{equation}{0}
	
	\subsection{A/B Testing}\label{apply-ab.sec}
	A/B testing has become the gold standard for testing new ideas and building product launch plans in high-tech firms. This section illustrates the application of our SMART procedure to A/B testing using a real-world experiment at Snap Inc.. The quantity of interest is the average treatment effect (ATE) between the control and treatment arms (each arm consists of several hundred users). The ATEs are calculated on a daily basis for thousands of metrics. To save the study costs and speed up the decision process, we adopt an adaptive design to gradually narrow down (day by day) the set of metrics to be investigated. To check whether this application provides a suitable scenario for applying our SMART procedure, note that (a) it is reasonable to assume that the expected value of the ATE for a given metric is fixed over time, and (b) the ATEs, which are allowed to vary across testing units, are approximately normally distributed according to the central limit theorem. We conclude that the conditions (fixed $\mu_i$ over different stages and asymptotic normality of the summary statistic) required by our methodology have been fulfilled. 
	
	
	
	We analyze a data set from a multi-stage experiment with a running duration of 22 days. As the data are collected on a daily basis, it is natural to view that the experiment has 22 stages. At every stage, the ATEs are computed for a total of 1702 metrics and standardized as z-scores. We set the null distribution of the ATE statistic as $N(0,1)$. Under the alternative, the ATE statistic at unit $i$ has mean $\mu_i$. We assume that $\mu_i$ is fixed at all stages $j=1, 2, \cdots$, and allow $\mu_i$ to vary across $i$. 
	The sparsity parameter $\hat{\pi}$ is estimated to be $\hat{\pi}\approx0.15$ by applying Jin and Cai's method \citep{JinCai07} to the data collected on Day 1. The pre-specified FDR and MDR levels are $0.05$ and $0.2$, respectively. 
	
	\begin{figure}[h]
		\includegraphics[width=1\textwidth]{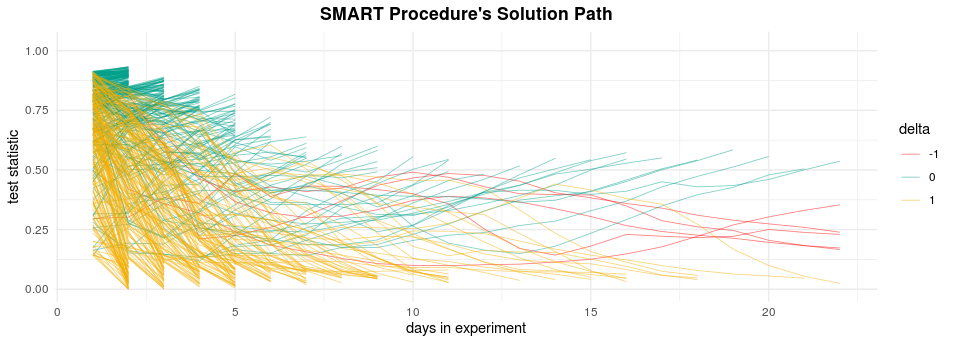}	
		\caption{\small The solution path of our SMART procedure (data source: Snap Inc.): $\delta=1/0/-1$ respectively indicates rejected/accepted/undecided hypotheses.}\label{smart-ab.fig}
	\end{figure}
	
	The solution path of the SMART procedure is shown in Figure \ref{smart-ab.fig}. We can see that by using the adaptive design, definitive decisions on a large number of metrics can be made during earlier periods of the multi-stage experiment. This translates directly to great savings in data storage and pipeline calculations. Specifically, under the traditional A/B testing paradigm, one needs to collect data on 1,702 metrics throughout the 22-day duration. This requires a total of 44,252 measurements (each measurement corresponds to the cost of measuring one metric per day). By contrast, the SMART procedure only requires 3,541  measurements to localize important metrics with effective error rates control. We conclude that the new algorithm promises to save the study costs and greatly speed up the decision making process. Such benefits enable the company to quickly narrow down the list; hence the efforts and resources can be focused on improving a few core metrics that are pivotal for business.

	\subsection{Applications to HTS studies}\label{apply-hts.sec}

	The goal of the HTS study conducted by McKoy et al. (2012) is to identify novel inhibitors of the amyloid beta peptide ($A\beta$), whose aggregation is believed to be a trigger of the Alzheimer's disease. In the study, a total of $p=51,840$ compounds are tested, with three measurements recorded for each compound. We use the observed data set as a pilot data set and simulate observations in later stages to illustrate how to design a multistage sampling and inference procedure for identifying useful compounds. 
	
	We compare the performances of different methods in two ways: (i) the total sample sizes needed to achieve pre-specified error rates, and (ii) the actual error rates achieved for a fixed total sample size. 
	We first obtain z-scores based on the average of the three measurements and then estimate the non-null proportion and null distribution using the method in \cite{JinCai07}. The estimated non-null proportion is $\hat{\pi}\approx 0.0007$, and the estimated null distribution (referred to as the \emph{empirical null distribution}, \cite{Efr04}) is $N(\hat{\mu}_0,\hat{\sigma}_0^2)$ with $\hat{\mu}_0=0.2459,\hat{\sigma}_0=0.6893$. Next, we choose the largest $100\hat{\pi}\%$ of the data and use their average as the signal amplitude $\hat{\mu}=3.194$. The observations in later stages will be generated based on the estimated parameters.  
	
	We set both the FDR and MDR at level $0.1$, apply SMART and record the total sample size. We then apply DS with the recorded sample size by SMART. The results are summarized below. Since DS always eliminates half of the locations at each step, for this particular instance DS requires at least $1.5p$ observations. We can see that with the sample size of $1.5p$, the DS method does not offer proper error rate control. The false discovery proportion (FDP) is much higher than the nominal level.
	
	\begin{table}[ht]
		\centering
		\begin{tabular}{rrrrr}
			\hline
			&Methods & FDP & MDP & Total Observation \\ 
			\hline
			&SMART	& 0.083333 & 0.1081081 & 56926 \\
			\hline
			&DS 		& 0.9971195 & 0 & 77641
		\end{tabular}
	\end{table}
	
	Next, we run DS up to 10 stages. The recorded {FDP} and missed discovery proportion ($\mbox{MDP}({\pmb \delta}) =\frac{\sum_{i=1}^p \theta_i(1-\delta_{i})}{(\sum_{i=1}^p \theta_i)}$) levels are 0.23 and 0.027, respectively. Next we apply SMART by setting the nominal {FDR} and MDR levels as 0.23 and 0.027 and compare the required sample sizes. The results are summarized below. We can see that SMART control the {FDP} and MDP below those of DS. Meanwhile, the required sample size is significantly smaller compared to DS.
	
	\begin{table}[ht]
		\centering
		\begin{tabular}{rrrrr}
			\hline
			&Methods & FDP & MDP & Total Observation \\ 
			\hline
			&DS	 & 0.2340426 & 0.02702703 & 104308 \\ 
			\hline
			&SMART   & 0.1487284 & 0.02162162 & 67850
			
		\end{tabular}

	\end{table}

	\subsection{Application to the Phoenix Deep Survey}\label{apply-image.sec}
	
	In astronomical surveys, a common goal is to separate sparse targets of interest (stars, supernovas, or galaxies) from background noise. We consider a dataset from Phoenix Deep Survey (PDS), a multi-wavelength survey of a region over 2 degrees diameter in the southern constellation Phoenix. The data set is publicly available. 
	Fig. 5(a) shows a telescope image from the PDS. It has $p=616\times 536=330,176$ pixels, among which $1131$ pixels exhibit signal amplitude of at least $2.98$. In practice we monitor the same region for a fixed period of time. After taking high resolution images, it is of interest to narrow down the focus quickly using a sequential testing procedure so that we can use limited computational resources to explore certain regions more closely. The image is converted into gray-scale with signal amplitudes standardized. Fig. 5(b) depicts a contaminated image with simulated Gaussian white noise. 
	
	For the first comparison, we apply SMART by setting both the FPR and MDR at $5\%$. The total number of measurements is recorded as 368,796. As a comparison, we apply DS with the recorded sample size, which is approximately $1.5p$. We can see that SMART control the error rates precisely. The resulting images for SMART and DS are demonstrated in Fig. 5(c) and (d), respectively. We can see SMART produces much sharper images than DS.
	
	\begin{table}[H]
		\centering
		\begin{tabular}{rrrrr}
			\hline
			&Methods & FDP & MDP & Total Observation \\ 
			\hline
			&SMART	& 0.06321335 & 0.05658709 & 368796 \\ 
			\hline
			&DS   & 0.9864338 & 0.00265252 & 495866
		\end{tabular}
	\end{table}
	
	For the second comparison, we first implement DS up to 12 stages. The corresponding FDP and MDP levels are recorded as 0.07 and 0.014.  We then apply SMART by setting the nominal FPR and MDR at these recorded error rates. The corresponding true error rates and required sample sizes of the two methods are summarized below.
	
	\begin{table}[ht]
		\centering
		\begin{tabular}{rrrrr}
			\hline
			&Methods & {FDP} & MDP & Total Observation \\ 
			\hline
			&DS	 & 0.07237937 & 0.01414677 & 670331 \\ 
			\hline
			&SMART   & 0.03192407 & 0.00795756 & 479214 \\
		\end{tabular}
	\end{table}
	
	We can see that SMART controls both the FDP and MDP below the nominal levels (0.07 and 0.014). The required sample size is also much smaller than DS. The resulting images for SMART and DS are shown in Fig. 5(e) and (f), we can see SMART produces slightly sharper images than its competitor DS. 
	
	\begin{figure}[ht]
		\centering
		\begin{subfigure}[b]{.5\textwidth}
			\centering
			\includegraphics[width=.7\linewidth]{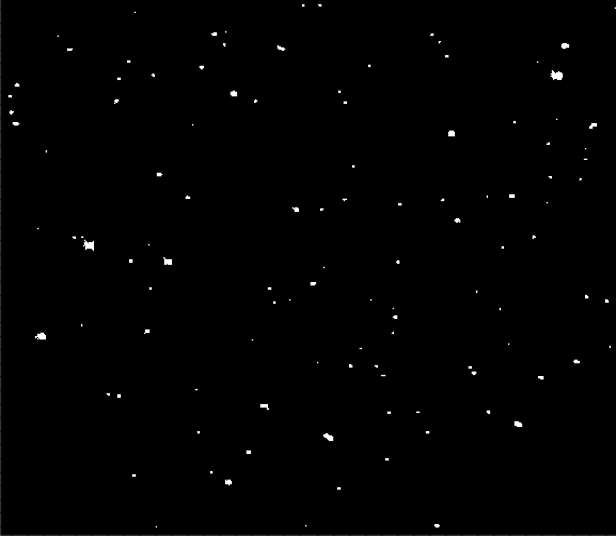}
			\caption{Original Data}
			\label{fig:Origin}
		\end{subfigure}%
		\begin{subfigure}[b]{.5\textwidth}
			\centering
			\includegraphics[width=.7\linewidth]{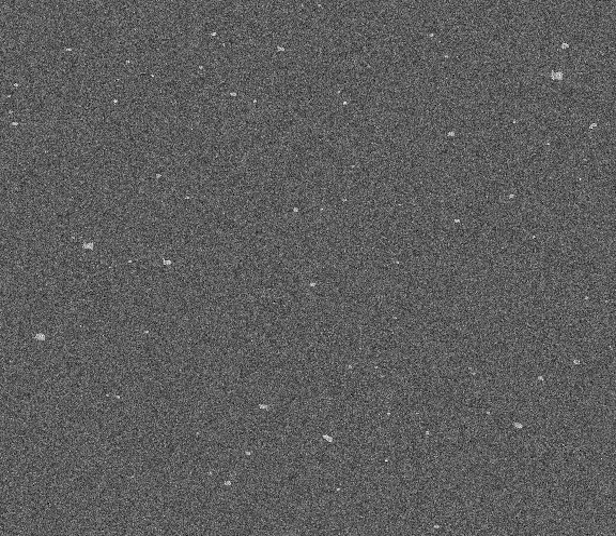}
			\caption{Noisy Observation}
			\label{fig:Noisy}
		\end{subfigure}
		
		\begin{subfigure}[b]{.5\textwidth}
			\centering
			\includegraphics[width=.7\linewidth]{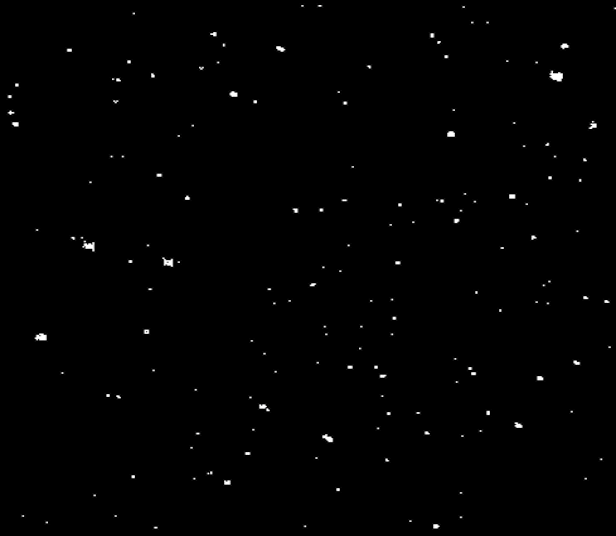}
			\caption{SMART result 1}
			\label{fig:MAT_1}
		\end{subfigure}%
		\begin{subfigure}[b]{.5\textwidth}
			\centering
			\includegraphics[width=.7\linewidth]{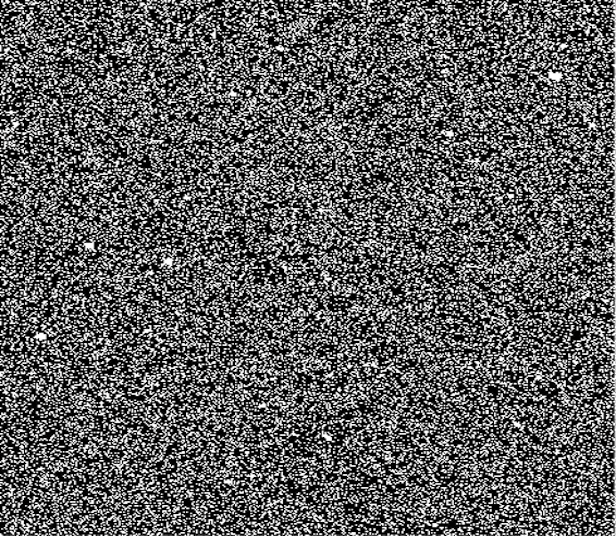}
			\caption{DS result 1}
			\label{fig:DS_1}
		\end{subfigure}
		
		\begin{subfigure}[b]{.5\textwidth}
			\centering
			\includegraphics[width=.7\linewidth]{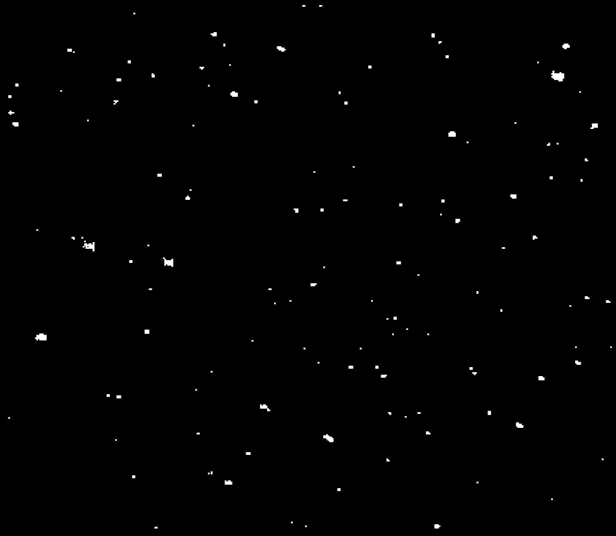}
			\caption{SMART result 2}
			\label{fig:Origin}
		\end{subfigure}%
		\begin{subfigure}[b]{.5\textwidth}
			\centering
			\includegraphics[width=.7\linewidth]{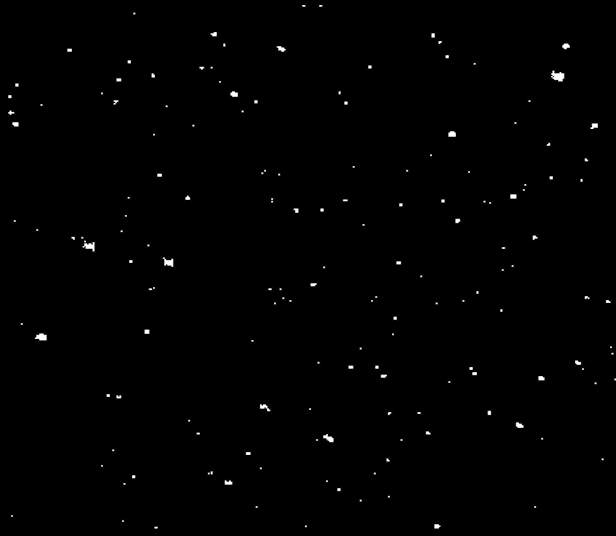}
			\caption{DS result 2}
			\label{fig:Noisy}
		\end{subfigure}
		\caption{SMART and DS comparison. Fig. (a) and (b) show the original radio telescope image and tainted image with white noise, respectively. Fig. (c) and (d) compare SMART and DS when the total number of observations are about the same.  Fig. (f) shows the resulting image when implementing DS for 12 stages, whereas Fig. (e) shows the image produced by SMART when using the recorded error rates from the 12-stage DS.}
		\label{fig:app}
	\end{figure}
	
	\section{Discussion} \label{discuss:sec}

\subsection{Open issues}
	
We mention some potential limitations of the proposed SMART procedure and discuss directions for future research. First, SMART assumes the effect size at a certain coordinate is fixed over time. This is a reasonable assumption for the three applications we considered in this article, where repeated measurements are taken on the same unknown effect sizes over time. However, in application scenarios such as multiple stage clinical trials, the end points or treatment effect sizes may vary from one stage to another. Second, the issue on multiple testing dependence needs much research. Our limited simulation results show that SMART remains to be valid under weak and positive dependences. It would be of great interest to justify such results theoretically. Moreover, the optimality issue under dependence remains unknown. We hope to pursue these directions in future research. 

The accuracy and effectiveness of sparse recovery, which involves both the FPR and MDR control, is a complicated issue that depends on several factors jointly. First, the reliable estimation of the stage-wise FPR/MDR requires a relatively large number of rejections, which in turn requires that the signals cannot be too sparse or too weak. Second, the consistent estimation of the non-null proportion (sparsity level) requires that the detection and discovery boundaries \citep{CaiSun17} must be achieved. The impact of sparsity on the finite sample performance of SMART is investigated in Appendix \ref{app:numspar} of the Supplementary Material. Finally, the quality of density estimators depends on the sample size and smoothness of the underlying function. Our algorithm is designed to capitalize on copious data in ways not possible for procedures intended for moderate amounts of data, and SMART is most useful in {large-scale} testing scenarios where the structural information can be learned from data with good precision. An important future research direction is the development of precise estimation methods, which are instrumental for constructing powerful multi-stage testing procedures. 
	
If we allow multiple samples at each stage and the number of samples are prefixed, then SMART operates essentially in the same way, and the theory on FPR and MDR control for the oracle SMART procedure still holds. However, the problem becomes complicated when the number of samples is allowed to be data-driven. Hence the optimal policy must be cast as a dynamic programming problem. We conjecture that the open-loop feedback control (OLFC) algorithm \citep{WeiHer13} may be incorporated into SMART but more research is needed. However, there are two complications. First, OLFC aims to optimize the estimation accuracy, whereas we aim to minimize the total sensing efforts subject to the constraints on the FPR and MDR. Second, OLFC can only handle two-stage designs and the extension to more stages is non-trivial. 

\subsection{Comparison with the method in \cite{SunCai07}}

We mention a few important distinctions between the methodology presented by \cite{SunCai07} and our adaptation. Firstly, while Sun and Cai's focus is on controlling the FDR, the objective of our SMART algorithm is \emph{signal recovery}. Specifically, Sun and Cai (2007) aims to maximize the power subject to constraints on the FDR, whereas SMART aims to minimize the expected sample sizes (ESS) subject to constraints on both the FPR and MDR. Secondly,  SMART is a multi-stage method that is more intricate than Sun and Cai's one-stage approach. While Sun and Cai' method only conducts signal selection, SMART involves both noise elimination and signal selection. Finally, the optimality theories developed in \cite{SunCai07} and our paper differ substantially. \cite{SunCai07} demonstrated that their procedure has the largest power asymptotically amongst all FDR procedures at the nominal level $\alpha$ for \emph{static data}. However, in the sequential setting with \emph{dynamic data streams}, our primary focus is to optimize sampling efficiency, measured by ESS, subject to the FPR and MDR constraints. This has necessitated the development of new techniques for deriving a new optimality theory, which presents significantly greater theoretical challenges than those encountered by \cite{SunCai07}.

\subsection{Connection to e-values and multi-armed bandits}

In various statistical applications, e-variables naturally arise as sequential likelihood ratios under the null: $\mathcal{L}_{i,j} = f(\pmb X_i^j | \theta_i = 1) / f(\pmb X_i^j | \theta_i = 0)$. One potential solution to our problem is to incorporate e-variables into sequential testing by using the e-BH procedure \citep{wang2022false}. However, this approach comes with limitations that require further investigation in future research. Firstly, the power of e-BH may be significantly lower than that of the SMART approach, which is likely due to the robustness of e-BH, which remains valid under arbitrary dependence. Secondly, e-BH only controls the FDR, while the signal recovery problem requires the control of both the FPR and MDR, with the latter being an unresolved issue under the e-value framework. Lastly, theoretical bounds on signal recovery using e-values require further exploration. 

Next we discuss a related line of research on multi-armed bandits. Both multi-armed bandits and SMART algorithms aim to efficiently identify a small number of arms/units from many potential candidates.
The multi-arm bandits problem has many variations that require different approaches. For instance, should the focus be on ``exploring'' or ``exploiting''? If the goal is to ``explore,'' should the emphasis be on finding the top M arms or identifying arms with mean parameters that exceed some threshold? Is the agent given a fixed budget beforehand, or is the aim to identify ``good'' arms with high probability? While these problems may seem similar, each has its unique nuances. Adapting solutions from one problem to another is a highly non-trivial task. Consequently, we find that solutions developed for fixed budget settings or identifying the top M arms are unsuitable for our problem without significant modification. In what follows, we discuss several important differences between our work and existing literature on the problem.


First, We have tackled the problem from a statistical perspective; the focus has been on the risk quantification of sequential decisions, and particularly the control of the error rates in multiple hypothesis testing. By contrast, the multi-armed bandits formulation solves the problem from an operational perspective. For example, the algorithms in \cite{degenne2019non, chen2015optimal, karnin2013almost, mason2020finding, katz2020true} focus on identifying the top $M$ arms, while the associated error rates or statistical risks in sequential decisions are not investigated along the way. 
		
Second, the sampling and operational schemes in the two lines of works are different. In the multi-armed bandits literature, the algorithm pulls one arm at a time. However in our setup, we collect samples from multiple locations at the same time (stage), meanwhile we narrow down the focus by identifying or eliminating multiple locations at once.

In \cite{HBB2021}, the agent is allowed to draw samples from multiple locations in each round. However, it is important to note that while their objective is to minimize regret by emphasizing ``exploitation,’’ our goal is to efficiently identify non-null locations while providing theoretical guarantees on both the FPR and MDR control. The operation of SMART has necessitated both exploration and exploitation. It is unclear how the algorithm and theory presented in \cite{HBB2021} can be adapted to address our specific problem. Moreover, under the fixed budget regime (\cite{MalNow14a, locatelli2016optimal, mukherjee2017thresholding}), an upper bound on the number of samples is pre-specified, and the theoretical analyses focus on whether it is possible to develop an algorithm that can identify arms above a certain threshold effectively. In contrast, we adopt the setup in sequential hypothesis testing, where the sampling process continues until definitive decisions are reached at all locations. An important feature of our algorithm is that the error rates (FPR and MDR) are guaranteed to fall below the target levels asymptotically in the entire sampling-decision process. 
		 
Third, the error rates considered in the two lines of works are different. The analysis of multi-armed bandits algorithms focuses on (a) the probability of finding the best arm \citep{chen2015optimal}, (b) the probability of finding all ``good'' arms \citep{mason2020finding}, or (c) the probability that all identified arms are ``good'' and none are ``bad'' \citep{locatelli2016optimal, katz2020true}. However, the goals in (b) and (c) can be too stringent for the large-scale setting. It is extremely difficult to identify \emph{all} good arms or providing guarantees that \emph{all} arms are good when thousands or more arms are being considered. 
 Concretely, in large-scale testing problems, researchers are happy as long as most, say 90\% of the good arms (missed discovery rate MDR is controlled at 0.1) can be identified correctly, or 90\% of the arms that we identified are good (false positive rate FPR is controlled at 0.1). In such scenarios the FPR and MDR adopted in our formulation seem to provide more appropriate error rate targets.

Finally The operational characteristics of the algorithms in the two lines of works are different. Under the conventional framework of multi-armed bandits, it is difficult to tailor algorithms according to user-specified error rates/probabilities. By contrast, we develop a convenient user interface where data-driven thresholds can be determined adaptively according to user-specified error rates. Our algorithms aim to control the error rates below the pre-specified levels at all decision points, while minimizing the total study budget (in terms of the total samples to be collected). 

\subsection{Connection to compound sequential decision theory}
The sparse recovery problem  \eqref{optimal-seq-proc.def} is closely related to compound sequential decision theory.  Suppose we are interested in classifying $\theta_i$ based on observed data. Consider the following loss function: 
\begin{equation}\label{loss.func}
	L_{\lambda_1, \lambda_2}(\pmb\theta, \pmb d)=\textstyle
	\lambda_1\left\{\sum_{i=1}^p (1-\theta_i) \delta_{i}\right\}+\lambda_2\left\{\sum_{i=1}^p\theta_i(1-\delta_{i})\right\}+\sum_{i=1}^p N_i,
\end{equation}
where $\lambda_1$ and $\lambda_2$ are the relative costs. The sum of the first two terms in (\ref{loss.func}) corresponds to the total decision errors, and the last term gives the total sampling costs. The optimal solution to this weighted classification problem is the \emph{Bayes sequential procedure} $\pmb d^\pi$ that minimizes the expected loss
\begin{equation}\label{optimal-class-rule.def}
	E\{L_{\lambda_1, \lambda_2}(\pmb\theta, \pmb d^\pi)\}=\textstyle\inf_{\pmb d} E\{ L_{\lambda_1, \lambda_2}(\pmb\theta, \pmb d) \}.
\end{equation}
Our oracle procedure in Theorem \ref{thm:theoopt} is inspired by a classical result in \cite{Ber85} which states that $\pmb d^\pi$ is a thresholding rule based on $T_{OR}^{i,j}$. The intuition is that with appropriately selected tuning parameters $\lambda_1$ and $\lambda_2$ the FPR/MDR problem in \eqref{optimal-seq-proc.def} is equivalent to the problem in \eqref{loss.func}. A proof that the FDR control problem is equivalent to a weighted classification problem can be found in \cite{SunCai07}.

\begin{remark}
	\cite{ouhamma2021online} considered the following loss function 
	\begin{equation}\label{eqou}
		L_T=\sum_{i=1}^{p}a_i \mathbb{I}(\theta_i\neq \delta_i),
	\end{equation}
	where the subscript $T$ denotes the total budget. The loss \eqref{eqou} is similar to \eqref{loss.func} in that the weight $a_i$ can be selected to reflect the relative severity of making a false positive and a false negative decision. However, \eqref{eqou} is designed for the ``fixed budget'' regime, which aims to minimize the total number of errors while adhering to a budget constraint of $T$. Under such a constraint, it may be infeasible to achieve the objective of simultaneously controlling both the FPR and MDR at user-specified levels. In contrast, the loss function \eqref{loss.func} considers the trade-off between the total sampling cost $\sum_{i=1}^p N_i$ and the total decision errors $\lambda_1\left\{\sum_{i=1}^p (1-\theta_i) \delta_{i}\right\}+\lambda_2\left\{\sum_{i=1}^p\theta_i(1-\delta_{i})\right\}$. This explicit trade-off is crucial in deriving the SMART algorithm, and it has not been reflected in Equation \eqref{eqou} due to the fixed budget constraint $T$.
\end{remark}

To gain more insights on the overall structure of the problem, we rewrite \eqref{loss.func} as the sum of stage-wise losses:
\beq\label{loss.func2}
L_{\lambda_1,\lambda_2}(\pmb \theta,\pmb d)= \sum_{j=1}^{N} \left[\sum_{i\in 
	\mathcal S_j}\left\{\lambda_1(1-\theta_i)\delta_{i}+\lambda_2\theta_i(1-\delta_{i})\right\}+  \mbox{Card}(\mathcal{A}_j)\right],
\eeq
where $N=\max\{N_i: 1\leq i\leq p\}$ is the total number of stages and $\mathcal S_j=\{i: N_i=j\}$ is the subset of coordinates at which we stop sampling and make terminal decisions at stage $j$; the remaining locations will become the active set for the next stage $\mathcal A_{j+1}=\mathcal A_j\setminus \mathcal S_j$, on which new observations are taken. We then proceed to make further decisions on $\mathcal A_{j+1}$. The process will be continued until the active set becomes empty.

This view from the simultaneous testing perspective motivates us to employ the idea of \emph{ranking and thresholding} in designing the SMART procedure in Algorithm 1. Ranking and selection has been widely used in the multiple testing literature. For example, the Benjamini-Hochberg procedure \citep{BenHoc95} first orders all $p$-values from the smallest to largest, and then uses a step-up method to choose a cutoff along the $p$-value ranking. 

\section{Derivation of thresholds}\label{derivation-thresholds.subsec}
In this section we provide details about the derivations of the approximation formulae in \eqref{approx.formula}
We need the following assumption in our derivation. The assumption has been commonly adopted in the literature on SPRT (e.g. \cite{Ber85, Sie85}).

\begin{assumption}\label{assumption-zi}
	Consider $Z_{i, 1}$ defined in Assumption 1. For all $i$, we have $\mathbb{P}_{\theta_i}(Z_{i,1}=0)<1$, $\mathbb{P}_{\theta_i}(|Z_{i,1}|<\infty)=1$, $\mathbb{P}_{\theta_i}(Z_{i,1}<0)>0$, and $\mathbb{P}_{\theta_i}(Z_{i,1}>0)>0$. Moreover, $M_{\theta_i}(t)=\mathbb{E}_{\theta_i}[e^{tZ_{i,1}}]$ exists for all $t$.
\end{assumption}

We start our derivation by noting that $T_{OR}^{i, j}$ is a monotone function of the likelihood ratio statistic $\mathcal{L}_{i,j}$:
\beq\label{TOR-SPR}
T_{OR}^{i,j} = \mathbb{P}\left(\theta_i=0|\pmb {X}_i^j\right) =1/\left(1+\frac{\pi}{1-\pi}\mathcal L_{i,j}\right).
\eeq
Hence $\pmb d^\pi(t_l, t_u)$ can be expressed as a thresholding rule based on $\mathcal{L}_{i,j}$:
\begin{eqnarray*}\label{SPRT.proc}
	& \mbox{stops sampling for unit $i$ at time $N_i=\min\left\{j\geq 1: \mathcal{L}_{i, j}\leq A \;\mbox{or} \; \mathcal{L}_{i,j} \geq B\right\}$,} \nonumber\\ & \mbox{deciding $\delta_{i, N_i}=0$ if $\mathcal{L}_{i, j}\leq A$ and $\delta_{i, N_i}=1$ if $\mathcal{L}_{i, j}\geq B$.}
\end{eqnarray*}
We first solve $(A,B)$ for a given pair $(\alpha,\gamma)$, then transform $(A, B)$ to $(t_l, t_u)$. The technique used in our derivation is similar to the classical ideas when deriving the upper and lower thresholds for SPRT. Since all testing units operate independently and have the same thresholds, it is sufficient to focus on the operation of SPRT on a generic testing unit. Hence, for simplicity, we drop index $i$ and denote $\mathcal{L}_{i,N_i}$, $\theta_i$ and $Z_{i,k}$ as $\mathcal{L}_{N}$, $\theta$ and $Z_{\cdot, k}$, respectively.

Under the random mixture model, the FPR and MDR of the SPRT with thresholds $(A, B)$ can be calculated as 
$$
\text{FPR}= \frac{(1-\pi)\mathbb{P}\left(\mathcal{L}_{N}>B|\theta=0\right)}{\mathbb{P}\left(\mathcal{L}_{N}>B\right)},\;
\text{MDR}= \mathbb{P}\left(\mathcal{L}_{N}<A|\theta=1\right).
$$
Let $S_{N} = \sum_{k=1}^{N} Z_{\cdot,k} =\log \mathcal{L}_{N}$. Denote $a=\log A$ and $b= \log B.$
Under Assumption 2, $\mathbb{P}_{\theta}(N<\infty)=1$ and all moment of $N$ exist. There exists a unique nonzero number $t_{\theta}$ for which $M_{\theta}(t_{\theta}) = 1$ (\cite{Ber85}).
This fundamental identity then implies
\begin{eqnarray*}
	1 &=& \mathbb{E}_{\theta}\left\{\exp(t_{\theta}S_{N})M_{\theta}(t_\theta)^{-N}\right\} \\ &= & \mathbb{E}_{\theta}\left\{\exp(t_{\theta}S_{N})\right\}\\
	&\approx & \exp (t_\theta a)\mathbb{P}_\theta\left(S_{N} \leq a\right) + \exp(t_\theta b)\mathbb{P}_{\theta}\left(S_{N} \geq b\right).
\end{eqnarray*}
In the above approximation, we ignore the overshoots and pretend that $S_{N}$ hits the boundaries $a$ and $b$ exactly. In this idealized situation, $S_{N}$ has a two-point distribution $\mathbb{P}_{\theta}^*$:
\begin{eqnarray*}
	\mathbb{P}_{\theta}^*(S_{N}=a)& = & \mathbb{P}_{\theta}(S_{N} \leq a),\\
	\mathbb{P}_{\theta}^*(S_{N}=b)& = & \mathbb{P}_{\theta}(S_{N} \geq b).
\end{eqnarray*}

Moreover, Assumption 2 implies that
\begin{align*}
	1 &= \mathbb{P}_{\theta}(N <\infty) = \mathbb{P}_{\theta}(\mathcal{L}_{N}\leq A) + \mathbb{P}_{\theta}(\mathcal{L}_{N} \geq B)\\
	&=\mathbb{P}_{\theta}(S_{N} \leq a) + \mathbb{P}_{\theta}(S_{N} \geq b).
\end{align*}
Thus we can solve from the above that
$$\mathbb{P}_\theta(\mathcal{L}_{N} \geq B) = \mathbb{P}_{\theta}(S_{N}\geq b) \approx \frac{-\exp(t_\theta a)}{\exp(t_\theta b) - \exp(t_\theta a)}. $$
According to Assumption 2, $\mathbb{P}_{\theta}(|Z_{.,k}| <\infty) = 1 $ for $\theta=0,1$. Then $t_{\theta=0}=1,t_{\theta=1}=-1$ (P493, \cite{Ber85}).
It follows that
\begin{eqnarray*}
	\text{FPR} & \approx & \frac{\left(1-\pi\right)\frac{1-A}{B-A}}{\mathbb{P}\left(\mathcal{L}_{N} \geq B\right) } = \frac{(1-\pi)\frac{1-A}{B-A}}{(1-\pi)\frac{1-A}{B-A} + \pi\frac{1-{1}/{A}}{{1}/{B} - {1}/{A}}} \\ &= & \frac{1-\pi}{1-\pi+\pi B},
	\\
	\text{MDR} & \approx & 1-\frac{1-\exp(-a)}{\exp(-b)-\exp(-a)} = 1- \frac{1-{1}/{A}}{{1}/{B} - {1}/{A}} \\ & = & \frac{A(B-1)}{B-A}.
\end{eqnarray*}
Setting $\mbox{FPR}=\alpha$, $\mbox{MDR}=\gamma$ and solving for $A$ and $B$, we have 
$$A \approx \frac{({\alpha}^{-1} -1)(1-\pi)\gamma}{({\alpha}^{-1} -1)(1-\pi) - \pi+\pi\gamma},\;B \approx \frac{({\alpha}^{-1}-1)(1-\pi)}{\pi}.$$
The relationship \eqref{TOR-SPR} implies that 
\beq
A = \frac{(1-\pi)(1-t_u)}{\pi t_u},\; B = \frac{(1-\pi)(1-t_l)}{\pi t_l}.
\eeq
Transforming from $\mathcal{L}_{i,j}$ to $T_{OR}^{i,j}$, the corresponding thresholds can be obtained as:
$$
t_{OR}^l = \alpha \quad \mbox{and} \quad 
t_{OR}^u = \frac{\pi\alpha\gamma + 1-\pi-\alpha}{\pi\gamma+1-\pi-\alpha}.
$$
To ensure an effective MDR control, we choose a more stringent upper threshold: 
$$
t_{OR}^u=\frac{1-\pi}{\pi\gamma+1-\pi}\geq \frac{\pi\alpha\gamma + 1-\pi-\alpha}{\pi\gamma+1-\pi-\alpha},\;\forall \alpha\geq0.
$$

	\section{Proofs}\label{Proof.sec}

\subsection{Proof of Theorem~\ref{thm:theoopt}}
\begin{proof}
	
	{\bf Part (a)} 	According to Assumption 1, $\mathbb{P}_{\theta_i}(N_i<\infty)=1$ for all $i$; see \cite{Ber85} for a proof. Since $\mathbb{P}_{\theta_i}(N_i<\infty)=1$ for all $i$, and $p$ is finite, we claim that $\mathbb{P}(\max{N_i}<\infty)=1$, i.e. the oracle procedure has a finite stopping time.
	Denote $\pmb d^{\pi}(t_l, t_u)=\{(N_i, \delta_i): 1\leq i\leq p\}$. Using the definition of $T_{OR}^{i, N_i}$, we have
	$$
	\mathbb{E}\left\{\sum_{i=1}^p (1-\theta_i)\delta_i\right\} = 
	\mathbb{E}_{\pmb X} \mathbb{E}_{\pmb \theta|\pmb X}\left\{\sum_{i=1}^p (1-\theta_i)\delta_i\right\}=\mathbb{E}_{\pmb X}\left(\sum_{i=1}^pT_{OR}^{i, N_i}\delta_i\right).
	$$
	Then the FPR is
	$Q_{OR}(t_l,t_u)={\mathbb{E}\left(\sum_{i=1}^pT_{OR}^{i, N_i}\delta_i\right)}/{\mathbb{E}\left(\sum_{i=1}^p\delta_i\right)}.
	$
	It follows that
	$$
	\mathbb{E}\left[\sum_{i=1}^p\left\{T_{OR}^{i,N_i}-Q_{OR}(t_l,t_u)\right\}\mathbb{I}\left(T_{OR}^{i,N_i}\leq t_l\right)\right]=\mathbb{E}\left[\sum_{i:T_{OR}^{i,N_i}\leq t_l}\left\{T_{OR}^{i,N_i}-Q_{OR}(t_l,t_u)\right\}\right]=0.
	$$
	The above equation implies that $Q_{}(t_l,t_u)\leq t_l$; otherwise every term on the LHS must be negative, resulting in a contradiction.

	Next we prove that for a fixed $t_u$, $Q_{OR}(t_l,t_u)$ is non-decreasing in $t_l$. Let $Q_{OR}(t_{l,j},t_u)=\alpha_j$ for $j=1,2$. We only need to show that if $t_{l,1}<t_{l,2}$, then $\alpha_1\leq \alpha_2$. 
	Denote $N_{i,1}$ and $N_{i,2}$ the stopping times for location $i$ corresponding to thresholds $(t_{l,1},t_u)$ and $(t_{l,2},t_u)$, respectively. If $t_{l,1}<t_{l,2}$, then it is easy to see that for any particular realization of the experiment, we must have $N_{i,1}\geq N_{i,2}$.  
	
	We shall show that if $t_{l,1}<t_{l,2}$ and $\alpha_1>\alpha_2$, then we will have a contradiction. To see this, note that
	\begin{eqnarray*}
		&&\left(T_{OR}^{i,N_{i,2}}-\alpha_2\right)\mathbb{I}\left(T_{OR}^{i,N_{i,2}}\leq t_{l,2}\right)\nonumber\\
		&=&\left(T_{OR}^{i,N_{i,2}}-\alpha_2\right)\mathbb{I}\left(T_{OR}^{i,N_{i,2}}\leq t_{l,1}\right)+\left(T_{OR}^{i,N_{i,2}}-\alpha_2\right)\mathbb{I}\left(t_{l,1}<T_{OR}^{i,N_{i,2}}\leq t_{l,2}\right)\nonumber\\
		&=&\left (T_{OR}^{i,N_{i,1}}-\alpha_2 \right)\mathbb{I}\left(T_{OR}^{i,N_{i,1}}\leq t_{l,1}\right)+\left(T_{OR}^{i,N_{i,2}}-\alpha_2\right)\mathbb{I}\left(t_{l,1}<T_{OR}^{i,N_{i,2}}\leq t_{l,2}\right)\nonumber\\
		&\geq & \left(T_{OR}^{i,N_{i,1}}-\alpha_1\right)\mathbb{I}\left(T_{OR}^{i,N_{i,1}}\leq t_{l,1}\right)+(\alpha_1-\alpha_2)\mathbb{I}\left(T_{OR}^{i,N_{i,1}}\leq t_{l,1}\right)\nonumber\\
		&&+\left(T_{OR}^{i,N_{i,2}}-\alpha_1\right)\mathbb{I}\left(t_{l,1}<T_{OR}^{i,N_{i,2}}\leq t_{l,2}\right).
	\end{eqnarray*}
	The second equality holds because if $T_{OR}^{i,N_{i,2}}<t_{l,1}$, then we must have $N_{i, 1}=N_{i, 2}$. 
	Taking expectations on both sides, we have
	\begin{eqnarray*}	
		\mathbb{E}\left\{\sum_{i=1}^p\left(T_{OR}^{i,N_{i,2}}-\alpha_2\right)\mathbb{I}\left(T_{OR}^{i,N_{i,2}}\leq t_{l,2}\right)\right\} & = & 0, \quad \mbox{and}  \\
		\mathbb{E}\left\{\sum_{i=1}^p\left(T_{OR}^{i,N_{i,1}}-\alpha_1\right)\mathbb{I}\left(T_{OR}^{i,N_{i,1}}\leq t_{l,1}\right)\right\} & = &  0.
	\end{eqnarray*}
	However, since $\alpha_1>\alpha_2$ and $\alpha_1\leq t_{l,1}$ as shown previously, we must have
	\begin{eqnarray*}
		\mathbb{E}\left\{\sum_{i=1}^p(\alpha_1-\alpha_2)\mathbb{I}\left(T_{OR}^{i,N_{i,1}}\leq t_{l,1}\right)\right\} & >0, & \quad \mbox{and} \\
		\mathbb{E}\left\{\sum_{i=1}^p\left(T_{OR}^{i,N_{i,2}}-\alpha_1\right)\mathbb{I}\left(t_{l,1}<T_{OR}^{i,N_{i,2}}\leq t_{l,2}\right)\right\} & \geq 0. &
	\end{eqnarray*}
	This leads to a contradiction. Therefore, we conclude that $Q_{OR}(t_l,t_u)$ is non-decreasing in $t_l$ for a fixed $t_u$.
	
	Next, we prove that $\tilde Q_{OR}(t_l,t_u)$ is non-increasing in $t_u$ for a fixed $t_l$. By the definition of MDR and similar arguments for the FPR part, we have 
	$$
	\mathbb{E}\left[\sum_{i=1}^p \left(1-T_{OR}^{i,N_i}\right)\left\{\mathbb{I}\left(T_{OR}^{i,N_i}\geq t_u\right)-\tilde Q_{OR}(t_l,t_u)\right\}\right]=0. 
	$$ 
	Since our model has a finite stopping time, naturally we have 
	$$
	\left\{i:T_{OR}^{i,N_i}\geq t_u\right\}\cup\left\{j:T_{OR}^{j,N_j}\leq t_l\right\}=\left\{1,2,3,\cdots,p\right\}.
	$$
	It follows that 
	$$
	\mathbb{E}\left[\textstyle\sum_{\left\{i:T_{OR}^{i,N_i}\geq t_u\right\}}\left(1-T_{OR}^{i,N_i}\right)\left\{1-\tilde Q_{OR}(t_l,t_u)\right\}\right]=\mathbb{E}\left\{\textstyle\sum_{\left\{j:T_{OR}^{j,N_j}\leq t_l\right\}}\left(1-T_{OR}^{j,N_j}\right)\tilde Q_{OR}(t_l,t_u)\right\}.
	$$
	We have
	\beq\label{Q-tl-tu}
	\frac{1-\tilde Q_{OR}(t_l,t_u)}{\tilde Q_{OR}(t_l,t_u)}=\frac{\mathbb{E}\left\{\sum_{\left\{j:T_{OR}^{j,N_j}\leq t_l\right\}}\left(1-T_{OR}^{j,N_j}\right)\right\}}{\mathbb{E}\left\{\sum_{\left\{i:T_{OR}^{i,N_i}\geq t_u\right\}}\left(1-T_{OR}^{i,N_i}\right)\right\}}.
	\eeq
	Consider two thresholds $t_{u,1}>t_{u,2}$. Denote $N_{i,1}$ and $N_{i,2}$ the corresponding stopping times at location $i$. The operation of the thresholding procedure implies that 
	$N_{i, 1}\geq N_{i, 2}$,  $\left\{i:T_{OR}^{i,N_{i,1}}\geq t_{u,1}\right\} \subset \left\{i:T_{OR}^{i,N_{i,2}}\geq t_{u,2}\right\}$, and $\left\{j:T_{OR}^{j,N_{j,2}}\leq t_l\right\}\subset \left\{j:T_{OR}^{j,N_{j,1}}\leq t_l\right\}$. Therefore,
	\begin{align*}
		&\mathbb{E}\left\{\textstyle\sum_{\left\{i: T_{OR}^{i,N_{i,2}}\geq t_{u,2}\right\}}\left(1-T_{OR}^{i,N_{i,2 }}\right)\right\}\\
		=&\mathbb{E}\left\{\textstyle\sum_{\left\{i: T_{OR}^{i,N_{i,2}}\geq t_{u,1}\right\}}\left(1-T_{OR}^{i,N_{i,2}}\right)\right\}+\mathbb{E}\left\{\textstyle\sum_{\left\{i: t_{u,1}>T_{OR}^{i,N_2}\geq t_{u,2}\right\}}\left(1-T_{OR}^{i,N_{i,2}}\right)\right\}\\
		\geq & \mathbb{E}\left\{\textstyle\sum_{\left\{i:T_{OR}^{i,N_{i,1}}\geq t_{u,1}\right\}}\left(1-T_{OR}^{i,N_{i,1}}\right)\right\}.
	\end{align*}
	We have shown that $\left\{j:T_{OR}^{j,N_{j,2}}\leq t_l\right\}\subset \left\{j:T_{OR}^{j,N_{j,1}}\leq t_l\right\}$. Moreover, on the set $\left\{j: T_{OR}^{j,N_{j,2}}\leq t_l\right\}$, we have $N_{i, 1}=N_{i, 2}$. It follows that 
	$$
	\mathbb{E}\textstyle\left\{\sum_{\left\{j:T_{OR}^{j,N_{j,1}}\leq t_l\right\}}\left(1-T_{OR}^{j,N_{j,1}}\right)\right\}\geq \mathbb{E}\textstyle\left\{\sum_{\left\{j:T_{OR}^{j,N_{j,2}}\leq t_l\right\}}\left(1-T_{OR}^{j,N_{j,2}}\right)\right\}.
	$$
	Combining the above results, we have 
	$$
	\frac{\mathbb{E}\textstyle\left\{\sum_{\left\{j:T_{OR}^{j,N_{j,1}}\leq t_l\right\}}\left(1-T_{OR}^{j,N_{j,1}}\right)\right\}}{\mathbb{E}\left\{\textstyle\sum_{\left\{i:T_{OR}^{i,N_{i,1}}\geq t_{u,1}\right\}}\left(1-T_{OR}^{i,N_{i,1}}\right)\right\}}\geq \frac{\mathbb{E}\textstyle\left\{\sum_{\left\{j:T_{OR}^{j,N_{j,2}}\leq t_l\right\}}\left(1-T_{OR}^{j,N_{j,2}}\right)\right\}}{\mathbb{E}\left\{\textstyle\sum_{\left\{i: T_{OR}^{i,N_{i,2}}\geq t_{u,2}\right\}}\left(1-T_{OR}^{i,N_{i,2 }}\right)\right\}}.
	$$
	Hence if $t_{u,1}>t_{u,2}$, then it follows from \eqref{Q-tl-tu} that
	$$
	\frac{1-\tilde Q_{OR}(t_l,t_{u,1})}{\tilde Q_{OR}(t_l,t_{u,1})}\geq\frac{1-\tilde Q_{OR}(t_l,t_{u,2})}{\tilde Q_{OR}(t_l,t_{u,2})}.
	$$
	Therefore $\tilde Q_{OR}(t_l,t_{u,1})\leq \tilde Q_{OR}(t_l,t_{u,2})$. We conclude that $\tilde Q_{OR}(t_l,t_u)$ is non-increasing in $t_u$ for a fixed $t_l$.

	\medskip
	
	{\bf Part (b).} The proof is divided into two parts. The first part describes a process that identifies a unique pair of oracle thresholds $(t_{OR}^l, t_{OR}^u)$. The second part shows that $\pmb d^\pi(t_{OR}^l, t_{OR}^u)$ has the largest power among all eligible procedures. 
	
	\medskip
	\noindent{\bf (1). Oracle thresholds.} Let $Q_{OR}(1, 1)=\bar\alpha$ be the theoretical upper bound corresponding to the FPR when \emph{all} hypotheses are rejected. A prespecified FPR level $\alpha>0$ is called \emph{eligible} if $\alpha< \bar\alpha$. Let $\mathcal R_{\alpha}=\{t_u: Q_{OR}(t_u,t_u) > \alpha\}$. We can see that $\mathcal R_{\alpha}$ is nonempty if $\alpha$ is eligible, since $Q_{OR}(0, 0)=0$ and $Q_{OR}(1, 1)=\bar\alpha$. 
	Consider $t_u \in \mathcal R_{\alpha}$. Note that $Q_{OR}(0, t_u)=0$ for all $t_u$, the following threshold is well defined: 
	\beq\label{t-l-or}
	t_{OR}^l(t_u)=\sup\{t_l:Q_{OR}(t_l, t_u)\leq \alpha\}.
	\eeq
	We claim that $Q\{t_{OR}^l(t_u), t_u\}=\alpha$. 
	
	We prove by contradiction. 
	We first note $Q_{OR}(t_l, t_u)$ is continuous in $t_l$ and $t_u$. To see that, note $X_{ij}$ follows a continuous distribution and $T_{OR}^{i,j}$ is a continuous function of $X_{ij}$. Therefore, the distribution of $T_{OR}^{i,j}$ is also continuous. By definition $\delta_i=\mathbb{I}(T_{OR}^{i,j}\leq t_l)$. Hence $\mathbb{E}(\delta_i)=\mathbb{P}(T_{OR}^{i,j}\leq t_l)$ is a continuous function of $t_l$. Next, the distribution of 
	$T_{OR}^{i,j}\delta_i$ is proportional to the distribution of $T_{OR}^{i,j}$ truncated at $t_l$. Since $T_{OR}^{i,j}$ follows a continuous distribution, it follows that $\mathbb{E}(T_{OR}^{i,j}\delta_i)$ is also continuous in $t_l$. Continuity of $Q_{OR}(t_l, t_u)$ in $t_u$ can be shown similarly.
	Now, according to the continuity of $Q_{OR}(t_l, t_u)$, for every $t_u\in R_\alpha$, we can find $t_l^*(t_u)$ such that $Q\left(t_l^*(t_u), t_u \right)=\alpha$ [since $Q_{OR}(0, t_u)=0$ and $Q_{OR}(t_u, t_u)>\alpha$]. If not the equality does not hold, i.e. we have $$Q\{t_{OR}^l(t_u), t_u\}<\alpha,$$ then the monotonicity of $Q_{OR}(t_l, t_u)$ implies that $t_l^*(t_u)>t_{OR}^l(t_u)$, which contradicts the definition of $t_{OR}^l(t_u)$. The above construction shows that, for every $t_u \in \mathcal R_{\alpha}$, we can always identify a unique $t_{OR}^l(t_u)$ such that $Q_{OR}\left\{t_{OR}^l(t_u), t_u\right\}=\alpha$. 
	
	We say $(\alpha, \gamma)$ constitute an \emph{eligible pair} of prespecified error rates if $\alpha$ is eligible, and for this $\alpha$, $\gamma$ satisfies 
	$$
	0<\gamma < \sup\left\{\tilde{Q}_{OR}\left(t^l_{OR}(t_u),t_u\right): t_u\in R_\alpha \right\}.
	$$
	In the above definition, the eligibility of $(\alpha, \gamma)$ only depends on the model, but not any given $t_u$.  
	Now consider an eligible pair $(\alpha, \gamma)$. The continuity of $\tilde{Q}_u(t_u)\equiv\tilde{Q}\{t^l_{OR}(t_u),t_u\}$ implies that we can find $t_u^*$ such that $\tilde{Q}\left\{t^l_{OR}(t_u^*),t_u^*\right\}=\gamma$. Let 
	$$
	t_{OR}^u=\inf\{t_u \in R_\alpha: \tilde{Q}_{OR}(t_{OR}^l(t_u),t_u)=\gamma\}.
	$$
	The pair of oracle thresholds are thus given by 
	$$
	(t^l_{OR}, t^u_{OR})\equiv \{t^l_{OR}(t_{OR}^u),t^u_{OR}\}.
	$$

	
	

	\medskip
	\noindent{\bf (2). Proof of optimality.} Denote $\pmb{d}_*=(\pmb{N}_*,\pmb{\delta}_*)$ a sequential procedure that satisfies
	$
	\mbox{FPR}(\pmb{d}_*)=\alpha_*\leq \alpha,\;\mbox{MDR}(\pmb{d}_*)=\gamma_*\leq \gamma,
	$
	where $\pmb{N}_*=(N_*^1,\cdots,N_*^p)$ and $\pmb{\delta}_*=(\delta_*^1,\cdots,\delta_*^p)$
	are the corresponding stopping times and decision rules. Denote ESS$(\pmb{d}_*)$ the expected average stopping times. By definition, we have
	$$
	\mathbb{E}\left\{\sum_{i=1}^p(T_{OR}^{i,N_*^i}-\alpha_*)\delta_*^i\right\}=0,\quad 
	\mathbb{E}\left\{\sum_{i=1}^p(1-T_{OR}^{i,N_*^i})(1-\delta_*^i-\gamma_*)\right\}=0.
	$$

Next we present a hypothetical scenario to justify the operational characteristics of all sequential rules and a mechanism aimed at enhancing any suboptimal rule that does not adhere to the desired order. Specifically, we demonstrate that if decisions on any two coordinates are inconsistent with the specified order, those decisions can be uniformly improved by swapping them.

	Suppose we sort $T_{OR}^{i,N_*^i}$ as $$T_{OR}^{(1),N_*^{(1)}}\leq T_{OR}^{(2),N_*^{(2)}}\leq\cdots\leq T_{OR}^{(p),N_*^{(p)}}$$ with their corresponding decisions $\delta_*^{(1)},\delta_*^{(2)},\cdots,\delta_*^{(p)}.$
	If $\pmb{\delta}_*$ does not take the following form of decision rule
	\beq\label{form-dec}
	\mbox{there exists a $k$}, \mbox{ such that } \delta_*^{(i)}=
	\begin{cases}
		1 & i \leq k\\
		0 & k <i \leq p	
	\end{cases},
	\eeq
	then we can always modify $\pmb{\delta_*}$ into such a form with the same ESS and smaller FPR and MDR. Specifically, suppose that there exists $l_1<l_2$ such that $\delta_*^{(l_1)}=0$ and $\delta_*^{(l_2)}=1$, then we swap these two decisions. Such operation can be iterated until the decision rule takes the form as \eqref{form-dec}. 
	Denote the new decision rule by $\pmb{d}_*^\prime=(\pmb{N}_*,\pmb{\delta}_*')$. Since $T_{OR}^{(l_1),N_*^{(l_1)}}\leq T_{OR}^{(l_2),N_*^{(l_2)}}$ in each swapping, we can reduce the FPR and MDR:
	\begin{eqnarray*}
		\alpha_*' & = & \frac{\sum_{i=1}^p\left(T_{OR}^{i,N_*^i}\delta_*'^{i}\right)}{\mathbb{E}\left(\sum_{i=1}^p\delta_*'^i\right)}\leq\frac{\sum_{i=1}^p\left(T_{OR}^{i,N_*^i}\delta_*^{i}\right)}{\mathbb{E}\left(\sum_{i=1}^p\delta_*^i\right)}=\alpha_*, \\
		\gamma_*'&= & \frac{\sum_{i=1}^p\left\{(1-T_{OR}^{i,N_*^i})(1-\delta_*'^i)\right\}}{p\pi}
		\\ & & \leq\frac{\sum_{i=1}^p\left\{(1-T_{OR}^{i,N_*^i})(1-\delta_*^i)\right\}}{p\pi}=\gamma_*.
	\end{eqnarray*}
	Expressing $\pmb{d}_*^\prime$ in the form of \eqref{d-proc}, we can find $t_l^\prime$ and $t_u^\prime$ such that 
	$$
	{\delta}_{i,N_i}=
	\begin{cases}
		1 & T_{OR}^{i,N_i}\leq t_l^\prime\\
		0 & T_{OR}^{i,N_i}\geq t_u^\prime
	\end{cases},
	$$
	where $Q_{OR}(t_l^\prime,t_u^\prime)=\alpha_*',\;\tilde{Q}_{OR}(t_l^\prime, t_u^\prime)=\gamma_*'$.

	Now we claim that $t_l^\prime\leq t_{OR}^l(t_{OR}^u)$ and $t_u^\prime\geq t_{OR}^u$. We prove by contradiction.
	First, if we have $t_{OR}^u>t_u'$, then by the definition of $t_{OR}^u$, we have 
	$$
	\tilde{Q}_{OR}\left(t_{OR}^l(t_u'),t_u'\right)>\gamma,\;Q_{OR}\left(t_{OR}^l(t_u'),t_u'\right)=\alpha.
	$$
	However, we also have
	$
	\tilde{Q}_{OR}\left(t_l',t_u'\right)=\gamma_*'\leq \gamma.
	$
	By the definition of $\tilde{Q}_{OR}(t_l,t_u)$, with the same $t_u$, only a larger $t_l$ could result in a strictly smaller MDR level. Together with the monotonicity of $\tilde{Q}_{OR}(t_l,t_u)$ for a fixed $t_u$, we claim that $t_{OR}^l(t_u')<t_l'$. Since $Q_{OR}(t_l,t_u)$ is non-decreasing in $t_l$ for a fixed $t_u$, we have
	$Q\left(t_{OR}^l(t_u'),t_u'\right)=\alpha$. It follows that $Q_{OR}\left(t_l',t_u'\right)>\alpha,$ contradicting the fact that $Q_{OR}(t_l',t_u')=\alpha_*'\leq \alpha$. Therefore we must have $t_u'\geq t_{OR}^u$.
	
	Next, assume that $t_l'>t_{OR}^l(t_{OR}^u)$. Then by the definition of $t_{OR}^l(t_u)$ and monotonicity of $Q_{OR}(t_l, t_u)$, we have $Q\left(t_l',t_{OR}^u\right)>\alpha$.
	Given the fact that $Q_{OR}(t_l',t_{OR}^u)\leq t_l'$, we must have $\alpha<t_l'$, claiming that $t_l'$ is always bounded below by $\alpha$ regardless of $\alpha_*'$, which leads to a contradiction since we always have $Q_{OR}(t_l',t_u')=\alpha_*'<t_l'$. Hence we must have $t_l'\leq t_{OR}^l(t_{OR}^u)$. Therefore
	$$\text{ESS}(\pmb{d}_*')=\text{ESS}(\pmb{d}_*)\geq \text{ESS}(\pmb{d}_{OR})$$ and the desired result follows. 
	
\end{proof}

\subsection{Proof of Theorem~\ref{threshold.thm}}

\begin{proof}
	The goal is to show that the pair $t_{OR}^l=\alpha$ and  $t_{OR}^u=\frac{1-\pi}{\pi\gamma+1-\pi}$ control the FPR and MDR. The FPR part is straightforward since according to the definition of $T_{OR}^{i, N_i}$, we have
	$$
	\text{FPR}(\tilde{\pmb{d}}_{OR})=\frac{\mathbb{E}\left\{\sum_{i=1}^pT_{OR}^{i,N_i}\mathbb{I}(T_{OR}^{i,N_i}\leq \alpha)\right\}}{\mathbb{E}\left\{\sum_{i=1}^p\mathbb{I}(T_{OR}^{i,N_i}\leq \alpha)\right\}}\leq \frac{\mathbb{E}\left\{\sum_{i=1}^p \alpha \cdot \mathbb{I}(T_{OR}^{i,N_i}\leq \alpha)\right\}}{\mathbb{E}\left\{\sum_{i=1}^p\mathbb{I}(T_{OR}^{i,N_i}\leq \alpha)\right\}}= \alpha. 
	$$
	Similarly for the FDR control, we have
	\begin{eqnarray*}
		\text{FDR}(\tilde{\pmb{d}}_{OR})=\mathbb{E}\left\{\frac{\sum_{i=1}^pT_{OR}^{i,N_i}\mathbb{I}(T_{OR}^{i,N_i}\leq \alpha)}{\sum_{i=1}^p\mathbb{I}(T_{OR}^{i,N_i}\leq \alpha)\vee 1}\right\}\leq \frac{\mathbb{E}\left\{\sum_{i=1}^p \alpha \cdot \mathbb{I}(T_{OR}^{i,N_i}\leq \alpha)\right\}}{\mathbb{E}\left\{\sum_{i=1}^p\mathbb{I}(T_{OR}^{i,N_i}\leq \alpha) \vee 1\right\}} 
		\leq \alpha. 
	\end{eqnarray*}
	
	\begin{remark}
		From the proof we can see that the choice of $t_{OR}^l=\alpha$, derived based on Wald's approximation, can be conservative in practice.
	\end{remark}

	To show the MDR part, we first carry out an analysis of the false negative rate (FNR), which is defined as
	$$
	\text{FNR}(\tilde{\pmb{d}}_{OR})=\frac{\mathbb{E}\left\{\sum_{i=1}^p\theta_i(1-\delta_i)\right\}}{\mathbb{E}\left\{\sum_{i=1}^p(1-\delta_i)\right\}}.
	$$
	According to the operation of $\tilde{\pmb{d}}_{OR}$, the FNR can be further calculated as
	\begin{eqnarray*}
		\text{FNR}(\tilde{\pmb{d}}_{OR}) & = & \frac{\mathbb{E}\left\{\sum_{i=1}^p(1-T_{OR}^{i,N_i})\mathbb{I}(T_{OR}^{i,N_i}\geq t_u)\right\}}{\mathbb{E}\left\{\sum_{i=1}^p\mathbb{I}(T_{OR}^{i,N_i}\geq t_u)\right\}}
		\\ & \leq & 1-t_u=\frac{\pi\gamma}{\pi \gamma+1-\pi}.
	\end{eqnarray*}
	Denote $\tilde{\pmb d}_{OR}^i=(\tilde N_{OR}^i, \tilde \delta_{OR}^i)$. We have shown that $\text{FPR}(\tilde{\pmb{d}}_{OR})\leq \alpha$.  Suppose the actual FPR level is $\tilde\alpha\leq \alpha$. Then 
	$\mathbb{E}\left\{\sum_{i=1}^p(1-\theta_i)\tilde{\delta}_{OR}^{i} \right\} = \tilde\alpha(\sum_{i=1}^p\tilde{\delta}_{OR}^{i})$. It follows that
	\beq\label{tilde-alpha}
	(1-\tilde\alpha)\mathbb{E}\left(\textstyle\sum_{i=1}^p \tilde{\delta}_{OR}^{i}\right)=\mathbb{E}\textstyle\left\{\sum_{i=1}^p\theta_i\tilde{\delta}_{OR}^{i}\right\}.
	\eeq
	
	Meanwhile, our analysis of the FNR shows that  
	\beq\label{FNR-eq}
	\mathbb{E}\left\{\sum_{i=1}^p \theta_i(1-\tilde{\delta}_{OR}^{i})\right\}\leq \frac{\pi\gamma}{\pi\gamma+1-\pi}\cdot \mathbb{E}\left\{\sum_{i=1}^p(1-\tilde{\delta}_{OR}^{i})\right\}.
	\eeq
	Combining \eqref{tilde-alpha} and \eqref{FNR-eq}, we obtain
	$$
	\left(\pi\gamma+1-\pi\right)\left\{p\pi-(1-\tilde\alpha)\mathbb{E}\left(\textstyle\sum_{i=1}^p\tilde{\delta}_{OR}^{i}\right)\right\}\leq \pi\gamma\left\{p-\mathbb{E}\left(\textstyle\sum_{i=1}^p\tilde{\delta}_{OR}^{i}\right) \right\}.
	$$
	It follows that
	$$
	\mathbb{E}\left(\sum_{i=1}^p\tilde{\delta}_{OR}^{i} \right)\geq \frac{p\pi(1-\pi)(1-\gamma)}{-\pi\gamma\tilde\alpha+(1-\pi)(1-\tilde\alpha)}\geq \frac{p\pi(1-\gamma)}{(1-\tilde\alpha)}
	$$
	Using \eqref{tilde-alpha} and  $p\pi=\mathbb{E}(\sum_i\theta_i)$, 
	we have
	$$
	\mathbb{E}(\textstyle\sum_{i=1}^p\theta_i\tilde{\delta}_{OR}^{i}) \geq \mathbb{E}(\sum_{i=1}^p\theta_i)-\gamma\mathbb{E}(\sum_{i=1}^p\theta_i). 
	$$
	Therefore $\mathbb{E}\{\sum_{i=1}^p\theta_i(1-\tilde{\delta}_{OR}^{i})\}\leq \gamma \mathbb{E} (\sum_{i=1}^p\theta_i)$ and the desired result follows.  
\end{proof}

\subsection{Proof of Theorem~\ref{thm:VALID}}\label{proof-thm3.subsec}

\begin{proof}
	
	{ {\bf Part (i). The FPR and FDR control.} To show the validity of SMART for FPR control, we use the idea in \cite{Efr08a, CaiSun09} for group-wise testing. Define stage-wise false positive rate $\text{sFPR}_j$ as
		$$
		\text{sFPR}_j := \frac{\mathbb{E}\left\{\sum_{i\in \mathcal S_j}(1-\theta_i)\delta_{i}\right\}}{\mathbb{E}\sum_{i\in \mathcal S_j} \delta_{i}},
		$$
		where $\text{sFPR}_j$ is the ratio of the expected number of false rejections at stage $j$ over the expected number of all rejections at stage $j$. The first step is to show that SMART controls all stage-wise FDRs at level $\alpha$. The second step is to show that the global FDR is controlled at level $\alpha$ by combining  hypotheses rejected from all stages.  By our definition of $\text{sFPR}_j$, we have 
		$$
		\text{sFPR}_j = \frac{\mathbb{E}\left(\sum_{i=1}^{k_j^s}T_{OR}^{(i),j}\right)}{\mathbb{E}\left(k_j^s\right)} \leq \frac{\mathbb{E}\left(k_j^s \alpha\right)}{\mathbb{E}\left(k_j^s\right)} = \alpha.
		$$
		Therefore SMART controls the sFPR  at level $\alpha$ across all stages.
		
		Next we show that if $\text{sFPR}_j$ is controlled universally at pre-specified levels across all stages, then the global FPR will be controlled at the same level. Let $N_{SM}$ denote the total number of stages that $\pmb d_{SM}$ has. It follows that
		\begin{eqnarray*}
			\text{FPR}(\pmb d_{SM}) &= & \frac{\mathbb{E}\left\{\sum_{j=1}^{N_{SM}}\sum_{i\in \mathcal{S}_j}\left(1-\theta_i\right)\delta^i_{SM}\right\}}{\mathbb{E}\left(\sum_{j=1}^{N_{SM}}\sum_{i\in \mathcal{S}_j}\delta^i_{SM}\right)}
			\\ & \leq  & \frac{\mathbb{E}\left(\sum_{j=1}^{N_{SM}}\alpha\sum_{i\in \mathcal{S}_j}\delta^i_{SM}\right)}{\mathbb{E}\left(\sum_{j=1}^{N_{SM}}\sum_{i\in \mathcal{S}_j}\delta^i_{SM}\right)}=\alpha.
		\end{eqnarray*}

		Consider the quantity $\mathbb{E}\left\{\sum_{j=1}^{N_{SM}}\sum_{i\in \mathcal{S}_j}\left(1-\theta_i\right)\delta^i_{SM}\right\}$. Note that $i\in \mathcal S_j$ indicates that (i) a total of $j$ data points $\pmb x_i^j=(x_i^1, \cdots, x_i^j)$ are eventually collected for the $i$th unit and (ii) a decision for unit $i$ is made at stage $j$. It follows that $\pmb \delta_{SM}^i$ only depends on $\pmb x_i^j$ and can be factored out:
		\begin{eqnarray}
			\mathbb{E}\left\{\sum_{j=1}^{N_{SM}}\sum_{i\in \mathcal{S}_j}\left(1-\theta_i\right)\delta^i_{SM}\right\} & = & \mathbb{E}\left\{\sum_{j=1}^{N_{SM}}\sum_{i\in \mathcal{S}_j}\EE\left[\left(1-\theta_i\right)\delta^i_{SM}|\pmb x_i^j\right]\right\}\nonumber
			\\ & = & \mathbb{E}\left\{\sum_{j=1}^{N_{SM}}\sum_{i\in \mathcal{S}_j}\delta^i_{SM}\EE\left[\left(1-\theta_i\right)|\pmb x_i^j\right]\right\}\nonumber
		\end{eqnarray}
		According to the definition of the oracle statistic, We have
		$$
		\mathbb{E}\left\{\sum_{j=1}^{N_{SM}}\sum_{i\in \mathcal{S}_j}\left(1-\theta_i\right)\delta^i_{SM}\right\}  =  \mathbb{E}\left\{\sum_{j=1}^{N_{SM}}\sum_{i\in \mathcal{S}_j}\delta^i_{SM}T_{OR}^{i,j}\right\}\leq \mathbb{E}\left\{\sum_{j=1}^{N_{SM}}\alpha\sum_{i\in \mathcal{S}_j}\delta^i_{SM}\right\}.
		$$
		The last inequality is due to the operation of SMART, which ensures that at every stage $j$, we always have $
		\sum_{i\in \mathcal{S}_j}T_{OR}^{i, j}\delta^i_{SM}\leq \alpha \sum_{i\in \mathcal{S}_j}\delta^i_{SM}.
		$
		Therefore we have
		\begin{eqnarray*}
			\text{FPR}(\pmb d_{SM})  \leq   \frac{\mathbb{E}\left(\sum_{j=1}^{N_{SM}}\alpha\sum_{i\in \mathcal{S}_j}\delta^i_{SM}\right)}{\mathbb{E}\left(\sum_{j=1}^{N_{SM}}\sum_{i\in \mathcal{S}_j}\delta^i_{SM}\right)}=\alpha.
		\end{eqnarray*}
		
		Similarly, we can prove the global FDR control by jointly evaluating the performance SMART at separate stages 
		\begin{eqnarray*}
			\text{FDR}(\pmb d_{SM}) & = &  \mathbb{E}\left\{\frac{\sum_{j=1}^{N_{SM}}\sum_{i\in \mathcal{S}_j}\left(1-\theta_i\right)\delta^i_{SM}}{(\sum_{j=1}^{N_{SM}}\sum_{i\in \mathcal{S}_j}\delta^i_{SM})\vee 1}\right\}
			\\ & = &   \mathbb{E}\left[\frac {\sum_{j=1}^{N_{SM}}\sum_{i\in \mathcal{S}_j}T_{OR}^{i, j}\delta^i_{SM}}{\left(\sum_{j=1}^{N_{SM}}\sum_{i\in \mathcal{S}_j}\delta^i_{SM}\right)\vee 1}
			\right] \leq  \alpha
		\end{eqnarray*}  
		The last inequality is due to the operation of SMART.
		\medskip
		
		{\bf Part (ii). MDR control.} Unlike the FDR analysis, stage-wise MDR control does not imply global MDR control. We introduce an intermediate quantity, the false non-discovery rate (FNR) and divide the proof into two steps: the first step shows that stage-wise FNR control implies global FNR control; the second step establishes the relationship between the global MDR and global FNR.  } 
	
	Define stage-wise false non-discovery rate $\text{sFNR}_j$ as
	$$
	\text{sFNR}_j := \frac{\mathbb{E}\left\{\sum_{i\in \mathcal S_j} \theta_i(1-\delta_{i})\right\}}{\mathbb{E}\sum_{i\in \mathcal S_j} (1-\delta_{i})},
	$$
	where $\text{sFNR}_j$ is the ratio of the expected number of false non-discoveries over the expected number of all non-discoveries. It follows that
	$$
	\text{sFNR}_j=\frac{\mathbb{E}\left\{\sum_{i=0}^{k_j^e-1}\left(1-T_{OR}^{k_j-i,j}\right)\right\}}{\mathbb{E}\left(k_j^e\right)}\leq \frac{\mathbb{E}\left(k_j^e\frac{\pi\gamma}{\pi\gamma+1-\pi}\right)}{\mathbb{E}\left(k_j^e\right)}=\frac{\pi\gamma}{\pi\gamma+1-\pi}.
	$$
	Therefore SMART controls the  sFNR at level $\frac{\pi\gamma}{\pi\gamma+1-\pi}$ across all stages.
	Next, we show that if $\text{sFNR}_j$ is controlled universally at pre-specified levels across all stages, then the global FNR will be controlled at the same level. 
	\begin{eqnarray*}
		\text{FNR}&= & \frac{\mathbb{E}\left\{\sum_{j=1}^N\sum_{i\in \mathcal{S}_j}\theta_i\left(1-\delta^i_{SM}\right)\right\}}{\mathbb{E}\left\{\sum_{j=1}^N\sum_{i\in \mathcal{S}_j}\left(1-\delta^i_{SM}\right)\right\}} 
		\\
		& \leq & \frac{\mathbb{E}\left\{\sum_{j=1}^N\frac{\pi\gamma}{\pi\gamma+1-\pi}\sum_{i\in \mathcal{S}_j}\left(1-\delta^i_{SM}\right)\right\}}{\mathbb{E}\left\{\sum_{j=1}^N\sum_{i\in \mathcal{S}_j}\left(1-\delta^i_{SM}\right)\right\}}\\
		&=& \frac{\pi\gamma}{\pi\gamma+1-\pi}.
	\end{eqnarray*}
	Finally, according to the proof of Theorem 2 (Appendix), the MDR and FNR satisfy the following relationship
	$$
	\mbox{$\mbox{MDR}\leq \gamma$ if $\text{FNR} \leq \frac{\pi\gamma}{\pi\gamma+1-\pi}$}.
	$$
	We conclude that the MDR is controlled at level $\gamma$, completing the proof. 
	
\end{proof}

\subsection{Proof of Theorem~\ref{limit.thm}}
\begin{proof}
	For a symmetric decision procedure $\pmb{d}$, denote its Type I and Type II errors on unit $i$ by 
	$\alpha^\prime=\mathbb P_{H_{i, 0}}(\mbox{Reject $H_{i, 0}$})$ and $\gamma^\prime=\mathbb P_{H_{i, 1}}(\mbox{Accept $H_{i, 0}$}).
	$
	It can be shown that the corresponding global error rates are given by
	\begin{equation}\label{FPR-T1}
		\text{FPR}(\pmb d)=\frac{(1-\pi)\alpha^\prime}{(1-\pi)\alpha^\prime+\pi(1-\gamma^\prime)} \quad \mbox{and} \quad \text{MDR}(\pmb d)=\gamma^\prime,
	\end{equation}
	respectively. Our result largely follows from the lower bound derived in \cite{MalNow14a} on family-wise error rate (FWER); we only highlight the main steps on how to go from the FWER paradigm to the FPR/MDR paradigm, which essentially involves exploiting the relationship \eqref{FPR-T1}.  
	From Thm. 2.39 in \cite{Sie85}, we have
	\begin{align*}
		\tau_1&\geq \frac{\alpha^\prime\log(\frac{\alpha^\prime}{1-\gamma^\prime})+(1-\alpha^\prime)\log(\frac{1-\alpha^\prime}{\gamma^\prime})}{D(\mathbb{F}_0||\mathbb{F}_1)} \quad \mbox{and}\\
		\tau_2&\geq \frac{(1-\gamma^\prime)\log{(\frac{(1-\gamma^\prime}{\alpha^\prime})}+\gamma^\prime\log{(\frac{\gamma^\prime}{1-\alpha^\prime})}}{D(\mathbb{F}_1||\mathbb{F}_0)},
	\end{align*}
	where $\tau_1$ and $\tau_2$ are the expected stopping times for null and non-null locations, respectively. Furthermore, from \cite{MalNow14a}, we have
	$$
	\tau_1\geq \frac{(1-\alpha^\prime)\log(\gamma^{\prime})^{-1}-\log{2}}{D(\mathbb{F}_0||\mathbb{F}_1)},\;\tau_2\geq \frac{(1-\gamma^\prime)\log(\alpha^{\prime})^{-1}-\log{2}}{D(\mathbb{F}_1||\mathbb{F}_0)}.
	$$
	Using the KL divergence $D_{KL}(F_0,F_1) =\max\left\{D(\mathbb{F}_0|\mathbb{F}_1),D(\mathbb{F}_1|\mathbb{F}_0)\right\}$, the average stopping time of all locations satisfies
	\begin{eqnarray}\label{KL-result}
		\tau & = & \frac{(p-p\pi)\tau_1+p\pi\tau_2}{p}\nonumber
		\\ & \geq & \frac{(1-\pi)(1-\alpha^\prime)\log(\gamma^{\prime})^{-1}+\pi(1-\gamma^\prime)\log(\alpha^{\prime})^{-1}-\log{2}}{D_{KL}(F_0,F_1)}.
	\end{eqnarray}
	We consider two situations. If $\alpha^\prime\leq \gamma^\prime$, then 
	$$
	\tau\geq(1-\pi)(1-\gamma^\prime)\log(\gamma^{\prime})^{-1}+\pi(1-\gamma^\prime)\log(\gamma^{\prime})^{-1}-\log{2}.
	$$
	Note that for $0\leq x\leq 1$, $\{x\log x: x\in (0, 1)\}$ reaches its minimum when $x=e^{-1}$. It follows that $2^{-1}<e^{-1/e}\leq{\gamma^\prime}^{\gamma^\prime}\leq 1$. Therefore 
	$$
	(1-\gamma^\prime)\log(\gamma^\prime)^{-1}\geq \log(2\gamma^\prime)^{-1}.
	$$
	Together with \eqref{KL-result}, we have
	$
	\tau\geq \frac{\log{(4\gamma^\prime)^{-1}}}{D_{KL}(F_0,F_1)}.
	$
	According to our constraint on $\tau$, we conclude that 
	$$
	{\log({4\eta})^{-1}}\geq \tau{D_{KL}(F_0,F_1)}\geq {\log(4\gamma^\prime)^{-1}}.
	$$
	Therefore, $\gamma^\prime\geq \eta$ and $R^*(\pmb{d})\geq \eta$. 
	
	If $\gamma^\prime\leq \alpha^\prime$, then we can similarly show that
	$$
	\tau\geq\frac{(1-\alpha^\prime)\log{\alpha^\prime}^{-1}-\log{2}}{D_{KL}(F_0,F_1)}\geq \frac{\log{\frac{1}{2\alpha^\prime}}-\log{2}}{D_{KL}(F_0,F_1)} \geq \frac{\log{\frac{1}{4\alpha^\prime}}}{D_{KL}(F_0,F_1)}.
	$$
	It follows that 
	$$
	\log({4\eta})^{-1}\geq \tau {D_{KL}(F_0,F_1)} \geq \log({4\alpha^\prime})^{-1},
	$$
	which implies $\alpha^\prime\geq \eta$. Under the assumption that $\pi<\frac{1}{3}$ and $\eta\leq \frac{1}{2}$, we have $\eta\leq \frac{1}{2}\leq \frac{1-2\pi}{1-\pi}$. Consider the function $k(x)=\frac{\left(1-\pi\right)x}{\left(1-\pi\right)x+\pi}$. It is easy to see that $k(x)$ is monotonically increasing in $x$. Hence
	\begin{align*}
		R^*(\pmb d)&\geq\frac{(1-\pi)\alpha^\prime}{(1-\pi)\alpha^\prime+\pi(1-\gamma^\prime)}
		\geq \frac{(1-\pi)\alpha^\prime}{(1-\pi)\alpha^\prime+\pi}\\
		&\geq\frac{(1-\pi)\eta}{(1-\pi)\eta+\pi}\geq\frac{(1-\pi)\eta}{(1-\pi)\frac{1-2\pi}{1-\pi}+\pi}=\eta,
	\end{align*}
	completing the proof.
\end{proof}

\subsection{Proof of Theorem~\ref{optimal.thm}}
\begin{proof}
	We have already shown that when $t_l=\alpha$ and $t_u=\frac{1-\pi}{\pi\gamma+1-\pi}$, the SMART procedure controls the FDR and MDR at level $\alpha$ and $\gamma$, respectively. Let 
	$$\alpha=\gamma=\frac{1}{{\xi}(p)^{1+\epsilon}}.$$ With the choice of $t_l$ and $t_u$ mentioned above, we have
	$$
	\lim_{p\rightarrow \infty}R^*(\pmb d)=\lim_{p\rightarrow \infty}\frac{2}{{\xi}(p)^{1+\epsilon}}=0,
	$$
	which proves the first part of the theorem.

	Next we establish the upper bound. Consider $p$ simultaneous SPRTs with the same threshold $t_l$ and $t_u$. The operation of our SMART procedure uses these thresholds for the moving averages; hence the SPRT approach with the same $t_l$ and $t_u$ will always take more samples. It is sufficient to show that the result holds for simultaneous SPRTs.

	We first use the relationship \eqref{TOR-SPR} to convert thresholds $t_l$ and $t_u$ to the thresholds for SPRTs:
	$$
	A = \frac{(1-\pi)(1-t_u)}{\pi t_u,}, \quad B = \frac{(1-\pi)(1-t_l)}{\pi t_l}.
	$$
	Under our specifications, we further have 
	$$
	A=\frac{\gamma(1-\pi)}{\pi\gamma+1-\pi},\quad B=\frac{(1-\pi)(1-\alpha)}{\pi\alpha}.
	$$
	From \cite{Sie85} (Equations 9-10, P10), we have 
	\begin{eqnarray*}
		\alpha^\prime & \leq  & B^{-1}(1-\gamma^\prime)\leq B^{-1},\\
		\gamma^\prime & \leq & A(1-\alpha^\prime)\leq A. 
	\end{eqnarray*}
	According to Assumption \eqref{eqn:C1}, we have 
	\begin{eqnarray*}
		\mathbb{E}(\log\mathcal{L}_{i,N_i}|\log\mathcal{L}_{i,N_i}<\log A) & \geq & \log A-C_1, \\ \mathbb{E}(\log\mathcal{L}_{i,N_i}|\log\mathcal{L}_{i,N_i}>\log B) & \leq & \log B+C_2
	\end{eqnarray*}
	for some positives constants $C_1$ and $C_2$. Consider $\mathbb{E}_{\theta_i=0}\left(\log\mathcal{L}_{i,N_i}\right)$. Then
	\begin{eqnarray*}
		&& -\mathbb{E}_{\theta_i=0}\left(\log\mathcal{L}_{i,N_i}\right)
		\\ & = & -(1-\alpha)\mathbb{E}_{\theta_i=0}\left(\log\mathcal{L}_{i,N_i}|\log\mathcal{L}_{i,N_i}<\log A\right)-\alpha\mathbb{E}_{\theta_i=0}(\log\mathcal{L}_{i,N_i}|\log\mathcal{L}_{i,N_i}>\log B)\\
		&\leq & (1-\alpha)(\log A^{-1}+C_1)-\alpha\log B \\
		&\leq & (1-\alpha)(\log A^{-1}+C_1)\leq \log A^{-1}+C_1.
	\end{eqnarray*}
	Likewise, we can show that 
	$$
	\mathbb{E}_{\theta_i=1}(\log \mathcal{L}_{i,N_i})\leq \log B +C_2.
	$$

	Let $C_3=(1-\pi)C_1+\pi C_2$. According to Wald's identity (P490, \cite{Ber85}), we have 
	\begin{eqnarray*}
		\mathbb{E}_{\theta_i=1}(N_i) & = & \frac{\mathbb{E}_{\theta_i=1}\left(\log\mathcal{L}_{i,N_i}\right)}{\mathbb{E}_{\theta_i=1}\left(\log\mathcal{L}_{i,1}\right)}=\frac{\mathbb{E}_{\theta_i=1}\left(\log\mathcal{L}_{i,N_i}\right)}{D({F}_1|{F}_0)}, \\
		\mathbb{E}_{\theta_i=0} (N_i) & = & \frac{\mathbb{E}_{\theta_i=0}\left(\log\mathcal{L}_{i,N_i}\right)}{\mathbb{E}_{\theta_i=0}\left(\log\mathcal{L}_{i,1}\right)}=\frac{-\mathbb{E}_{\theta_i=0}\left(\log\mathcal{L}_{i,N_i}\right)}{D({F}_0|{F}_1)}.
	\end{eqnarray*}
	It follows that
	\begin{eqnarray} \label{thm5.eq0}
		\lim_{p\rightarrow \infty}\frac{\tau}{\log {\xi}(p)}& = & \lim_{p\rightarrow \infty}\frac{(1-\pi)\mathbb{E}_{\theta_i=0}(N_i)+\pi\mathbb{E}_{\theta_i=1}(N_i)}{\log {\xi}(p)} \nonumber
		\\
		&\leq & \lim_{p\rightarrow \infty}\frac{(1-\pi)(\log{A}^{-1}+C_1)}{\log {\xi}(p)D({F}_0|{F}_1)}+\lim_{p\rightarrow \infty}\frac{\pi(\log{B}+C_2)}{\log {\xi}(p)D({F}_1|{F}_0)}\nonumber \\
		&= &\lim_{p\rightarrow \infty}\frac{(1-\pi)\log{\frac{\pi\gamma+1-\pi}{\gamma(1-\pi)}}+\pi\log{\frac{(1-\pi)(1-\alpha)}{\pi\alpha}}+C_3}{\log {\xi}(p)\min\left\{(D({F}_0|{F}_1),D({F}_1|{F}_0)\right\}}\\
		&= & \lim_{p\rightarrow \infty}\frac{(1-\pi)\log({\gamma}^{-1})+\pi\log{\frac{(1-\pi)}{\pi\alpha}}+C_3}{\log {\xi}(p)\min\left\{(D({F}_0|{F}_1),D({F}_1|{F}_0)\right\}}\label{thm5.eq1}\\
		&= & \lim_{p\rightarrow \infty}\frac{\log(\alpha^{-1})}{\log {\xi}(p)\min\left\{(D({F}_0|{F}_1),D({F}_1|{F}_0)\right\}}\label{thm5.eq2}\\
		&= & \frac{1+\epsilon}{\min\left\{(D({F}_0|{F}_1),D({F}_1|{F}_0)\right\}}\nonumber.
	\end{eqnarray}
	From \eqref{thm5.eq0} to \eqref{thm5.eq1}, we have used the fact $\alpha$ and $\gamma$ are  error rates converging to zero; hence 
	\begin{eqnarray*}
		\frac{\pi\gamma+1-\pi}{\gamma(1-\pi)} & = & \frac 1\gamma \{1+o(1)\} \\
		\frac{(1-\pi)(1-\alpha)}{\pi\alpha}& = &\frac{(1-\pi)}{\pi\alpha}\{1+o(1)\}.
	\end{eqnarray*}
	Equation \eqref{thm5.eq2} uses the fact that $\alpha=\gamma$.
	The desired result follows.
\end{proof}


\subsection{Proof of Proposition \ref{prop1}}\label{app:error}
Let $\phi(\cdot,\mu,\sigma^2)$ be the density function of $N(\mu,\sigma^2)$. 	Observe that
$$
\dfrac{   \prod_{t=1}^{j}\phi(x_{it},\eta_k,\hat{\sigma}^2)   }{\prod_{t=1}^{j}\phi(x_{it},0,\hat{\sigma}^2)  }=\exp\left\{   \dfrac{1}{2\hat{\sigma}^2}\left[  j\eta_k(2\bar{x}^j_i-\eta_k)   \right]    \right\}
,$$ where $\bar{x}^j_i=\frac{1}{j}\sum_{t=1}^j x_{it}$. We first consider the case $x_{it}\sim N(\mu_{i},\sigma^2)$, 
\cite{howard2021time} proves the following result:
\begin{equation}\label{eq2}
	\mathbb{P}\left(\forall\ j:|\bar{x}^j_i-\mu_i|< V(j,\alpha) \right)>1-\alpha, 
\end{equation}
where
$$V(j,\alpha)=1.7\sqrt{\sigma\dfrac{\log\log (2j)+0.72\log (5.2/\alpha)}{j}}. $$	 
Hence, if $|\eta_{k}-\mu_i|<|\mu_i|/2-\epsilon$ and $\eta_{k}\mu_{i}>0$ then $\exists K$ depending on $\alpha$ such that 
\begin{equation}\label{eq1}
	\mathbb{P}\left(\forall\ j>K:\dfrac{   \prod_{t=1}^{j}\phi(x_{it},\eta_k,\hat{\sigma}^2)   }{\prod_{t=1}^{j}\phi(x_{it},0,\hat{\sigma}^2)  }>\exp (c_\alpha j)\right)>1-\alpha,
\end{equation}
for some $c_\alpha$ depends on $\sigma,\ \hat{\sigma},\ \alpha$. Recall $g(\cdot)=\sum_{s=1}^{m_1}w_s\delta_{\eta_s}(\cdot)$, we have
$$
T^{i,j}_{OR}:=\mathbb{P}(\theta_i=0|\pmb{X}^j_i)=\dfrac{(1-\pi)\prod^j_{t=1}\phi(x_{it},0,\sigma^2) }{(1-\pi)\prod^j_{t=1}\phi(x_{it},0,\sigma^2)+\pi\sum_{s=1}^{m_1} w_s\prod_{t=1}^j\phi(x_{it},\eta_s,\sigma^2) }.
$$ 
Since $w_s>\epsilon$ for some $\epsilon$, and $\mu_{i}=\eta_s$ for some $\eta_s$, by \eqref{eq2} we have 
\begin{equation}\label{eq3}
	\mathbb{P}\left(\forall\ j>K:T^{i,j}_{OR}<\exp (-c_\alpha j)\right)>1-\alpha,
\end{equation}

Let $\hat{g}(\cdot)=\sum_{s=1}^{m_2}\hat{w}_s \delta_{\hat{\eta}_s}(\cdot)$. Note that $\hat{\mu}^j_i$ can be written as
$$
\hat{\mu}^j_i=\sum_{s=1}^{m}p^j_s\hat{\eta}_s,\quad \text{where}\ \ p^j_s=\dfrac{   \hat{w}_s\prod_{t=1}^{j}\phi(x_{it},\hat{\eta}_s,\hat{\sigma}^2)   }{\sum_{k=1}^{m_2}\hat{w}_k\prod_{t=1}^{j}\phi(x_{it},\hat{\eta}_k,\hat{\sigma}^2)}.
$$
Let $\hat{\eta}_s$ be the point in $\{\hat{\eta}_1,\ldots,\hat{\eta}_{m_2}  \}$ that is closest to $\mu_i$ then use similar argument as above, we can show that there exist $K'$ depending on $\alpha$ such that 
$$
\mathbb{P}\left(\forall\ j>K:|1-p_s^j|<\exp(-c_\alpha j)\right)>1-\alpha.
$$
As a result, there exist $K$ such that 
$$
\mathbb{P}\left(\forall\ j>K:|\hat{\mu}^j_i-\mu_i|< |\mu_{i}|/2-\epsilon\right)>1-\alpha.
$$
Recall 
$$
\hat{T}^{i,j}_{OR}:=\dfrac{(1-\hat{\pi})\prod^j_{t=1}\phi(x_{it},0,\hat{\sigma}^2) }{(1-\hat{\pi})\prod^j_{t=1}\phi(x_{it},0,\hat{\sigma}^2)+\hat{\pi}\prod_{t=1}^j\phi(x_{it},\hat{\mu}^j_{i},\hat{\sigma}^2) }.
$$
By \eqref{eq1} we also have there $\exists K$ depending on $\alpha$ such that 
$$
\mathbb{P}\left(\forall\ j>K: \hat{T}^{i,j}_{OR}<\exp (-c_\alpha j)\right)>1-\alpha.
$$
Combine this with \eqref{eq3} we have the desired result. The case for $x_{it}\sim N(0,\sigma^2)$ is similar.

	\section*{Acknowledgement}
	We thank the editor, AE and referees for their thorough and constructive feedback, which significantly contributes to the enhancement of the paper's presentation. W.Sun thanks Prof. Aaditya Ramdas for the helpful reference of \cite{howard2021time} and valuable insights into the problem formulation. B. Gang’s research was supported by NSFC grant 12201123 and STCSM grant 22YF1403000.

	\bibliography{myrefs}{}
	\bibliographystyle{imsart-nameyear}

	\newpage
	\appendix
	


	\begin{center}\LARGE
		Supplementary Numerical Results for ``Sparse Recovery With Multiple Data Streams: A Sequential Adaptive Testing Approach''
	\end{center}
	
	\setcounter{equation}{0}
	
	\setcounter{figure}{0}
	

	\section{The impact of sparsity levels on SMART}\label{app:numspar}
	
	This section conducts simulation studies to investigate the effectiveness of FPR/MDR control at different sparsity levels. The numerical results show that sparsity has substantial impact on the finite-sample performance of the data-driven procedure. 
	
\begin{figure*}[ht]
		\centering
		\includegraphics[scale=0.6]{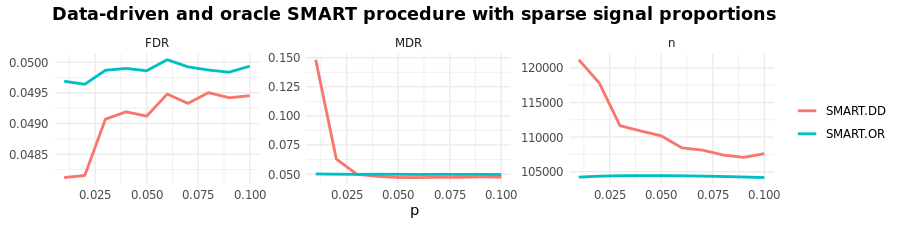}	
		\caption{Error rates and total samples for data-driven and oracle SMART procedures with varying sparse signal proportions.}\label{sparsity-impact.fig}
	\end{figure*}

	Consider the two group mixture model $(1-p)N(0,1)+pN(\mu,1)$ with $\mu=3.5$. 
	We generate $m=100,000$ observations and vary $p$ from 0.01 to 0.1 with step-size 0.01. The FDR and MDR are computed by averaging the results in 100 data sets. The FPR, MDR and total sample sizes are at varied sparsity levels are displayed in \ref{sparsity-impact.fig}. the estimation accuracy is negatively affected when signals become more sparse: SMART has mildly inflated FPR, conservative MDR, and requires a larger sample size to effectively recover the signals. 
	
	Finally, we emphasize that the conclusions from numerical studies are limited and preliminary. The effectiveness of FPR/MDR control is a complicated issue that depends on several factors jointly.
	\begin{description}
		
		\item (i) The reliable estimation of the stage-wise FPR/MDR requires a relatively large number of rejections, which in turn requires that the signals cannot be too sparse or too weak.
		
		\item (ii) The consistent estimation of the non-null proportion requires that the detection and discovery boundaries must be achieved.
		
		\item (iii) The quality of density estimators depends on the sample size and smoothness of the underlying function. 
	\end{description}
	
	We conclude that an important future research direction is the development of precise estimation methods, which is instrumental for constructing powerful multi-stage testing procedures. 
	
	\section{Comparison of realized FPR and FDR levels}\label{app:numfpr}
	
	We present the comparison of the realized FPR and FDR levels for the settings in Section 5.1. We can see that the FDR and FPR for each method (OR.SM and DD.SM) are very similar in all scenarios. However, the FPR and FDR can be very different when signals are rare and weak. See discussions in the next section.
	
	\begin{figure}
		\begin{tabular}{c}
			\includegraphics[width=4.5in, height=2.5in]{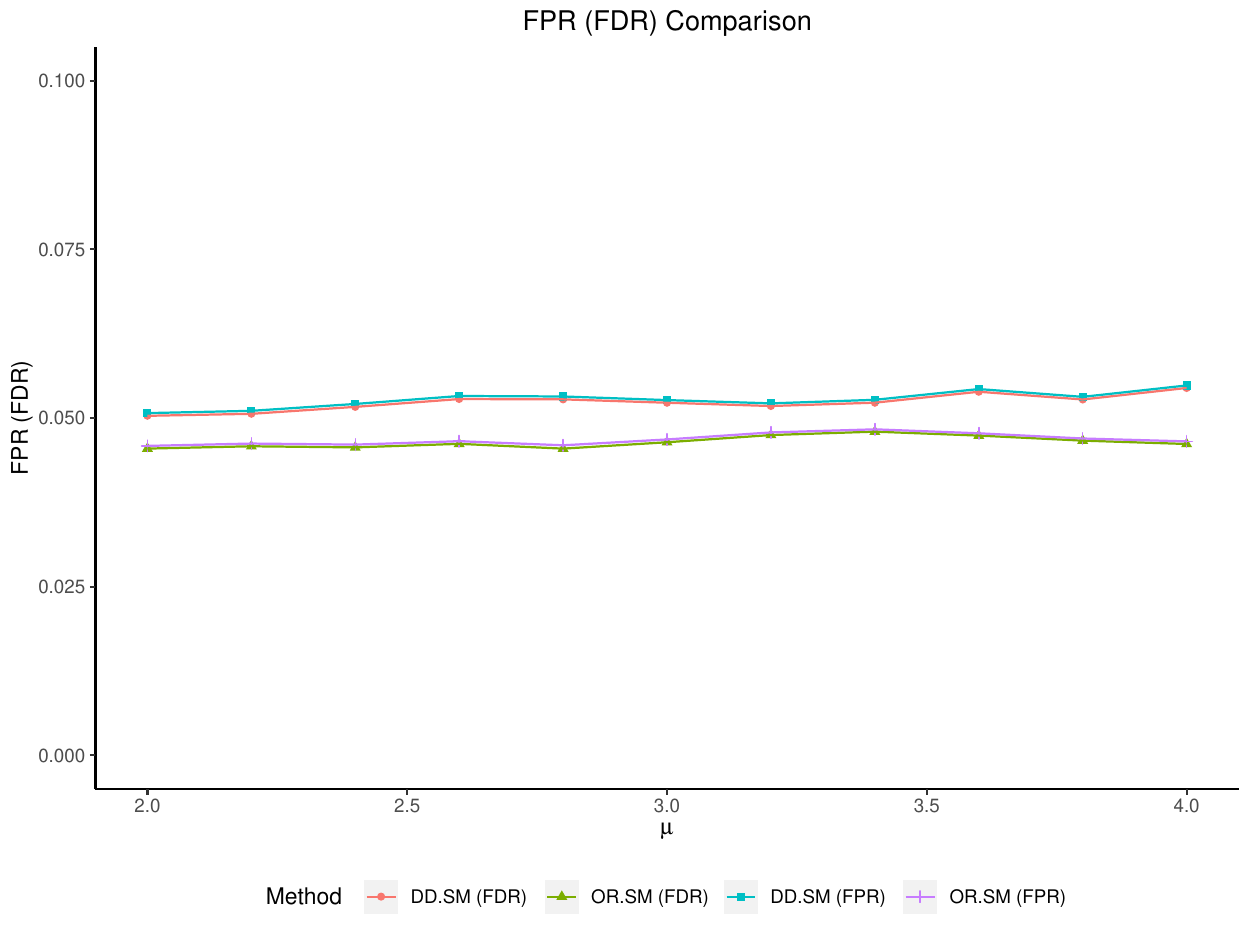}	\\
			\includegraphics[width=4.5in, height=2.5in]{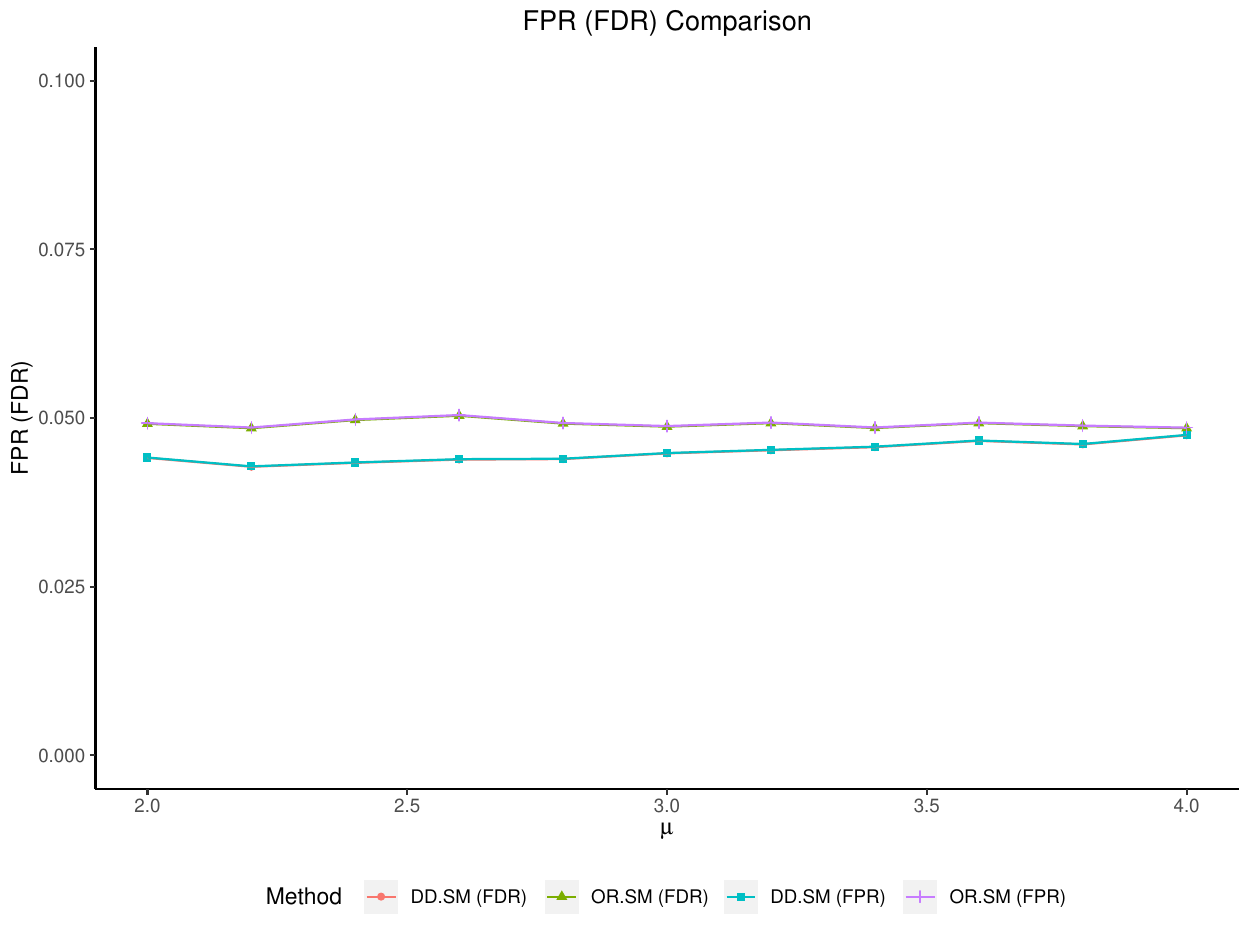}\\
			\includegraphics[width=4.5in, height=2.5in]{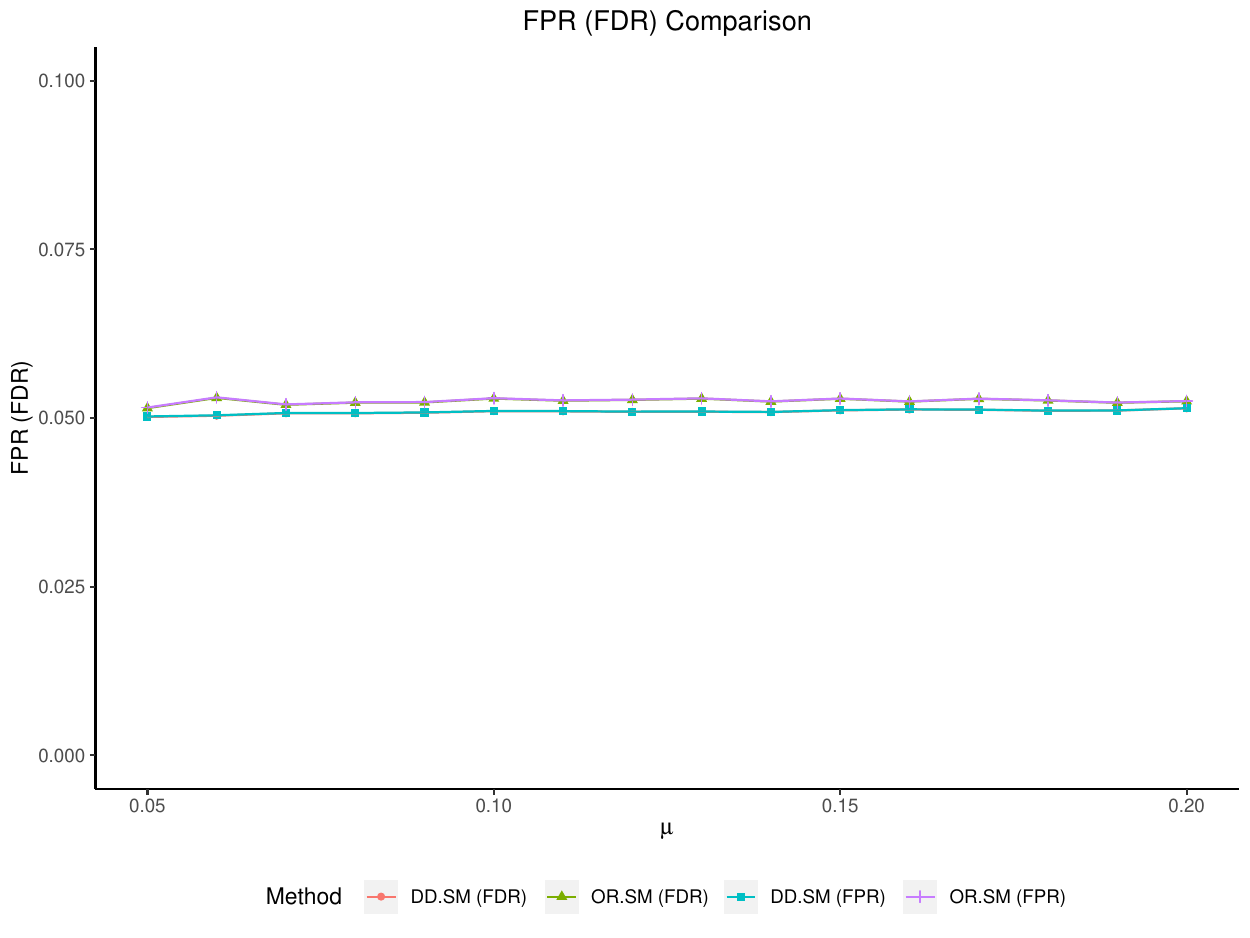}	
		\end{tabular}
		\vspace{-0.5cm} \caption{Dependent tests: Settings 1-3 are described in the main text. SMART controls the error rates at the nominal levels in most settings.}\label{Dependent.fig}
	\end{figure}

	\section{Some further investigations of FPR and FDR}\label{app:numfuther}
	
	We compare the FPR and FDR under the setting where signals are sparse and weak. We simulate $100$ data sets, each with $m=10000$ observations obeying the following two-group mixture model 
	$$
	X_i\sim (1-p)N(0,1)+pN(\mu,1).
	$$ 
	In Setting 1, we fix $p=0.05$ and vary $\mu$ from 0 to 2 with step size 0.1. In Setting 2, we vary  $p$ from 0.01 to 0.2 with step size 0.01 and fix $\mu=1$. The range of 0 to 2 is considered as the ``weak'' signal setting\footnote{Under the calibration $\mu_m=\sqrt{2r\log m}$, the signals are considered to be relatively weak when $r<0.5$.}. 
	
\begin{figure*}[!t]
		\centering
		\includegraphics[scale=0.35]{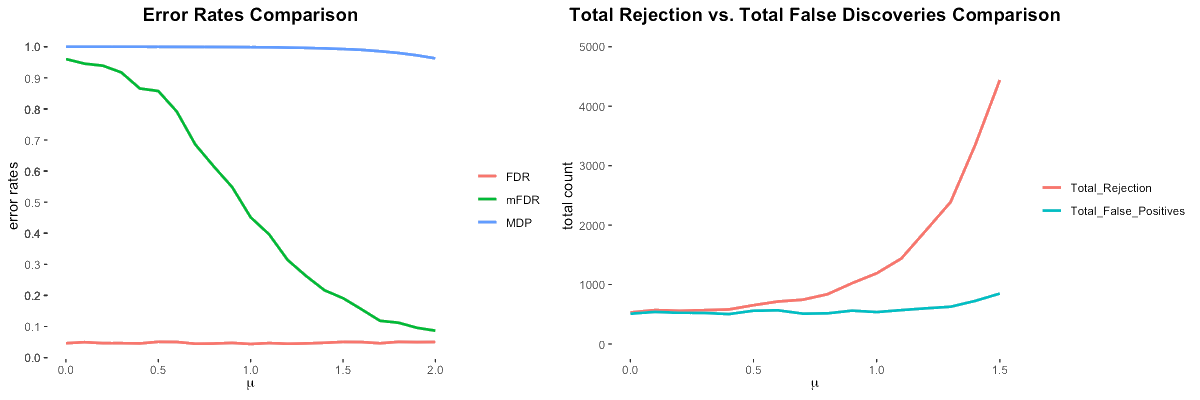}	
		\includegraphics[scale=0.35]{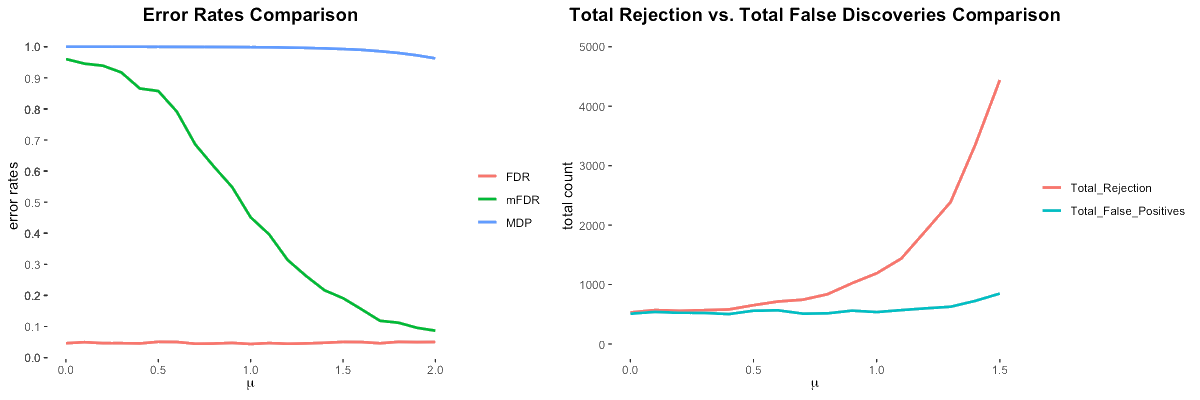}	
		\caption{Error rates comparison between mFDR and FDR with varying signal strength and proportions.}\label{FDR-FPR.fig}
	\end{figure*}

	We apply the BH procedure at $\alpha=0.05$ at varied sparsity levels and inspect the corresponding FPR (or the marginal FDR, mFDR) and FDR levels. Let $R^k$ and $V^k$ be the number of rejections and number of false positives in data set $k$, $k=1, \cdots, 100$. Then the FDR and mFDR are estimated as
	\begin{eqnarray*}
		\widehat{\mbox{FDR}}   =  \frac{1}{100}\sum_{k=1}^{100} \frac{V_k}{R_k\vee 1}; \quad 
		\widehat{\mbox{mFDR}}  =  \frac{\sum_{k=1}^{100}{V_k}}{\sum_{k=1}^{100}R_k}. 
	\end{eqnarray*}
	The simulation results are summarized in Figure \ref{FDR-FPR.fig}, where we also display the total number of rejections ($\sum_{k=1}^{100}{R_k}$) and total number of false positives ($\sum_{k=1}^{100}{V_k}$).  The following observations can be made. 
	\begin{enumerate}
		\item [(a)]  The FDR levels are equal to the nominal level approximately in all settings. The mFDR is very different from FDR when the signals are sparse and weak. 
		
		\item [(b)] From the very left parts of the two panels on the right column, we can see that almost all rejections made by BH in the 100 data sets are false positives when signals are weak and sparse. However, this is not reflected by the FDR since $P(\mbox{FDP}=0)$ is high. By contrast, the mFDR correctly reveals that most of our discoveries are false positives.
	\end{enumerate}

\end{document}